\documentclass[a4paper,aps,preprintnumbers,showpacs,twocolumn,superscriptaddress,nofootinbib]{revtex4-2}
\usepackage{amsmath}
\usepackage{amssymb}
\usepackage{graphicx}
\usepackage{epsfig}
\usepackage{xcolor}
\usepackage{url}
\usepackage{bm}
\usepackage{mathrsfs}
\usepackage[utf8]{inputenc}
\usepackage{hyperref}
\usepackage{enumerate}
\usepackage{amsthm}
\usepackage{bbm}
\usepackage[normalem]{ulem}
\usepackage{upgreek}
\usepackage{tensor}
\usepackage{siunitx}
\usepackage{orcidlink}
\usepackage[caption=false]{subfig}
\usepackage{float}

\newcommand{\beq}{\begin{equation}}
\newcommand{\eeq}{\end{equation}}
\newcommand{\bea}{\begin{eqnarray}}
\newcommand{\eea}{\end{eqnarray}}
\newcommand{\bit}{\begin{itemize}}
\newcommand{\eit}{\end{itemize}}
\newcommand{\ben}{\begin{enumerate}}
\newcommand{\een}{\end{enumerate}}
\newcommand{\nn}{\nonumber}

\newcommand{\coord}{\tau, \sigma, \theta,\varphi}

\newcommand{\bgmetric} {g_{\mu\nu}}

\newcommand{\rstar}{r_{*}}

\definecolor{colour1}{HTML}{0571b0} 
\definecolor{colour2}{HTML}{92c5de} 
\definecolor{colour3}{HTML}{f4a582} 
\definecolor{colour4}{HTML}{ca0020} 
\definecolor{colour5}{HTML}{fe4a49} 

\hypersetup{colorlinks=true, linkcolor=colour1, citecolor=colour1,
filecolor=colour1, urlcolor=colour1}

\def\scri{\mathscr{I}}
\makeatletter
\newcommand{\subalign}[1]{%
  \vcenter{%
    \Let@ \restore@math@cr \default@tag
    \baselineskip\fontdimen10 \scriptfont\tw@
    \advance\baselineskip\fontdimen12 \scriptfont\tw@
    \lineskip\thr@@\fontdimen8 \scriptfont\thr@@
    \lineskiplimit\lineskip
    \ialign{\hfil$\m@th\scriptstyle##$&$\m@th\scriptstyle{}##$\hfil\crcr
      #1\crcr
    }%
  }%
}
\makeatother
\renewcommand{\d}[1]{\,\textnormal{d}#1}

\newcommand{\UCD}{\affiliation{School of Mathematics and Statistics, University College Dublin, Belfield, Dublin 4, Ireland, D04 V1W8}}
          
\theoremstyle{plain}

\begin{document}
\title{Hyperboloidal method for frequency-domain self-force calculations}
\author{Rodrigo Panosso Macedo\,\orcidlink{0000-0003-2942-5080}}
\affiliation{STAG Research Centre, University of Southampton, University Road SO17 1BJ, Southampton, UK}
\affiliation{CENTRA, Departamento de F\'{\i}sica, Instituto Superior T\'ecnico -- IST, Universidade de Lisboa -- UL, Avenida Rovisco Pais 1, 1049 Lisboa, Portugal}
\affiliation{School of Mathematical Sciences, Queen Mary, University of
  London, \\ Mile End Road, London E1 4NS, United Kingdom}
\author{Benjamin Leather\,\orcidlink{0000-0001-6186-7271}}
\UCD
\author{Niels Warburton\,\orcidlink{0000-0003-0914-8645}}
\UCD
\author{Barry Wardell\,\orcidlink{0000-0001-6176-9006}}
\UCD
\author{An\i l Zengino\u{g}lu\,\orcidlink{0000-0001-7896-6268}}
\affiliation{Institute for Physical Science and Technology, University of Maryland, College Park, MD 20742, USA}
\date{\today}
\begin{abstract}
Gravitational self-force theory is the leading approach for modeling gravitational wave emission from small mass-ratio compact binaries.
This method perturbatively expands the metric of the binary in powers of the mass ratio.
The source for the perturbations depends on the orbital configuration, calculational approach, and the order of the perturbative expansion.
These sources fall into three broad classes: (i) distributional, (ii) worldtube, and (iii) unbounded support.
The latter, in particular, is important for emerging second-order (in the mass ratio) calculations.
Traditional frequency domain approaches employ the variation of parameters method and compute the perturbation on standard time slices 
with numerical boundary conditions supplied at finite radius from series expansions of the asymptotic behavior.
This approach has been very successful, but the boundary conditions calculations are tedious, and the approach is not well suited 
to unbounded sources where homogeneous solutions must be computed at all radii. This work develops an alternative approach where 
hyperboloidal slices foliate the spacetime, and compactifying coordinates simplify the boundary treatment.  We implement this approach 
with a multi-domain spectral solver with analytic mesh refinement and use the scalar-field self-force on circular orbits around a 
Schwarzschild black hole as an example problem.  The method works efficiently for all three source classes encountered in self-force 
calculations and has distinct advantages over the traditional approach.  For example, our code efficiently computes the perturbation 
for orbits with extremely large orbital radii ($r_{p}>10^5M$) or modes with very high spherical harmonic mode index ($\ell \ge 100$).
Our results indicate that hyperboloidal methods can play an essential role in self-force calculations.\end{abstract}
\maketitle
\section{Introduction}
Observations of gravitational waves are providing new insights into the population statistics of compact binaries \cite{LIGOScientific:2020kqk} and enabling tests of Einstein's general relativity (GR) in strong-field, dynamical spacetimes \cite{LIGOScientific:2020tif}.
As present detectors are upgraded and new detectors come online, a wider range of systems will appear.

One particularly interesting class of sources are compact binaries where the mass ratio, $\epsilon$, of the smaller to the larger mass is small.
For example, extreme mass-ratio inspirals (EMRIs) with $\epsilon \lesssim 10^{-4}$ are sources for the future space-based LISA detector \cite{amaroseoane2017laser}.
Another example are intermediate mass-ratio inspirals (IMRIs) with $10^{-4} \lesssim \epsilon \lesssim 10^{-1}$ which are sources for both ground- and space-based detectors \cite{Amaro-Seoane:2018gbb}.
Searching for and estimating the parameters of these binaries requires precise theoretical waveform templates to compare against the detector data stream.

The small mass ratio of E/IMRIs lends itself to a perturbative treatment through black hole perturbation theory, and in particular, the gravitational self-force approach \cite{Poisson:2011nh,Barack:2018yvs,Pound:2021qin}.
In this approach one expands the spacetime metric of the binary as ${\sf g_{\mu\nu}} = \bgmetric + \epsilon h^{(1)}_{\mu\nu} + \epsilon^{2} h^{(2)}_{\mu\nu} + \mathcal{O}(\epsilon^{3})$, where $g_{\mu\nu}$ is the metric of primary, and the $h^{(n)}_{\mu\nu}$ are $n$-th order perturbative corrections.
Taking this expansion through second-order in the mass ratio [$\mathcal{O}(\epsilon^2)$] is important for precision tests of GR with EMRIs \cite{Hinderer:2008dm}, and enables efficient modeling of IMRIs \cite{Wardell:2021fyy}.
The equations governing the metric perturbations $h^{(n)}_{\mu\nu}$ are obtained by substituting the expansion above into the Einstein field equations and solving order-by-order along with appropriate regularization schemes to handle the behavior of the metric perturbation near the secondary \cite{Poisson:2011nh,Pound:2012nt,Gralla:2012db}.
These equations can then be solved in the time- or frequency domains, typically after decomposing the perturbation onto a spherical or spheroidal harmonic basis.

The majority of self-force calculations have been carried out in the frequency domain \cite{Diaz-Rivera:2004nim, Warburton:2011hp, Akcay:2013wfa, Akcay:2010dx, Merlin:2014qda, vandeMeent:2016pee, vandeMeent:2017bcc} where computing the perturbation reduces to solving a set of ordinary differential equations (ODEs).
The source for each Fourier mode of the perturbation depends on the orbital configuration, calculational approach, and the order of the perturbative expansion.

These sources fall into three broad classes: (i) distributional, (ii) worldtube, and (iii) unbounded support.
Distributional sources are encountered at first-order (in the mass ratio) when using a point-particle model for the secondary moving on a fixed orbital radius \cite{Diaz-Rivera:2004nim,Akcay:2010dx,Warburton:2010eq}.
Eccentric orbits, which librate between a minimum and maximum radius, lead to worldtube sources \cite{Warburton:2011hp}.
This class of sources also arises when the secondary is modeled using an effective-source approach where the source is confined to a compact worldtube around the worldline \cite{Warburton:2013lea,Wardell:2015ada}.
Finally, sources with unbounded support appear in second-order calculations where a vital ingredient of the second-order source involves products of the first-order metric perturbation \cite{Miller:2020bft}.

The long-established approach for obtaining solutions for each Fourier mode, whether at first or second-order, is through the Green's function method of \emph{variation of parameters}.
To generate the physical solution, one constructs a basis of linearly independent homogeneous solutions that satisfy ingoing boundary conditions at the bifurcation horizon ($r \rightarrow 2M$) and outgoing boundary conditions at spatial infinity ($r \rightarrow \infty$).
The homogeneous solutions are typically computed by either constructing appropriate numerical boundary conditions at finite radii and numerically integrating into the spacetime or by using the semi-analytic Mano-Suzuki-Takasugi (MST) method \cite{Sasaki:2003xr}.
One then integrates these homogeneous solutions against the source term to construct the inhomogeneous solution.

This approach has been instrumental in previous frequency-domain self-force calculations, but it does have some drawbacks.
For the numerical integration method, the boundary conditions are formally straightforward to compute from Frobenius or asymptotic series expansions but deriving them is tedious work.
Furthermore, these series expansions of the boundary conditions must be evaluated in the wave zone to converge at large radii.
For low-frequency modes, which occur for large radius orbits and some modes of eccentric orbit calculations \cite{Akcay:2013wfa, Osburn:2014hoa}, the wave zone moves into the very weak field, which means the integration of the homogeneous solutions accumulates a lot of error from the many steps the numerical integrator must take to extend the solution into the strong field.

The MST method avoids these issues by writing the perturbation as a rapidly convergent series of hypergeometric functions that satisfy the boundary conditions by construction and can be evaluated at any radius.
The challenge with this approach is finding the coefficients in these series expansions.
For low-frequency modes, this can be done very efficiently \cite{Shah:2013uya} (or even analytically, e.g., \cite{Bini:2013zaa,Kavanagh:2015lva,Munna:2020iju}), but for modes with higher frequencies, numerically finding the coefficients and evaluating the many terms in the series can be computationally expensive and often requires the use of arithmetic beyond machine precision \cite{ThroweThesis}.
This makes the MST approach ill-suited to working with sources with unbounded support as the homogeneous solutions then need to be evaluated at all radii to employ the variations of parameters approach.
This class of sources is also challenging for the numerical integration method as the homogeneous solutions may not even be regular near the `opposite' boundary to where the boundary conditions are set.

This work develops a new approach to self-force calculations that resolves these challenges and works efficiently for all three classes of sources described above.
We first transform the field equation for the perturbation to hyperboloidal slices \cite{Zenginoglu:2007jw, Zenginoglu:2011jz, PanossoMacedo:2019npm, Miller:2020bft}.
These slices provide a smooth foliation instead of intersecting at the black hole horizon and spatial infinity. Compactifying the radial coordinate leads to a regular geometry allowing us to place both the future event horizon $\mathcal{H}^{+}$ and future null infinity $\scri^{+}$ on our numerical grid. We do not need to provide data on the grid boundaries because there are no incoming characteristics into the numerical domain. The resulting boundary conditions are behavioral instead of numerical.

This combination of hyperboloidal slicing and compactification has already proven very successful in time-domain black hole perturbation calculations \cite{zenginouglu2008hyperboloidal, zenginouglu2009gravitational, bizon2010saddle, Zenginoglu:2011zz, racz2011numerical, zenginouglu2012caustic, vega2013scalar, harms2014new, thornburg2017scalar, zhang2020object, ripley2021numerical}.
For our frequency-domain implementation, we efficiently solve the perturbation equations using the spectral methods developed in Refs.~\cite{Ansorg:2016ztf, PanossoMacedo:2018hab}.
These techniques, expanded to include the pseudospectrum of perturbations, have been applied successfully to the study of quasinormal modes \cite{jaramillo2021gravitational, destounis2021pseudospectrum, jaramillo2021pseudospectrum, gasperin2021physical, Ripley:2022ypi}.
We use the same coordinates employed in these papers to tackle the self-force problem. 

We demonstrate our approach on a scalar-field toy problem that captures all the key features of self-force calculations while avoiding additional complexity that arises in the gravitational case.
We show that our method works efficiently for distributional, worldtube, and unbounded support sources.
We also demonstrate that it performs well for very large radius circular orbits and very high spherical harmonic mode indices in combination with analytic mesh refinement.
The paper is organized as follows.
In Sec.~\ref{sec:SSF} we give the field equation and mode decomposition on standard $t$-slicing, and discuss the three classes of sources.
In Sec.~\ref{sec:Hyperboloidal} we transform the field equations to hyperboloidal slicing and compactify them.
In Sec.~\ref{sec:NumMeth} we give the details of the spectral numerical scheme. We present our results in Sec.~\ref{sec:results} for all three classes of sources and for large radius orbits. 
In this work we adopt the metric signature $(- + + +)$ and use geometrized units such that $G = c = 1$.


\section{Frequency domain self-force problem: Schwarzschild background}\label{sec:SSF}
The line element for the Schwarzschild solution with mass $M$ in standard coordinates $(t,r,\theta,\varphi)$ is
\beq
	\label{eq:Schwarzschild_metric}
	ds^2 = -f(r) dt^2 + f(r)^{-1} dr^2 + r^2 \left(d\theta^2 + \sin^2\theta \, d\varphi\right),
\eeq
with $f(r) = 1- 2M/r$. 
The frequency-domain field equations in the self-force problem for a field $\phi$ have the generic form \cite{Diaz-Rivera:2004nim,Warburton:2010eq,Warburton:2013lea,Akcay:2010dx,Zenginoglu:2011jz,Wardell:2015ada,Miller:2020bft}
\beq
\label{eq:FreqDomain}
\Delta \phi = {\cal S},
\eeq
where $\Delta$ is a second order derivative operator on the Schwarzschild background.
We discuss the specific form of the operator $\Delta$ and the source ${\cal S}$ for a scalar field example in the following sections.

\subsection{Scalar-field example}
We focus on a scalar self-force (SSF) toy model in this work.
This model captures all the essential features of self-force calculations while avoiding subtle technical issues in the gravitational case, such as gauge choices.
We follow Ref.~\cite{Warburton:2013lea} and consider a particle of mass $\mu$ with scalar charge $q$, moving on a geodesic with coordinates $x^\mu(\uptau)$ where $\uptau$ is the particle's proper time. 
In this toy model, the particle's motion gives rise to a scalar field, which acts back on the scalar charge to generate the SSF.  
The dynamics of the scalar field $\Phi(t,r,\theta,\varphi)$ is dictated by the wave equation in curved spacetime,
\beq
	\label{eq:WaveEquation}
	\square \Phi := \nabla^{\alpha}\nabla_{\alpha}\Phi = - 4\pi \rho,
\eeq
where $\nabla_{\alpha}$ is the covariant derivative with respect to the background Schwarzschild metric and $\rho$ is the particle's scalar density supported on the particle's worldline,
\begin{equation}\label{eq:deltaSource}
	\rho(t,r,\theta,\varphi) = q\int \delta^4(x^\mu - x_p^\mu(\tau)) [-g(x)]^{-1/2},
\end{equation}  
where $g = -r^4 \sin^2\theta$ is the metric determinant.
This equation is equivalent to the spin-$0$ Teukolsky equation \cite{Teukolsky:1973ha}.  
We must impose appropriate outgoing boundary conditions to obtain the retarded field, $\Phi^{\text{ret}}$, from Eq.~(\ref{eq:WaveEquation}). This retarded field, however, is divergent at the particle.
The backreaction on the particle is calculated from a residual field \cite{Quinn:2000wa,Detweiler:2002mi}
\beq
	\Phi^{\mathcal{R}}(x) = \Phi^{\text{ret}}(x) - \Phi^{\mathcal{P}}(x)
\eeq
where $\Phi^{\mathcal{P}}$ is a puncture field defined in a region around the particle that cancels the divergence in the retarded field. The equations of motion are then given by
\beq
	u^{\beta} \nabla_{\beta} (\mu \, u_{\alpha}) = F_{\alpha}(x_{p}) = \lim_{x \rightarrow x_{p}} q \nabla_{\alpha}
	\Phi^{\mathcal{R}}(x).
	\label{eq:EqnOfMotion}
\eeq

For reviews of self-force theory see Refs.~\cite{Poisson:2011nh,Barack:2018yvs}.
For this work, it is sufficient to know that the residual field can be calculated either by first computing the retarded field and then subtracting the singular contribution using the mode-sum approach \cite{Barack:1999wf}, or by reformulating Eq.~\eqref{eq:WaveEquation} to directly solve for the regular field using the effective-source approach \cite{Vega:2007mc,Barack:2007jh}. 
How these two approaches affect the source of Eq.~\ref{eq:WaveEquation} is discussed in Sec.~\ref{sec:Sources} below.

\subsection{The operator $\Delta$}
The operator on the left-hand side of Eq.~\eqref{eq:FreqDomain} follows from decomposing the scalar field into Fourier and spherical harmonic modes 
\beq \label{eq:fieldDecomp}
	\Phi(t,r,\theta,\varphi) = \int \sum_{lm} \phi_{\ell m}(r) Y_{\ell m}(\theta, \varphi) e^{-i\omega t}\, d\omega,
\eeq
where $Y_{\ell m}(\theta, \varphi)$ are the usual spherical harmonics normalized such that $\int Y_{\ell m} Y^*_{\ell' m'} \sin\theta \d\theta = \delta^{l'}_l \delta_m^{m'}$.
Substituting this into Eq.~\eqref{eq:WaveEquation} leads to separatable equations where for each $(\ell,m)$-mode the radial equation is governed by
\beq
	\label{eq:LaplaceOperator}
	\Delta_{\ell m} = \dfrac{d^2}{dr^2} + 2\dfrac{(1-M/r)}{r\, f} \dfrac{d}{dr} + \dfrac{1}{f}\left( \dfrac{\omega^2}{f} - \dfrac{\ell(\ell +1)}{r^2}\right).
\eeq
Appendix \ref{app:operators_spin} discusses the operator $\Delta_{\ell m}$ for the Bardeen-Press-Teukolsky (BPT) \cite{Bardeen:1973xb} and Regge-Wheeler-Zerilli (RWZ) \cite{Martel:2005ir} formulations of black-hole perturbation theory.
\subsection{The source ${\cal S}$}\label{sec:Sources}
We now discuss the most common source types in self-force calculations that appear on the right-hand side of the Eq.~\eqref{eq:FreqDomain}. In our examples, the perturbation is a particle of mass $\mu$ moving on a circular geodesic with radius $r_p$.
Circular geodesics can be parameterized by their energy ${\cal E}$, angular momentum ${\cal L}$, or azimuthal frequency 
$\Omega_\varphi$.
In terms of the orbital radius, they are given explicitly as
\beq
	{\cal E} = \dfrac{f_p}{\sqrt{ 1 - 3M/r_p }}, \, {\cal L} = \dfrac{\sqrt{r_p M}}{\sqrt{1-3M/r_p}}, \, \Omega_\varphi = \sqrt{\dfrac{M}{r_p^3}},
\eeq
where $f_p = f(r_p)$.
The mode frequency becomes $\omega = m \Omega_\varphi$ and the integral in Eq.~\eqref{eq:fieldDecomp} becomes a discrete sum over $m$ modes \cite{Diaz-Rivera:2004nim}.

\subsubsection{Distributional source}
The first case we consider has a distributional source with support on the particle's orbit. 
This case arises when we directly solve for the retarded field with a point-particle source, as is common in black hole perturbation theory.
The regular field can then be computed using the mode-sum approach \cite{Barack:1999wf}.

In our scalar-field example, the source for each mode is given by decomposing Eq.~\eqref{eq:deltaSource} into spherical harmonic and Fourier modes as in Eq.~\eqref{eq:fieldDecomp}.
The field equation takes the form
\beq
	\label{c}
	\Delta_{\ell m} \phi_{\ell m} = {\cal S}^{\rm d}_{\ell m},
\eeq 
where the distributional source is given by \cite{Diaz-Rivera:2004nim,Warburton:2013lea}
\beq
	\label{eq:Source_delta}
	{\cal S}^{\rm d}_{\ell m} = \kappa_{\ell m} \delta(r - r_p), \quad \kappa_{\ell m} = - \dfrac{4\pi q}{ {\cal E}_p r_p^2} \hat c_{\ell m} P^{m}_{\ell}(0),
\eeq
with $\hat c_{\ell m} = \sqrt{\dfrac{2\ell+1}{4\pi} \dfrac{(\ell - m)!}{ (\ell+m)!}}$ arising from the definition
of the defintion of the spherical harmonic function: $Y_{\ell m}(\theta, \varphi) = \hat{c}_{\ell m} P^{m}_{\ell}(\cos\theta)e^{i m \varphi}$, where $P^{m}_{\ell}(\cos\theta)$ is the associated Legendre Polynomial.
Note that solutions to Eq.~\eqref{c} are not unique. We must impose outgoing boundary conditions to obtain the retarded solution as we discuss in Sec.~\ref{sec:BCs}.

\subsubsection{Worldtube sources}\label{sec:worldtube}
The second scenario we consider has extended sources with compact support around the particle, i.e., cases in which the source functions are defined on the compact worldtube $r\in[r_{-}, r_{+}]$, with $r_{-}\le r_p \le r_{+}$.
These types of sources occur in eccentric orbit \cite{Barack:2008ms,Warburton:2011hp} and effective-source \cite{Warburton:2013lea,Wardell:2015ada} calculations.
We demonstrate our approach with the effective-source case, where we directly solve for the residual field.
For each spherical harmonic mode we write
\beq
	\label{eq:residual_field_def}
	\Phi^{\mathcal{R}}_{\ell m} = \phi^{\rm ret}_{\ell m} - \phi^{\mathcal{P}}_{\ell m}.
\eeq 
Applying the operator \eqref{eq:LaplaceOperator} to this equation, and using Eq.~\eqref{c}, we obtain the differential equation for the residual field $\Phi^{\mathcal{R}}_{\ell m}$
\beq
	\Delta_{\ell m} \Phi^{\mathcal{R}}_{\ell m} = {\cal S}^{\rm w}_{\ell m},
\eeq
with ${\cal S}^{\rm w}_{\ell m}$ an effective source defined within a worldtube around the particle
\beq
{\cal S}^{\rm w}_{\ell m} = 
\left\{ 
\begin{array}{ccc} 
0 & {\rm if} & r<r_{-}, \, r>r_{+}, \\
{\cal S}^{\rm d}_{\ell m} - \Delta_{\ell
m}\phi^{\mathcal{P}}_{\ell m} & {\rm if} & r_{-} \leq r \leq r_{+}.
\end{array}
\right.
\label{eq:EffectiveSource_FrequencyDomain}
\eeq
In the scalar toy-model, the distributional term ${\cal S}^{\rm d}_{\ell m}$ is given by Eq.~\eqref{eq:Source_delta} and the corresponding modes of the puncture field are given by~\cite{Warburton:2013lea}
\bea
	&& \Phi^{\mathcal{P}}_{\ell m} = \delta(\omega - m\Omega_\varphi) \bigg[\dfrac{\kappa_{\ell m}}{2} |\Delta r| + \chi_{\ell m} \Delta r + \xi_{\ell m} \bigg],
\eea
with $\Delta r = r - r_p$ and
\bea
\label{eq:chi}
&& \chi_{\ell m} = \dfrac{4 q \, Y_{\ell m}(\pi/2,0) }{(2\ell+1) r_p^2} \sqrt{\dfrac{1-3M/r_p}{f_p} }\, (E-2K), \\
\label{eq:xi}
&& \xi_{\ell m} = \dfrac{8 q \, Y_{\ell m}(\pi/2,0) }{(2\ell+1) r_p} \sqrt{\dfrac{1-3M/r_p}{f_p} }\, K.
\eea
The functions $K$ and $E$ are the complete elliptic integrals of first and second kind, respectively, with arguments $M/(r_p f_p)$. 
By construction of the puncture field, $\Phi^{\mathcal{P}}_{\ell m}{}_{, rr} = \kappa_{\ell m} \delta(r-r_p)$ and so it follows for $r\in[r_{-},r_{+}]$
\beq
\label{eq:EffectiveSource_FrequencyDomain_NoDelta}
	{\cal S}^{\rm w}_{\ell m} = - \Bigg[ 2\dfrac{(1-M/r)}{r\, f} \Phi^{\mathcal{P}}_{\ell m}{}_{,r} + \dfrac{1}{f}\left( \dfrac{\omega^2}{f} - \dfrac{\ell(\ell+1)}{r^2}\right)\Phi^{\mathcal{P}}_{\ell m} \Bigg].
\eeq
\subsubsection{Sources with unbounded support}\label{sec:extended sources}
Sources with unbounded support arise in various recent self-force calculations.
They appear in second-order GSF calculations where a contribution to the source for the second-order metric perturbation comes from the second-order Einstein tensor, which is computed from quadratic combinations of the first-order metric perturbation and its derivatives \cite{Miller:2020bft}.
The two-timescale approach to second-order calculations introduces ``slow-time derivatives'' of the first-order metric perturbation \cite{Miller:2020bft} and the calculation of these also introduces unbounded support source terms.
Further unbounded support sources appear when modeling hyperbolic orbits in the frequency domain \cite{Hopper:2017iyq}.

In this work we will use the slow-time derivative calculation to demonstrate how the hyperboloidal approach applies to sources with unbounded support.
For quasi-circular inspirals, the main computational challenge when calculating slow-time derivatives is to compute \cite{Miller:2020bft}
\beq 
	\label{eq:ret_psi}
	\psi^{\rm ret}_{\ell m} = \partial_{r_{p}} \phi^{\rm ret}_{\ell m}
\eeq
Hereafter we refer to $\psi^{\rm ret}_{\ell m}$ as the ``parametric derivative'' of the perturbation. 
Taking an $r_p$-derivative of Eq.~\eqref{c} and rearranging we find that $\psi^{\rm ret}_{\ell m}$ satisfies the equation
\beq
	\label{eq:Eq_psi}
	\Delta_{\ell m} \psi^{\rm ret}_{\ell m} = {\cal S}^{\rm u}_{\ell m},
\eeq
with the source
\bea
	\label{eq:Source}
	{\cal S}^{\rm u}_{\ell m} &=& \partial_{r_{p}}\kappa_{\ell m} \delta(r - r_p) - \kappa_{\ell m} \delta'(r - r_p)\nn \\
	&&- 2\dfrac{\omega\, \partial_{r_{p}}\omega }{f^2} \phi^{\rm ret}_{\ell m}.
\eea
Note that ${\cal S}^{\rm u}_{\ell m}$ has both Dirac-delta distributions and a term involving retarded field $\phi^{\rm ret}_{\ell m}$ which extends all over the spatial domain (unbounded support).
The distribution terms in this case are also more complicated as they involve both $\delta(r-r_p)$, and $\delta'(r-r_p)$.
Thus, both $\psi^{\rm ret}_{\ell m}$ and $\partial_r \psi^{\rm ret}_{\ell m}$ exhibit discontinuities at the particle's orbit fixed by $\kappa_{\ell m}$ and $\partial_{r_{p}}\kappa_{\ell m}{}$.

\subsection{The boundary conditions}\label{sec:BCs}
The physical boundary conditions are typically specified on $t=\;$constant hypersurfaces that intersect the bifurcation horizon ${\cal B}$ at $r=2M$, and at spatial infinity $i^0$ as $r \rightarrow \infty$ (see thin, dashed lines in Fig.~\ref{fig:Penrose}).
These boundary conditions pick the retarded solution whose energy radiates towards the black hole or to infinity. 
For compact sources the asymptotic form of the boundary conditions is given by
\beq
	\label{eq:BC_frequency}
	\phi^{\rm ret}_{\ell m}(r) \sim \dfrac{ e^{\pm i\omega r^*}}{r} , \quad r^*\rightarrow \pm \infty,
\eeq
where we have introduced the radial tortoise coordinate defined as $d\rstar/dr = f(r)^{-1}$.

For implementation with a numerical scheme, the oscillations along the $t$-slices means that compactification of the radial domain leads to an infinite resolution problem and is therefore avoided within the standard approach \cite{GroschOrszag77, zenginouglu2011hyperboloidal, zenginouglu2021null}.
Instead, the unbounded domain is truncated and the boundary conditions are imposed at a finite radius. 

To find the boundary conditions at a finite distance, one performs a series expansion.
For example, the outer boundary condition towards spatial infinity is often expanded at some large radius $r_{\text{out}}$ in the form
\begin{align}\label{eq:BC_expansion}
	\phi^{\rm ret}_{\ell m}(r_\text{out}) = e^{i\omega r^*}\sum_{k=0}^\infty a_{\ell m k} (\omega r_{\text{out}})^{-k}
\end{align}
The coefficients $a_{lm,k\ge1}$ are determined by substituting the expansion into the homogeneous equation $\Delta_{lm}\phi_{lm} = 0$ and solving the resulting recurrence relation.
For the scalar field, these recurrence relations can be found in, e.g., Appendix A of Ref.~\cite{Warburton:2013lea}.
Computing these relations is tedious work, which becomes substantially more involved for perturbation of Kerr spacetime 
(e.g., Appendix C of Ref.~\cite{Warburton:2010eq}) or for gravitational perturbations \cite{Akcay:2010dx,Osburn:2014hoa}.

For the expansion in Eq.~\eqref{eq:BC_expansion} to converge, we must have $\omega r_{\text{out}} \gg 1$.
This can be problematic when very low-frequency modes occur as the outer boundary must then move out very far.
The unbounded support source given in Eq.~\eqref{eq:Source} falls off sufficiently rapidly that the asymptotic boundary condition is given by just the $r_{p}$-derivative of Eq.~\ref{eq:BC_frequency}
\beq
	\label{eq:BC_slowderivative}
	\psi^{\rm ret}_{\ell m}(r) \sim \pm \dfrac{i \partial_{r_{p}}\omega \, r^* e^{\pm i\omega r^*}}{r}, \quad r^*\rightarrow \pm \infty.
\eeq
Constructing boundary conditions at a finite radius for the unbounded support source is more involved as now the recurrence relation for the coefficients involve coefficients of the expansion of the retarded field, 
$\phi^{\rm ret}_{\ell m}$, that appears in the source -- see Ref.~\cite{Hopper:2012ty} for an example where such boundary conditions are computed.

\section{Hyperboloidal method for self-force in frequency domain}\label{sec:Hyperboloidal}
Hyperboloidal surfaces are spacelike surfaces that behave like a spacetime hyperboloid near null horizons. The term hyperboloidal in the literature typically refers to null infinity \cite{friedrich1983cauchy, frauendiener2004conformal}. We expand the usage of the term to encompass also other null surfaces, such as the black hole horizon or the cosmological horizon. Horizon-penetrating coordinates, such as the original Eddington-Finkelstein or the Painlev\'e-Gullstrand coordinates, are hyperboloidal, which becomes clear when written with respect to the tortoise coordinate that pushes the black hole horizon to negative infinity. Naturally, first numerical implementations of hyperboloidal coordinates in black hole spacetimes also included horizon-penetrating coordinates \cite{zenginouglu2008hyperboloidal, zenginouglu2009gravitational, cruz2010numerical}. Therefore, it makes sense to use the term for both the black hole horizon and null infinity.

\begin{figure}[h]
	\centering
	\includegraphics[width=0.48\textwidth]{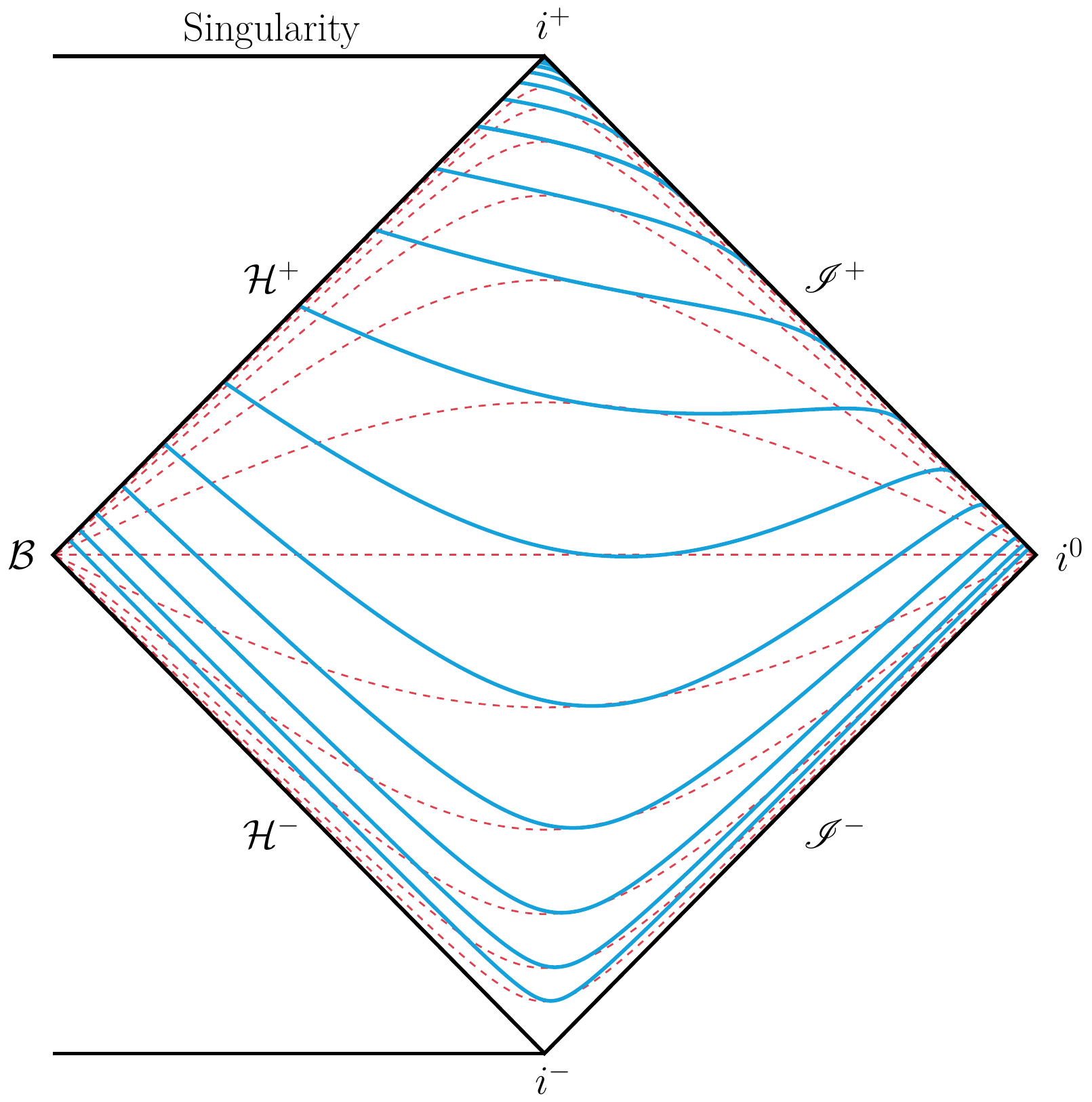}
\caption{Carter-Penrose diagram for the Schwarzschild exterior region. 
Thin, dashed lines depict standard Schwarzschild time surfaces $t=$constant extending between the bifurcation sphere 
$\cal B$ at the horizon $r=2M$ and space-like infinity $i^0$ as $r\rightarrow \infty$. The intersection of these time surfaces 
near $\cal B$ and $i^0$ imply a coordinate singularity. The domain must be truncated and boundary data must be imposed 
near $\cal B$ and $i^0$. Thick, solid lines depict hyperboloidal time surfaces $\tau=$constant extending between the 
black-hole horizon ${\cal H}^+$ at $\sigma=1$ and future null infinity ${\scri^+}$ as $\sigma=0$ given by 
Eq.~\eqref{eq:Hyper_r}. These coordinates provide a smooth foliation on the full exterior domain which means that 
both the horizon and null infinity can be included in the computational domain. No external boundary conditions are needed 
to study perturbations.}
\label{fig:Penrose}
\end{figure}

The similarity of hyperboloidal coordinates near null infinity and near the black hole horizon is also visible when viewed in a Penrose diagram (see Fig.~\ref{fig:Penrose} and \cite{Zenginoglu:2011jz}). Hyperboloidal coordinates foliate the (future) event horizon ${\cal H}^+$ instead of intersecting at the bifurcation sphere ${\cal B}$ at $r=2M$, and they foliate (future) null infinity $\scri^+$ instead of intersecting at spatial infinity $i^0$ when $r\rightarrow \infty$. 
Consequently, we can include the black hole horizon and null infinity in our computational domain, 
which removes the need for the complicated boundary conditions described in the previous section.
Another important advantage of the method is that the construction only depends on the background spacetime.
In contrast, boundary conditions must be computed separately for each problem with different sources 
or different formulations of the perturbations.

Among the many ways to construct hyperboloidal surfaces, a convenient and common
method is to fix the coordinate location of null infinity (scri) on the grid \cite{Zenginoglu:2007jw}.
Scri-fixing has the essential advantage of leaving the timelike Killing field
of stationary black holes invariant. Consequently, coefficients of equations describing
black hole perturbations are time-independent, and the event horizon and null infinity
are fixed at the numerical boundaries.
The scri-fixing method of Ref.~\cite{Zenginoglu:2007jw} to construct
hyperboloidal coordinates consists of three steps:
\begin{enumerate}
\item Introduce a time coordinate that respects the
timelike Killing field and satisfies certain asymptotic conditions.
\item Map the unbounded spatial domain to a compact domain.
\item Rescale the fields for regularity at the domain boundary.
\end{enumerate}
Level sets of the hyperboloidal time
coordinate $\tau$\footnote{Not to be confused with \emph{proper time}, which shall be denoted by $\uptau$ in this work.}
 penetrate the (future) black-hole horizon at $r=2M$, and
future null infinity $\scri^+$ as $r\rightarrow \infty$ as depicted on the Carter-Penrose diagram Fig.~\ref{fig:Penrose}. As both surfaces are
incoming null surfaces, no boundary data is prescribed. The boundary
conditions after the spatial mapping are behavioral as opposed to numerical in
the terminology of Boyd \cite{Boyd}. This implies trivial boundary
treatment in spectral methods after a suitable choice of function space.

There are many specific hyperboloidal coordinates using scri-fixing
(see \cite{PanossoMacedo:2019npm} for a review in the context of
Kerr
spacetime). Here, we follow
\cite{Ansorg:2016ztf,PanossoMacedo:2018hab,PanossoMacedo:2019npm,jaramillo2021gravitational} and work in the
so-called \emph{minimal gauge}. Specifically, the transformation between the original Schwarzschild coordinates 
$(t,r,\theta,\varphi)$ and the hyperboloidal coordinates $(\tau,\sigma,\theta,\varphi)$ reads
\beq
    t = \lambda \bigg(\tau - H(\sigma)\bigg), \quad r= \dfrac{2M}{\sigma}\,, \label{eq:Hyper_r}
\eeq
with $\lambda=4M$ and the height function
\bea
\label{eq:height_wrt_t}
H(\sigma) &=&\dfrac{1}{2} \bigg(\ln(1-\sigma) - \dfrac{1}{\sigma} + \ln \sigma \bigg).
\eea
Thus, along $\tau = $ constant, $\scri^+$ is located at $\sigma = 0$ and
the black-hole horizon is at $\sigma =1$. 

As discussed in Ref.~\cite{PanossoMacedo:2018hab,PanossoMacedo:2019npm}, this gauge retains the minimal structure in the coordinate transformation needed to construct hyperboloidal slices. Consequently, the corresponding equations on black-hole perturbation theory assume the most simple form. Figure \ref{fig:Penrose} shows the level sets $\tau=\,$constant in the Carter-Penrose diagram, where the desired properties become evident: the hypersurfaces penetrate the black-hole horizon ${\cal H}^+$, and they extend up to future null infinity $\scri^+$.
For regularity of the transformed equations, the asymptotic fall-off behavior of the unknown field must be taken into account \cite{zenginouglu2011hyperboloidal}. 
The rescaling that takes out the asymptotic fall-off is geometrically related to the conformal completion of the asymptotically flat background spacetime \cite{Penrose63}. 
In the frequency domain, the time transformation Eq.~\eqref{eq:Hyper_r} corresponds to a rescaling
\cite{Zenginoglu:2011jz,PanossoMacedo:2019npm, zenginouglu2021null}.
The scalar field rescales as
\beq
	\label{eq:HyperFieldReScale}
	\phi = Z\, \overline{\phi}, \quad Z = \Omega \, e^{s H}, \quad s = -i \omega \lambda.
\eeq
The conformal factor $\Omega=\sigma/\lambda$ accounts for the scalar field's fall-off behavior  $\sim 1/r$, whereas the exponential term naturally arises from the Fourier factor $e^{-i\omega t}$ when the time transformation in Eq.~\eqref{eq:Hyper_r} is taken into account. 
In this way, $Z$ automatically incorporates the boundary behavior~\eqref{eq:BC_frequency} via the geometrical interpretation of the height function from the spacetime perspective. 
Hereafter will denote the rescaled quantities with an overline, e.g., $\overline{\phi}$.

Equivalent to Eq.~\eqref{eq:FreqDomain}, the hyperboloidal field $\overline \phi$ satisfies
\beq
\label{eq:HypDiffEq}
{\boldsymbol A} \overline \phi = \overline {\cal S},
\eeq
with the operator ${\boldsymbol A}$ and source $\overline {\cal S}$ related to the original $\Delta$ and ${\cal S}$ via
\beq
\label{eq:HyperOperatorReScale}
\Delta \phi = {\cal F} {\boldsymbol A} \, \overline{\phi} \Longrightarrow \overline{\cal S} = {\cal F}^{-1} {\cal S}.
\eeq
We discuss the re-scaling factor ${\cal F}$ in the upcoming section. First, let us express the operator ${\boldsymbol A}$ as
\beq 
\label{eq:Hyperboloidal_Laplace_operator}
{\boldsymbol A} = \alpha_2 \dfrac{d^2}{d\sigma^2} + \alpha_1
\dfrac{d}{d\sigma} + \alpha_0.
\eeq
An important property is that the transformed operator ${\boldsymbol A}$ degenerates at the domain boundaries. In other words, the operator's principal part $\alpha_2$ vanishes at $\sigma = 0$ and $\sigma = 1$. Thus, the original
considerations about ingoing/outgoing boundary conditions are re-casted
into questions about the underlying solution's regularity. In practical terms, due to the vanishing of the coefficient
$\alpha_2$ at $\sigma=0$ and $\sigma=1$, the regularity conditions for a field $\overline{\phi}$ satisfying Eq.~\eqref{eq:HypDiffEq} reads
\beq
\label{eq:BC_hyp}
\left.\Bigg(\alpha_1 \partial_{\sigma}\overline{\phi} +
\alpha _0 \overline{\phi} \Bigg) \right|_{\substack{\sigma
=
0, \\ \sigma = 1}} = \left.\overline{\cal S}\right|_{\substack{\sigma
=
0, \\ \sigma = 1}}.
\eeq 
In this way, the boundary conditions follow {\em directly} from the equation,
and no external data is allowed if one seeks a regular solution. In the above considerations, we assume $\overline{\cal S}$ 
is finite at $\sigma=0$ and $\sigma=1$. As discussed, this is the case for the examples under consideration. A more detailed 
study on the regularity classes of $\overline{\cal S}$ is necessary for the sources on the two-time scale analysis
~\cite{Miller:2020bft}.
\subsection{The hyperboloidal operator $A$ and factor ${\cal F}$}
The operator ${\boldsymbol A}$ acting on the hyperboloidal scalar field $\overline{\phi}_{\ell m}$ follows from Eqs.~\eqref{eq:LaplaceOperator} and \eqref{eq:HyperOperatorReScale} via a factor~\cite{PanossoMacedo:2019npm}
\beq
\label{eq:F}
{\cal F} = \dfrac{Z}{r^2 f}.
\eeq
The original radial coordinate $r$ is understood as the function $r(\sigma)$ according to Eq.~\eqref{eq:Hyper_r}. The coefficients on Eq.~\eqref{eq:Hyperboloidal_Laplace_operator} are
\bea
\label{eq:Hyperboloidal_Laplace_operator_coefficients}
&\alpha_2 = \sigma^2 (1-\sigma), \quad \alpha_1 = \sigma(2-3\sigma) +
s(1-2\sigma^2) \nn \\
&\alpha_0 = - \bigg[ \ell(\ell+1) + \sigma + 2s\sigma + s^2(1+\sigma)\bigg].
\eea
The polynomial structure in $\sigma$ manifests the hyperboloidal minimal gauge's simplicity. With the explicit expressions above, it becomes evident that
$\boldsymbol{A}_{\ell m}$ is a degenerate operator, i.e., $\alpha_2=0$ at $\sigma=0$ and $\sigma=1$.
Appendix~\ref{app:operators_spin} discusses the factor ${\cal F}$ and operator $\boldsymbol{A}$ for fields with spin weight 
$p\neq 0$ in both BPT and RWZ formalisms.
\subsection{The hyperboloidal source $\overline{\cal S}$}
We now turn our attention to the transformation of the different types of source terms discussed in Sec.~\ref{sec:Sources}.
\subsubsection{Distributional sources}
The transformation of Eq.~\eqref{eq:Source_delta}, where the source term has delta-support on the particle's orbit, gives
\beq
	\overline{\cal S}^{\rm d}_{\ell m} = \overline\kappa_{\ell m} \, \delta(\sigma - \sigma_p). 
\eeq
The constant $\overline\kappa_{\ell m}$ relates to the original $\kappa_{\ell m}$ via
\bea
\label{eq:Hyperboloidal_kappa}
\overline \kappa_{\ell m}& =& \dfrac{\sigma_p^2}{2M {\cal F}}\kappa_{\ell m} \nn \\
&=& 2M \dfrac{f_p}{Z_p} \kappa_{\ell m}.
\eea
The first line in Eq.~\eqref{eq:Hyperboloidal_kappa} has a generic form, and the transformation incorporates two terms: a rescaling by ${\cal F}^{-1}$ from Eq.~\eqref{eq:HyperOperatorReScale}, and a change of coordinates in the delta function accomplished by 
\beq \delta(g(\sigma)) = \dfrac{\delta(\sigma -
	\sigma_p)}{|g'(\sigma_p)|}, \quad g(\sigma) = r(\sigma) - r_p.
\eeq
In the above expression, $r(\sigma)$ is given by Eq.~\eqref{eq:Hyper_r}. The second line in Eq.~\eqref{eq:Hyperboloidal_kappa} makes explicit use of the function ${\cal F}$ in Eq.~\eqref{eq:F}.
In this context, the (hyperboloidal) retarded field $\overline{\phi}^{\rm ret}_{\ell m}$ is the regular solution to the equation 
\beq 
	\label{eq:Hyperboloidal_Laplace_eq}
	\boldsymbol{A}_{\ell m} \overline{\phi}^{\rm ret}_{\ell m} = \overline{\cal S}^{\rm d}_{\ell m},
\eeq
i.e., $\overline{\phi}^{\rm ret}_{\ell m}$ must satisfy Eq.~\eqref{eq:BC_hyp} with the right-hand side $\overline{\cal S}=0$ at $\sigma=0$ and $\sigma=1$. Note that, as opposed to the standard case, the transformed equation does not allow for regular advanced solutions. The retarded behavior is not imposed through a separate boundary condition, but through the equation itself. 
In order to construct advanced solutions, one would need to change the causal nature of the slicing by changing the sign of the height function in \eqref{eq:height_wrt_t} so that the hyperboloidal surfaces extend between past event horizon and past null infinity.

The delta-function source in the right-hand side of
\eqref{eq:Hyperboloidal_Laplace_eq} imposes a jump in the field's first
derivative in the form
\beq
\label{eq:jump_modesum}
\left.\bigg(\partial_{\sigma }\overline{\phi}^{\rm ret }_{\ell m}{}_+{} -
\partial_{\sigma }\overline{\phi}^{\rm ret }_{\ell
m}{}_-\bigg)\right|_{\sigma=\sigma_p} = \overline{\rm J}_p.
\eeq
with 
\beq
\overline{\rm J}_p = \left. \dfrac{\overline \kappa_{\ell
m}}{\alpha_2} \right|_{\sigma=\sigma_p}.
\eeq
In the above expressions we have defined
\beq
\left.\bigg(\partial_{\sigma } \overline{\phi}^{\rm ret }_{\ell m}{}_\pm
\bigg)\right|_{\sigma=\sigma_p} = \lim_{\epsilon \rightarrow 0}
\partial_{\sigma } \overline{\phi}^{{\rm ret }}_{\ell m}(\sigma_p \pm
\epsilon).
\eeq 
\subsubsection{Worldtube sources}
For sources with compact support around the particle's orbit, the hyperboloidal residual field $\overline{\Phi}^{\mathcal{R}}_{\ell m}$ satisfies
\beq
\label{eq:HypEq_EffecSource}
\boldsymbol{A}_{\ell m} \overline{\phi}^{\mathcal{R}}_{\ell m} = \overline{\cal S}^{\rm w}_{\ell m},
\eeq
with $\overline{\cal S}^{\rm w}_{\ell m}$ defined within the worldtube $\sigma\in[\sigma_{-}, \sigma_{+}]$. Note that from Eq.~\eqref{eq:Hyper_r} one has $r_{+} = r(\sigma_{-})$ and $r_{-} = r(\sigma_{+})$.
Considering ${\cal S}^{\rm w}_{\ell m}$ given by
Eq.~\eqref{eq:EffectiveSource_FrequencyDomain}, we obtain the transformed expression 
\beq
\label{eq:HyperEffectSource}
\overline{\cal S}^{\rm w}_{\ell m} = \overline{\kappa}_{\ell m}
\delta(\sigma - \sigma_p) - {\boldsymbol A}_{\ell m} \overline{\Phi}^{\mathcal{P}}_{\ell
m}.
\eeq
As expected, $\alpha_2 \, \partial^2_{\sigma\sigma }
\overline{\Phi}^{\mathcal{P}}_{\ell m} = \overline{\kappa}_{\ell m} \delta(\sigma -
\sigma_p)$, so the delta-source cancels out in the right-hand-side of
Eq.~\eqref{eq:HyperEffectSource}. We are left with
\beq
\overline{\cal S}^{\rm w}_{\ell m} = \alpha_1 \, \partial_{\sigma }
\overline{\Phi}^{\mathcal{P}}_{\ell m} + \alpha_0 \, \overline{\Phi}^{\mathcal{P}}_{\ell m}.
\eeq
Alternatively, the rescaling from Eq.~\eqref{eq:HyperOperatorReScale} applies directly into the regularised expression~\eqref{eq:EffectiveSource_FrequencyDomain_NoDelta}.
As explained, Eq.~\eqref{eq:BC_hyp} fixes the regularity conditions for $\overline{\phi}^{\mathcal{R}}_{\ell m}$. Since $\overline{S}_{\ell m}=0$ at $\sigma=0$ and $\sigma=1$, the conditions reduce to same as for the retarded field. 
In fact, by definition one has 
\beq
	\label{eq:Transition_phiR}
	\overline {\Phi}^{\mathcal{R}}_{\ell m} = \left\{ 
	\begin{array}{cc}
			\overline {\phi}^{\rm ret}_{\ell m}, & \sigma\in[0,\sigma_{-}), \,\, \sigma\in(\sigma_{+}, 1] \\
			\overline {\phi}^{\rm ret}_{\ell m} - \overline {\phi}^\mathcal{P}_{\ell m}, &  \quad
			\sigma\in[\sigma_{-},\sigma_{+}]
	\end{array}
	\right. ,
\eeq
i.e., $\overline{\Phi}^{\mathcal{R}}_{\ell m}$ and $\overline{\phi}^{\rm ret}_{\ell m}$ coincide everywhere outside the worldtube.
Eq.~\eqref{eq:Transition_phiR} fixes the transition conditions at the boundaries $\sigma_{-, out}$. 
Specifically, Eq.~\eqref{eq:Transition_phiR} imposes
\bea
	&\left.\bigg(\overline{\Phi}^{\mathcal{R}}_{\ell m}{}_+ - \overline{\Phi}^{\mathcal{R}}_{\ell m}{}_-
	\bigg)\right|_{\sigma = \sigma_{-}} = -\overline{\phi}^{\mathcal{P}}_{\ell m}(\sigma_{-}), \\
	&\left.\bigg(\partial_\sigma \overline{\Phi}^{\mathcal{R}}_{\ell m}{}_+ -
	\partial_\sigma\overline{\Phi}^{\mathcal{R}}_{\ell m}{}_- \bigg)\right|_{\sigma =
	\sigma_{-}} = -\partial_\sigma \overline{\phi}^{\mathcal{P}}_{\ell m}(\sigma_{-}), \\
	&\left.\bigg(\overline{\Phi}^{\mathcal{R}}_{\ell m}{}_+ - \overline{\Phi}^{\mathcal{R}}_{\ell m}{}_-
	\bigg)\right|_{\sigma = \sigma_{+}} = \overline{\phi}^{\mathcal{P}}_{\ell
	m}(\sigma_{+}),\\
	&\left.\bigg(\partial_\sigma\overline{\Phi}^{\mathcal{R}}_{\ell m}{}_+ -
	\partial_\sigma\overline{\Phi}^{\mathcal{R}}_{\ell m}{}_- \bigg)\right|_{\sigma =
	\sigma_{+}} = \partial_\sigma \overline{\phi}^{\mathcal{P}}_{\ell m}(\sigma_{+}).
\eea
Finally, a unique solution follows by fixing $\overline{\Phi}^{\mathcal{R}}_{\ell m}$ at the particle's location via continuity conditions
\bea
	&\left.\bigg(\overline{\Phi}^{\mathcal{R}}_{\ell m}{}_+ - \overline{\Phi}^{\mathcal{R}}_{\ell m}{}_-
	\bigg)\right|_{\sigma = \sigma_{p}} = 0, \\
	&\left.\bigg(\partial_\sigma \overline{\Phi}^{\mathcal{R}}_{\ell m}{}_+ -
	\partial_\sigma\overline{\Phi}^{\mathcal{R}}_{\ell m}{}_- \bigg)\right|_{\sigma = \sigma_{p}} = 0.
\eea

\subsubsection{Unbounded support sources}
The transformation of Eq.~\eqref{eq:Eq_psi} follows similarly. 
By taking the derivative of Eq.~\eqref{eq:Hyperboloidal_Laplace_eq} with respect to $r_p$ one obtains the hyperboloidal parametric derivative field $\overline {\psi}^{\rm ret}_{\ell m} = \overline {\phi}^{\rm ret}_{\ell m,r_p}$
\bea
\label{eq:ExtendedSourceOperator_Hyp}
\boldsymbol{A}_{\ell m} \overline { \psi}^{\rm ret}_{\ell m}  = \overline {{\cal S}}^{\rm u}_{\ell m}.
\eea
Here, the extended hyperboloidal source reads
\bea
\label{eq:ExtendedSource_Hyp}
\overline {{\cal S}}^{\rm u}_{\ell m} &=&  \partial_{r_{p}}\overline\kappa_{\ell m}
\delta(\sigma - \sigma_p) +
\dfrac{\sigma_p^2}{2M} \overline \kappa_{\ell m} \, \delta'(\sigma -
\sigma_p) \nn \\
 	&& + \boldsymbol{C} \overline { \phi}^{\rm ret}_{\ell m} ,
\eea
with the operator $\boldsymbol{C} =\partial_{r_{p}} \boldsymbol{A}$ given by
\beq
\boldsymbol{C} = \partial_{r_{p}}s \Bigg( 2 \sigma + 2 s (1+\sigma) -(1-2 \sigma^2)
\partial_\sigma \Bigg).
\eeq
The relation between the original field $\psi^{\rm ret}_{\rm \ell m}$ and its hyperboloidal equivalent $ \overline { \psi}^{\rm ret}_{\ell m} $ does not follow from Eq.~\eqref{eq:HyperFieldReScale} in contrast to $\phi^{\rm ret}_{\rm \ell m}$ and $ \overline { \phi}^{\rm ret}_{\ell m} $.
Because $Z$ depends on $r_p$ through the frequency $s$, Eq.~\eqref{eq:HyperFieldReScale} leads to
\bea
	\label{eq:SlowTimeDervReScale}
	\psi^{\rm ret}_{\ell m} &=& Z\, \overline{\psi}^{\rm ret}_{\ell m} + \partial_{r_{p}}Z\,\overline{\phi}^{\rm ret}_{\ell m}.
\eea
The field $\overline{\psi}^{\rm ret}_{\ell m}$ is then uniquely determined via
the regularity conditions at $\sigma=0$ and $\sigma=1$, together with the jump conditions at the particle location. 
According to Eq.~\eqref{eq:BC_hyp} the regularity conditions read
 \beq
\Bigg( \left. \alpha_1 \partial_\sigma \overline{\psi}^{\rm ret}_{\ell m} +
\alpha _0 \overline{\psi}^{\rm ret}_{\ell m} \Bigg) \right|_{\substack{\sigma
=
0, \\ \sigma = 1}} = \left. \boldsymbol{C} \overline { \phi}^{\rm
ret}_{\ell m}  \right|_{\substack{\sigma = 0, \\ \sigma = 1}},
 \eeq
whereas the jump conditions at the particle location are
\bea
&\left.\bigg(\overline{\psi}^{\rm ret}_{\ell m}{}_+ - \overline{\psi}^{\rm
ret}_{\ell m}{}_- \bigg)\right|_{\sigma = \sigma_{p}} = \dfrac{\sigma_p^2\,
\overline{\rm J}_p}{2M}  \\
&\left.\bigg(\partial_\sigma \overline{\psi}^{\rm ret}_{\ell m}{}_+ -
\partial_\sigma\overline{\psi}^{\rm ret}_{\ell m}{}_- \bigg)\right|_{\sigma =
\sigma_{p}} = \partial_{r_{p}}\overline{\rm J}_p - \dfrac{\sigma_p^2\,\overline{\rm
J}_p}{2M} \left. \dfrac{\alpha_1}{\alpha_2}\right|_{\sigma=\sigma_p}.
\eea

\subsection{Energy flux and the self-force}

As a consistency check of our calculations, it is useful to use a flux-balance law and compare our results to those in the literature.
For these we need to compute the energy flux radiated to infinity and the horizon.
In the following subsections, we derive the balance law and show how to calculate the energy fluxes from data computed on the hyperboloidal slices.

\subsubsection{Flux Balance Law}
The total energy flux must balance the
work $\mathcal{W}$ done on the scalar charge by the SSF such that 
\beq
	\dot{E}_{\text{total}} = - \mathcal{W} = -\mu \dot{\mathcal{E}},
	\label{eq:work_done_relation}
\eeq
where $E_{\text{total}}$ is the total radiated (scalar) energy, overdot denotes a derivative with respect to coordinate time $t$, and
we have written the work done in terms of the rate of change of specific energy, $\mathcal{E}$, per unit time. The specific energy itself is given by 
$\mathcal{E} = - \xi^{\mu}_{(t)} u_{\mu} = -g_{\mu\nu} \xi^{\nu}_{(t)} u^{\nu}$, where $\xi^{\mu}_{(t)}$ is the timelike Killing vector field satisfying the Killing equation $\nabla_{\beta}\xi_{\alpha} + \nabla_{\alpha}\xi_{\beta} = 0$. To take advantage of this, we transform the derivative that 
appears on the right-hand-side to a derivative with respect to proper time,
\begin{align}
	\dot{\mathcal{E}} &= (u^{t})^{-1} u^{\alpha} \nabla_{\alpha} \mathcal{E} \nonumber \\ 
	&= -g_{\mu\nu} (u^{t})^{-1} \left( u^{\nu}u^{\alpha} \nabla_{\alpha} \xi^{\mu}_{(t)} 
	+ \xi^{\mu}_{(t)} u^{\alpha}\nabla_{\alpha} u^{\nu} \right)\nonumber \\
	&= -g_{\mu\nu} (u^{t})^{-1} \xi^{\mu}_{(t)} u^{\alpha}\nabla_{\alpha} u^{\nu}.
	\label{eq:t_derivative_specific_energy}
\end{align}
The term $u^{\nu}u^{\alpha} \nabla_{\alpha} \xi^{\mu}_{(t)}$ vanishes due to $\xi^{\mu}_{(t)}$ satisfying Killing's equation.  
Since the motion of our particle is determined by the (self-)forced equation of motion, Eq.~\eqref{eq:EqnOfMotion} and our timelike Killing vector is given by the Kronecker Delta,
$\xi^{\mu}_{(t)} = \delta^{\mu}_{t}$, we find from Eqs.~(\ref{eq:work_done_relation}) and (\ref{eq:t_derivative_specific_energy})
\beq
	F_{t} = \mu u^{t} \dot{E}_{\text{total}}. 
	\label{eq:flux_balance_law}
\eeq
Note that we have neglected the rate of change of the mass per unit proper time, $d\mu/d\uptau$, as we are in a stationary, circular orbit configuration.  
In more general setups, the mass of the scalar charge can vary due to the SSF component that is tangent to $u^{\alpha}$ such that $d\mu/d\uptau = - u^{\alpha}F_{\alpha}$ \cite{Quinn:2000wa,Warburton:2011hp}.

We compute the $r_{p}$-derivative of the self-force from our calculations involving sources with unbounded support.  
As with our original field equation ~\eqref{eq:Eq_psi}, one can take an $r_p$-derivative of both sides of Eq.~\eqref{eq:flux_balance_law} to find
\beq
\mathcal{D}_{r_{p}}  F_{t} = \mu \left( \partial_{r_{p}}u^{t}\dot{E}_{\text{total}} + u^{t} \partial_{r_{p}} 
	\dot{E}_{\text{total}} \right).
	\label{eq:noncompact_flux_balance_law}
\eeq
Note that one must carefully consider the $r_p$-derivative on the left-hand side of Eq.~\eqref{eq:noncompact_flux_balance_law}, since the operations $\partial_{r_{p}}$ and $\lim_{r\rightarrow r_p}$ do not commute with each other. More specifically, the right-hand side involves quantities evaluated at the black-hole horizon, and at future null infinity. Thus, $\partial_{r_p}$ accounts for the explicit parametric dependence on the particle's orbit. The left-hand side, however, must account for the parametric $r_p$-dependence, as well as the contribution from the field's value at $r_p$. Hence, for a given quantity $\displaystyle \varpi_{p} = \lim_{r\rightarrow r_p} \varpi(r)$, one obtains 
\beq
\label{eq:total_rp_derivative}
\mathcal{D}_{r_{p}} \varpi_{p}= \left.\Bigg(\partial_{r_{p}}\varpi(r) +  \varpi'(r) \Bigg)\right|_{r_p}.
\eeq

\subsubsection{Hyperboloidal Flux}
The total radiated energy can be evaluated from the energy flux vector
\beq
	\varepsilon^{\alpha} := -g^{\alpha\beta}T_{\beta\mu}\xi^{\mu}_{(t)},
\eeq
where $T^{\mu\nu}$ is the stress-energy tensor of the scalar field \cite{Poisson:2004}.
We wish to calculate the flux flowing to $\scri^{+}$ (future null infinity) and
down to the black hole. To do so let us consider a timelike hypersurface with
$r = r_{0}$ labelled $\Sigma^{0}$. The scalar-field energy flowing through an
infinitesimal surface element of the hypersurface, $d\Sigma^{0}$, that spans
a small time $dt$ is given by
\beq
	dE^{0} = \int_{\Sigma} \varepsilon^{\alpha} d\Sigma^{0}_{\alpha} 
	= \int_{\Sigma} \tensor{T}{^{\alpha}_{\mu}}\xi^{\mu}_{(t)}
	d\Sigma^{0}_{\alpha}.
	\label{eq:infinitesmal_flux}
\eeq
Here $d\Sigma^{0}_{\alpha}$ is an outward-pointing surface element on the
section of the hypersurface $d\Sigma^{0}$. Since our hypersurface is
timelike, the outward-pointing surface elements are expressed as $d\Sigma^{0}_{\alpha} = \sqrt{-h}\,
n_{\alpha} dt d\theta d\varphi$, where $h$ is the determinant of the induced metric on $\Sigma^{0}$ and
$n_{\alpha}$ is the radial unit normal vector to the hypersurface.  Explicitly in terms of the standard 
Schwarschild coordinates $n_{\alpha} = \delta^{r}_{\alpha}/\sqrt{f_{0}}$ and therefore $h = -f_{0} r_{0}^{2} \sin^{2}\theta$,
where subscript $``0"$ means the function is evaluated at $r = r_{0}$.  Bringing this all together and
substituting the coordinate form of the Killing tensor, one finds the flux of energy through the
hypersurface $\Sigma^{0}$ to be
\beq
	\dot{E}^{0} = \frac{dE^{0}}{dt}
	= f_{0} r_{0}^{2} \oint \tensor{T}{_{tr}} dw,
	\label{eq:flux_expression}
\eeq
where $dw$ is the standard differential solid angle.\footnote{This is written 
differently than the normal convention $d\Omega$, so as to not be confused with the conformal factor 
$\Omega$ introduced in Eq.~\eqref{eq:HyperFieldReScale}.}
Our aim is to write Eq.~\eqref{eq:flux_expression} in terms of our hyperboloidal
coordinates $x^{\alpha^{\prime}} = (\coord)$ and the hyperboloidal field $\overline{\phi}(\sigma)$ 
to evaluate the radiative flux at future null infinity $(\sigma = 0)$ and the horizon $(\sigma = 1)$.  
By transforming the stress-energy tensor of the scalar field $T_{\alpha\beta}$ into our coordinates, we find
\beq
	\tensor{T}{_{tr}} =  
	-\frac{\sigma^{2}}{2M \lambda} \left( \tensor{T}{_{\tau\tau}}
	H_{,\sigma} + \tensor{T}{_{\tau\sigma}} \right).
\eeq
It follows from Eq.~\eqref{eq:height_wrt_t}
\beq
	\partial_{\sigma} H = \frac{1 - 2\sigma^{2}}{2\sigma^{2}(1 - \sigma)}.
\eeq
As discussed in Appendix \ref{app:hyperboloidal_flux}, evaluating $\tensor{T}{_{\tau\tau}}$ and $\tensor{T}{_{\tau\sigma}}$
in terms of the conformal field and noting in our hyperboloidal coordinates, $f(\sigma) = (1 - \sigma)$, we find remarkably simple expressions for the flux integrands:
\begin{align}
	f_{0} \left( \tensor{T}{_{\tau\tau}}
	H_{,\sigma} + \tensor{T}{_{\tau\sigma}} \right) \big|_{\sigma_{0} = 0}  &= 
	\frac{1}{8\pi\lambda^{2}} (\partial_{\tau}\overline{\Phi})^{2},\\
	f_{0} \left( \tensor{T}{_{\tau\tau}}
	H_{,\sigma} + \tensor{T}{_{\tau\sigma}} \right)\big|_{\sigma_{0} = 1}  &= 
	-\frac{1}{8\pi\lambda^{2}} (\partial_{\tau}\overline{\Phi})^{2}.
\end{align}
Therefore our flux expressions become 
\begin{align}
	\dot{E}^{\scri^{+}} :=
	+&\dot{E}^{0} \big|_{\sigma_{0} = 0} 
	= \frac{1}{16\pi \lambda^{2}} \oint 
	(\partial_{\tau}\overline{\Phi})^{2} 
	\bigg|_{\sigma = 0} dw \\
	\dot{E}^{\mathcal{H}^{+}} :=
	-&\dot{E}^{0} \big|_{\sigma_{0} = 1} 
	= \frac{1}{16\pi \lambda^{2}} \oint 
	(\partial_{\tau}\overline{\Phi})^{2} 
	\bigg|_{\sigma = 1} dw,
	\label{eq:flux_expression_limits}
\end{align}
where our sign convention is chosen such that the outflow of energy towards $\scri^{+}$ is positive
and the inflow of energy towards the horizon is negative.  The Fourier and spherical harmonic mode decomposition of the conformal scalar
field given by
\beq
	\overline{\Phi}(\tau, \sigma, \theta, \varphi) = \sum_{\ell, m}
	\overline{\phi}_{\ell m}(\sigma) Y_{\ell m}(\theta, \varphi) e^{s\tau}
	:= \sum_{\ell m} \overline{\Phi}_{\ell m},
	\label{eq:scalar_field_decomposition}
\eeq
allows us to make the replacement $\partial_{\tau}\overline{\Phi}_{\ell m} =
s \overline{\Phi}_{\ell m}$.  If we substitute this into
Eq.~\eqref{eq:flux_expression_limits}, the integral is readily evaluated with the
standard spherical harmonic orthogonality relation, leaving us with succinct expressions for the flux at the horizon and
infinity,
\begin{align}
	\dot{E}^{\scri^{+}} &= \frac{1}{16\pi\, \lambda^{2}} \sum_{\ell m}
	\big| s \overline{\phi}_{\ell m} \big|^{2}_{\sigma = 0},
	\label{eq:flux_scri}\\
	\dot{E}^{\mathcal{H}^{+}} &= \frac{1}{16\pi\lambda^{2}} \sum_{\ell m}
	\big| s \overline{\phi}_{\ell m} \big|^{2}_{\sigma = 1}.
	\label{eq:flux_hor}
\end{align}
If we are to compare our results with the parametric derivative of the field, $\psi_{\ell m}$,
we need to compute the $r_{p}$-derivative of the flux. 
As our conformal field is complex, we find 
\begin{align}
	&\partial_{r_{p}}\dot{E}^{\scri^{+}} = 
	\frac{1}{16\pi \lambda^{2}} \nonumber \\
	&\times \sum_{\ell m}
	\,\text{Re} \bigg[ s \overline{\phi}_{\ell m} \big(\partial_{r_{p}}s\, 
	\overline{\phi}_{\ell m} + s \overline{\psi}_{\ell m}\big)^{*} 
	\bigg]_{\sigma = 0},\\
	&\partial_{r_{p}}\dot{E}^{\mathcal{H}^{+}} = 
	\frac{1}{16\pi \lambda^{2}} \nonumber \\
	&\times \sum_{\ell m}
	\,\text{Re} \bigg[ s \overline{\phi}_{\ell m} \big(\partial_{r_{p}}s\, 
	\overline{\phi}_{\ell m} + s \overline{\psi}_{\ell m}\big)^{*} 
	\bigg]_{\sigma = 1}.
\end{align}

\subsection{Self-force}\label{sec:SelfForceTheory}
To calculate the self-force within our hyperboloidal approach we start with the expression for the self-force in 
covariant form given in Eq.~\eqref{eq:EqnOfMotion}.  We first consider the $t$-component of the self-force in terms 
of conformal scalar field.  The transformation to conformal coordinates yields
\beq
	F^{\text{self}}_{t}
	= \frac{q}{\lambda} \lim_{x^{\mu} \rightarrow x^{\mu}_{p}}
	\Omega\, \partial_{\tau} \overline{\Phi}(x^{\mu}).
	\label{eq:self_force_t_transformation}
\eeq
We shall denote the $\ell$-mode contribution to the full self-force field by
$F_{\ell t}$.  With the help of Eq.~\eqref{eq:HyperFieldReScale}, 
substituting the decomposition from Eq.~\eqref{eq:scalar_field_decomposition} into 
Eq.~\eqref{eq:self_force_t_transformation} and taking the limit to the worldline we find
\beq
	F_{\ell t} = \dfrac{q}{\lambda} \sum^{\ell}_{m =-\ell} s\,   Z(\sigma_p) \overline{\phi}_{\ell m}(\sigma_p)
	Y_{\ell m}(\pi/2, 0).
	\label{eq:F_lt}
\eeq
This expression can be used to directly evaluate left-hand side of the balance law,
Eq.~\eqref{eq:noncompact_flux_balance_law}, by taking a $r_p$-derivative of both sides of 
Eq.~\eqref{eq:F_lt}:
\bea
\mathcal{D}_{r_{p}}&&F_{t} =  \dfrac{q}{\lambda} \sum^{\ell}_{m =-\ell} \Bigg[ \partial_{r_{p}}s\,   
	Z \overline{\phi}_{\ell m} \nn \\ && + s\, Z \bigg(\mathcal{D}_{r_{p}} \overline{\phi}_{\ell m}
	+ \mathcal{D}_{r_{p}}\ln \!Z \overline{\phi}_{\ell m}   \bigg)\Bigg]Y_{\ell m}(\pi/2, 0).
	\label{eq:dr0_F_lt}
\eea
Using Eq.~\eqref{eq:total_rp_derivative}, one obtains explicitly 
\bea
	\mathcal{D}_{r_{p}}\overline{\phi}_{\ell m}   &=& \overline{\psi}_{\ell m}(\sigma_p)    
    -\dfrac{2M}{r_p^2} \overline{\phi}_{\ell m}{}_{,\sigma}   \Bigg|_{\sigma=\sigma_{r_p}}, \\
	\mathcal{D}_{r_{p}}\ln \!Z &=& \partial_{r_{p}}s -\dfrac{2M}{r_p^2} \left.\Bigg( (\ln\Omega)_{,\sigma} 
    + sH_{,\sigma}\Bigg)\right|_{\sigma=\sigma_{r_p}}.
\eea
Obtaining the $r$-component of the SSF, meanwhile, is a bit more involved. Due to the coordinate 
transformation given in Eq.~\eqref{eq:Hyper_r}, one obtains 
\beq
	\partial_r = -\frac{\sigma^2}{2M}\left(\partial_\sigma + H_{,\sigma}\partial_\tau \right). 
\eeq
Therefore,
\beq
	F^{\text{self}}_{r} = -\frac{q}{2M} \lim_{x^\mu \rightarrow x_{p}^\mu} \sigma^{2}  \left[
	H_{,\sigma} \partial_{\tau} \Phi(x^\mu)
	+ \partial_{\sigma} \Phi(x^\mu) \right],
	\label{eq:self_force_r_transformation}
\eeq
which yields
\begin{multline}\label{eq:F_lr}
	F^{\pm}_{\ell r} = -q\frac{\sigma_{p}^{2}}{2M}
	\sum^{\ell}_{m =-\ell} Z(\sigma_p) \bigg[ 
	H_{,\sigma} s \overline{\phi}^{\pm}_{\ell m} \\
	+ \overline{\phi}^{\pm}_{\ell m}{}_{,\sigma}
	+ \frac{\overline{\phi}^{\pm}_{\ell m}}{\sigma}
	\bigg]_{\sigma=\sigma_p}
	Y_{\ell m}(\pi/2, 0).
\end{multline}
Here, $F^{+}_{\ell r}$ and $F^{-}_{\ell r}$ correspond to approaching the worldline from the range $r > r_{p}$ and $r < r_{p}$ respectively.
This distinction is necessary if we set $\overline{\phi}_{\ell m} = \overline{\phi}_{\ell m}^{\rm ret}$ above as then the derivatives of the scalar field $\partial_{\tau}\overline{\phi}$ and $\partial_{\sigma}\overline{\phi}$ at the particle location have two well-defined, but generally different one-sided limits.
In this case the left-hand side of Eq.~\eqref{eq:F_lr} represents the unregularized $\ell$-modes of the force. To compute the $r$-component of the SSF we use the mode-sum regularization formula \cite{Barack:1999wf}
\begin{align}\label{eq:mode-sum}
	F_{\ell r}^{\rm self} = F^\pm_{\ell r} \mp A_r \ell(\ell+1)  - B_r -  \sum_{n=1}^3 F^\ell_{r[2n]},
\end{align}
where $A_r,B_r, F^\ell_{r[2n]}$ are known as \textit{regularization parameters}.
The $A_r$ and $B_r$ act to regularize the self-force, and the $F^\ell_{r[2n]}$ act to accelerate the convergence of the $\ell$-mode sum \cite{Heffernan:2012su}.
If instead in Eq.~\eqref{eq:F_lr} we set $\bar{\phi}_{\ell m} = \bar{\phi}^{\mathcal{R}}_{\ell m}$ as computed from the effective-source approach then the limit is the same from both directions and the left-hand side of Eq.~\eqref{eq:F_lr} 
becomes $F_{\ell r}$.

\section{Numerical methods}\label{sec:NumMeth}
This section details the numerical methods providing highly accurate solutions to the equations transformed into compactified hyperboloidal coordinates. 
We follow the conceptual framework from Refs.~\cite{Ansorg:2003br,Ansorg:2006gd,Meinel:2008kpy,Ansorg13} employing a multi-domain spectral method \cite{canuto2007spectral,Boyd,GraNov07}, enhanced with analytic mesh refinement to improve the computation of solutions with steep gradients.

\subsection{Multi-domain spectral methods}
We use a collocation-point spectral method to solve the hyperboloidal equation on the compact domain $\sigma\in[0,1]$. 
Specifically, we employ the algorithms detailed in Ref.~\cite{Meinel:2008kpy} to find the numerical approximations $f^{(i_{\rm Field})}$, with $i_{\rm Field}=0\cdots N_{\rm Field}$, assuming $n_{\rm Field} = N_{\rm Field}+1$ real-valued functions. 
For instance, the scalar self-force field described in the previous section is a complex-valued function. 
Therefore, the numerical scheme must solve for a total of $n_{\rm Field} = 2$ unknown functions: the scalar field's real and imaginary part.

We divide the interval $[0,1]$ into $n_{\rm dom}$ sub-domains
\beq 
\label{eq:sigma_func_x}
\sigma \in \underbrace{ [\sigma_0,\sigma_1]}_{\text{domain }i_{\rm d}=1 } \!\! \cup
\cdots \cup \,\, \underbrace{[\sigma_{i_{{\rm d}-1}},\sigma_{i_{\rm d}}
]}_{\text{domain }i_{\rm d} } \cup \cdots \cup \underbrace{[\sigma_{n_{\rm
dom-1}},\sigma_{n_{\rm
dom}}]}_{\text{domain }i_{\rm d}=n_{\rm dom} }.
\eeq
In our coordinates, future null infinity is at $\sigma_0 = 0$ and the black-hole horizon is at $\sigma_{n_{\rm
dom}} = 1$.
It is convenient to map each sub-domain $\sigma \in[\sigma_{i_{{\rm d}-1}},\sigma_{i_{\rm d}} ]$, labelled by $i_{\rm d}=1\cdots n_{\rm dom}$, into a
coordinate $x\in[-1,1]$ via
\bea
\sigma &=& \dfrac{1}{2}\left[\sigma_{i_{\rm d}}(1+x) + \sigma_{i_{{\rm
d}-1}}(1-x) \right], \\
x &=& \dfrac{2\sigma - (\sigma_{i_{\rm d}} + \sigma_{i_{{\rm
d}-1}})}{\sigma_{i_{\rm d}} - \sigma_{i_{{\rm d}-1}}}.
\eea
At each domain $i_{\rm d}$, the numerical scheme approximates a given function, $f^{(i_{\rm d},i_{\rm Field})}(x)$, via the finite expansion
\beq
\label{eq:SpecAprox_func}
f^{(i_{\rm d},i_{\rm Field})}_{N_{i_{\rm d}}}(x) = \sum_{k=0}^{N_{i_{\rm d}}} c^{(i_{\rm d},i_{\rm Field})}_k T_k(x),
\eeq
with $N_{i_{\rm d}}$ the truncation order, and $T_k(x) = \cos[k \arccos(x)]$ the Chebyshev
polynomials of first kind. 
The Chebyshev coefficients $c^{(i_{\rm d},i_{\rm Field})}_i$ are
fixed by a collocation method. For this purpose, we discretise the interval
$x\in[-1,1]$ in terms of the Chebyshev-Lobatto grid
\beq
\label{eq:LobattoChebGrid_x}
x_i = \cos\left(\pi \frac{i}{N_{i_{\rm d}}} \right), \quad i = 0 \cdots N_{i_{\rm d}},
\eeq
and impose that the expression \eqref{eq:SpecAprox_func} {\em coincides} with
the exact function $f^{(i_{\rm d},i_{\rm Field})}(x)$ at the grid points. In other words,
the coefficients $c^{(i_{\rm d},i_{\rm Field})}_i$ follow from inverting the equation
\beq
f^{(i_{\rm d},i_{\rm Field})}_{N_{i_{\rm d}}}(x_i) = f^{(i_{\rm d},i_{\rm Field})}(x_i).
\eeq
The above considerations assume an {\it a priori} known function $f^{(i_{\rm
d},i_{\rm Field})}(x)$ from which we construct the approximation $f^{(i_{\rm d},i_{\rm Field})}_{N_{i_{\rm d}}}(x)$. In
practice, though, $f^{(i_{\rm d},i_{\rm Field})}(x)$ is not given, and we only have access
to the underlying differential equation the function must satisfy.
To obtain the function's values at the discrete grid points, we first
collect the unknown components from all different domains into the single vector. More specifically, let us define 
\beq 
f_i^{(i_{\rm d},i_{\rm Field})}=f_{N_{i_{\rm d}}}^{(i_{\rm d},i_{\rm Field})}(x_i)
\eeq 
as the function's value for a given field $i_{\rm Field}$, at the grid point $x_i$ within the domain $i_{\rm d}$. 
Then, we collect each of these values into the vector
\bea
\vec{X} = \left( f_i^{(i_{\rm d},i_{\rm Field})} \right)_{\subalign{& i_{\rm d} = 1 \cdots n_{\rm dom} \\  & i_{\rm Field} = 0\cdots N_{\rm Field} \\ & i = 0\cdots N_{i_{\rm d}}}} \, ,
\eea
which has a total of 
\beq 
n_{\rm total} = n_{\rm Field} \sum_{i_{\rm d}=1}^{n_{\rm dom}}(N_{i_{\rm d}}+1)
\eeq 
components.
Enforcing the differential equations, together with its boundary or transition conditions
at all domains and all collocation points leads to an algebraic system of $n_{\rm total}$ linear equations $\vec{F}(\vec{X})$. 
Recall that imposing the differential equation at the grid points requires calculating
approximations for the first and second derivatives, respectively,
$f'_i{}^{(i_{\rm d},i_{\rm Field})}$ and $f''_i{}^{(i_{\rm d},i_{\rm Field})}$. They
result from applying specific spectral differential
matrices to the vectors $\vec{X}$~\cite{Boyd,canuto2007spectral,trefethen2000spectral}. 
We solve the linear system $\vec{F}(\vec{X})$ for the vector $\vec{X}$ using an LU
decomposition. Thus, the algorithm scales as $n_{\rm total}^3$ and should be
sufficiently fast for low-to-moderate values of $n_{\rm dom}$ and $N_{i_{\rm d}}$.
\subsection{Convergence}
Spectral methods are very efficient when the underlying
function $f^{(i_{\rm d},i_{\rm Field})}(x)$ is analytic because the approximated numerical solution $f^{(i_{\rm d},i_{\rm Field})}_{N_{i_{\rm d}}}(x)$ converges exponentially to the exact solution as the numerical resolution $N_{i_{\rm d}}$ increases (see
\cite{Boyd,GraNov07,canuto2007spectral,trefethen2000spectral} and references
therein). 
Because we do not have access to an explicit expression for the
exact
solution $f^{(i_{\rm d},i_{\rm Field})}(x)$, the numerical error is estimated by fixing a
reference solution obtained with a given high accuracy $N_{i_{\rm d}}=N_{i_{\rm d}}^{\rm ref}$, and
measuring a relative error 
\beq 
\label{eq:RelError}
{\cal E}^{(i_{\rm d}, i_{\rm Field})}_{N_{i_{\rm d}}} = \left| 1- \dfrac{f^{(i_{\rm d},i_{\rm Field})}_{N_{i_{\rm d}}}(x) }{ f^{(i_{\rm d},i_{\rm Field})}_{N_{i_{\rm d}}^{\rm ref}}(x)} \right|, \quad N_{i_{\rm d}}<N_{i_{\rm d}}^{\rm ref}.
\eeq
 In particular, we are interested in measuring the error at the particle's location.
The Chebyshev coefficients $c^{(i_{\rm d},i_{\rm Field})}_k$ provide an efficient way to
estimate the error of a numerical solution at a fixed $N_{i_{\rm d}}$ because their
asymptotic behavior for $k \gg 1$ determines the rate at which the error
${\cal E}^{(i_{\rm d}, i_{\rm Field})}_{N_{i_{\rm d}}}$ decays to zero as $N_{i_{\rm d}}\rightarrow \infty$. Indeed, the exponential
convergence ${\cal E}^{(i_{\rm d}, i_{\rm Field})}_{N_{i_{\rm d}}}\sim {\cal C}^{-N_{i_{\rm d}}} $ for analytic
functions follows from a behavior $c^{(i_{\rm d},i_{\rm Field})}_k \sim \bar {\cal C}^{-k}$
 (with constants ${\cal C}$ and $\bar{\cal C}$).
 
Particular scenarios may jeopardize the fast convergence rate.
Clearly,
the exponential decay depends on the regularity of the underlying solution. If
the solution is known to be on a regularity class $C^l([-1,1])$, then the
convergence rate (as well as the behavior of the Chebyshev
coefficients) will be merely algebraic. We do not find these issues in the
scenarios studied here. 

An exponential decay does not always imply a
highly accurate solution for a small-to-moderate numerical resolution $N_{i_{\rm d}}$. The
error and the Chebyshev coefficients of functions with steep gradients may
decay with a relatively small exponential rate. As discussed in
the following sections, this is the case for large angular modes $\ell$ or large orbital radii $r_{p}$. 
In the next section, we describe the ``analytic mesh-refinement" (AnMR) technique, which introduces yet another coordinate mapping to increase the grid density around the steep region.

\subsection{Analytic mesh-refinement}\label{sec:AnMR}
\begin{figure}[ht!]
\centering
\includegraphics[width=0.48\textwidth]{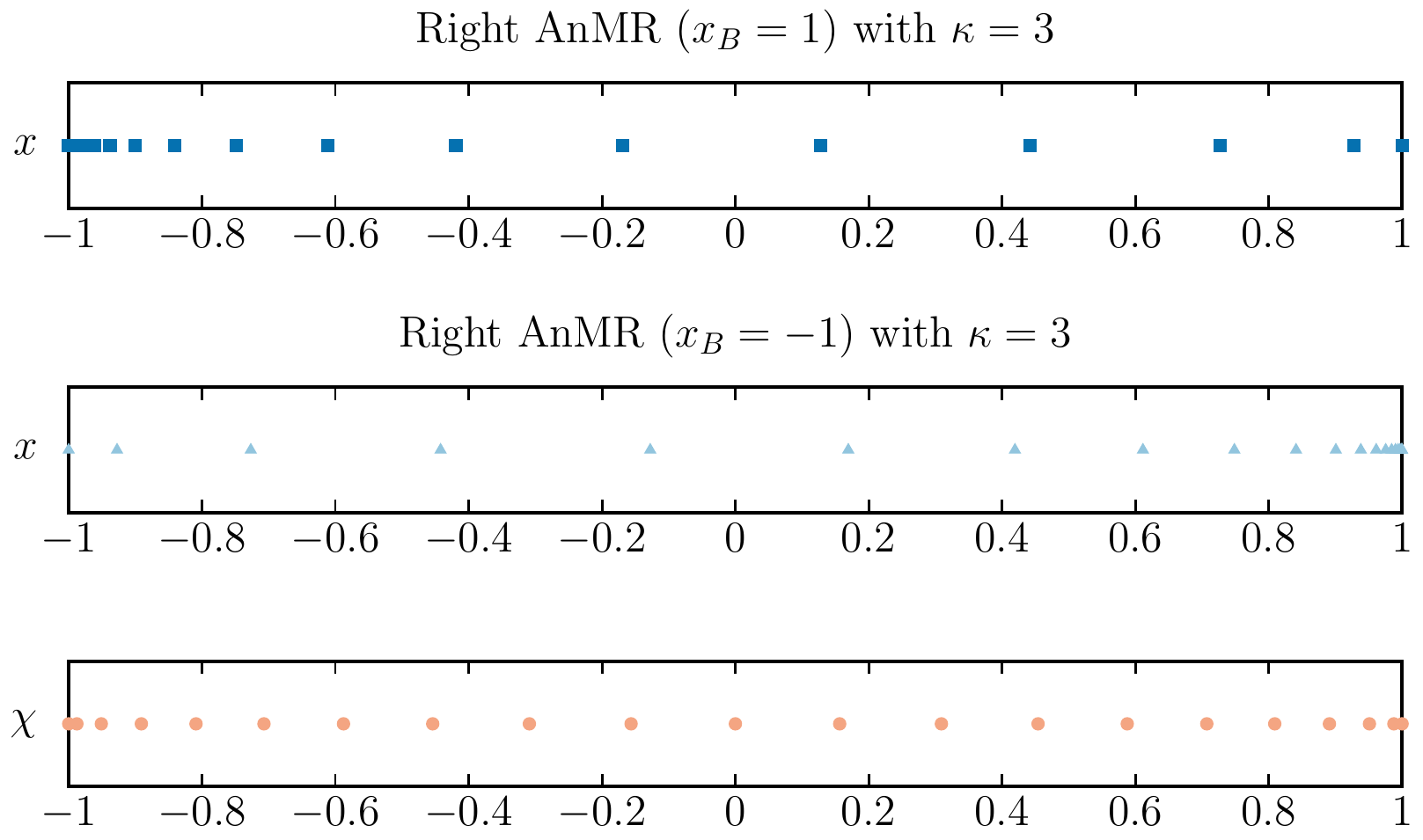}
\caption{Analytic mesh-refinement (AnMR) to better resolve functions with strong gradients. {\em Bottom panel:} A Chebyshev-Lobatto grid is considered for the coordinate $\chi\in[-1,1]$ according to Eq.~\eqref{eq:LobattoChebGrid_chi}. {\em Middle panel:} For $x_{\rm B} = -1$, the AnMR map \eqref{eq:AnMR} populates the grid points around the left boundary. {\em Top panel:} For $x_{\rm B} = 1$, the AnMR map \eqref{eq:AnMR} populates the grid points around the right boundary. Examples with AnMR-parameter $\kappa = 3$.}
\label{fig:AnMR}
\end{figure}
Within a given domain $i_{\rm d}$, we map the interval $[-1,1]$ into itself via
\beq
\label{eq:AnMR}
x = x_{\rm B}\Bigg( 1 - \dfrac{2 \sinh\left[ \kappa (1-x_{\rm B} \chi) \right] }{\sinh(2\kappa)} \Bigg), \quad {\chi \in [-1,1]},
\eeq
with a mesh-refinement parameter $\kappa \geq 0$. The limit $\kappa\rightarrow 0$ recovers the identity $x=\chi$. The parameter $x_{\rm B}$ indicates whether the steep region is around the left ($x_{\rm B}= -1$) or the right boundary ($x_{\rm B}= 1$). The AnMR technique discretises the grid $\chi\in[-1,1]$ --- as opposed to $x$ in Eq.~\eqref{eq:LobattoChebGrid_x} --- via 
\beq
\label{eq:LobattoChebGrid_chi}
\chi_i = \cos\left(\pi \frac{i}{N_{i_{\rm d}}} \right), \quad i = 0 \cdots N_{i_{\rm d}}.
\eeq
The grid $x_i$ follows from the AnMR mapping \eqref{eq:AnMR}, which then fixes the grid in the hyperboloidal radial coordinate $\sigma$ via Eq.~\eqref{eq:sigma_func_x}.  
The bottom panel of Fig.~\ref{fig:AnMR} displays the Chebyshev-Lobatto grid for the $\chi$ coordinate according to Eq.~\eqref{eq:LobattoChebGrid_chi}, whereas the middle and top panels reveal the effect of the mapping
\eqref{eq:AnMR} with a parameter $\kappa=3$. They demonstrate, respectively, the accumulation of grid points on either the left $(x_{\rm B} = -1)$ or right boundary $(x_{\rm B} = 1)$. As we shall demonstrate, the increase of point 
density in these regions allows us to accurately represent functions with steep gradients around the particle with a low-to-moderate spectral resolution $N_{i_{\rm d}}$.

\section{Results}\label{sec:results}
This section presents several numerical results that demonstrate the effectiveness of self-force calculations using compactified hyperboloidal coordinates combined with spectral methods.
In each subsection, we present results for examples from the three classes of sources commonly found in self-force calculations: distributional, worldtube, and unbounded support.
Computing self-force for large radius orbits is a challenging problem for all three classes and is therefore presented in a separate subsection.
At all steps of the code development, we compare the solution around the particle with the corresponding solution obtained from the Black Hole Perturbation Toolkit (BHPToolkit) \cite{BHPToolkit}. 
Such cross-checks attest to our results' correctness and allow us to perform convergence tests.

\subsection{Distributional sources}\label{sec:results_distributional_source}

We compute the retarded field $\overline{\phi}_{\ell m}^{\rm ret}$ using a distributional source solving Eq.~\eqref{eq:Hyperboloidal_Laplace_eq}.
We first examine individual modes and then present results for the self-force computed using the mode-sum approach.
In our computations we split the grid at the particle's location, $\sigma_p$, and employ the same spectral resolution in both domains, i.e., $N_1 = N_2 = N$. 

\begin{figure*}[ht!]
	 \centering
     \subfloat{
         \centering
         \includegraphics[width=0.48\textwidth]{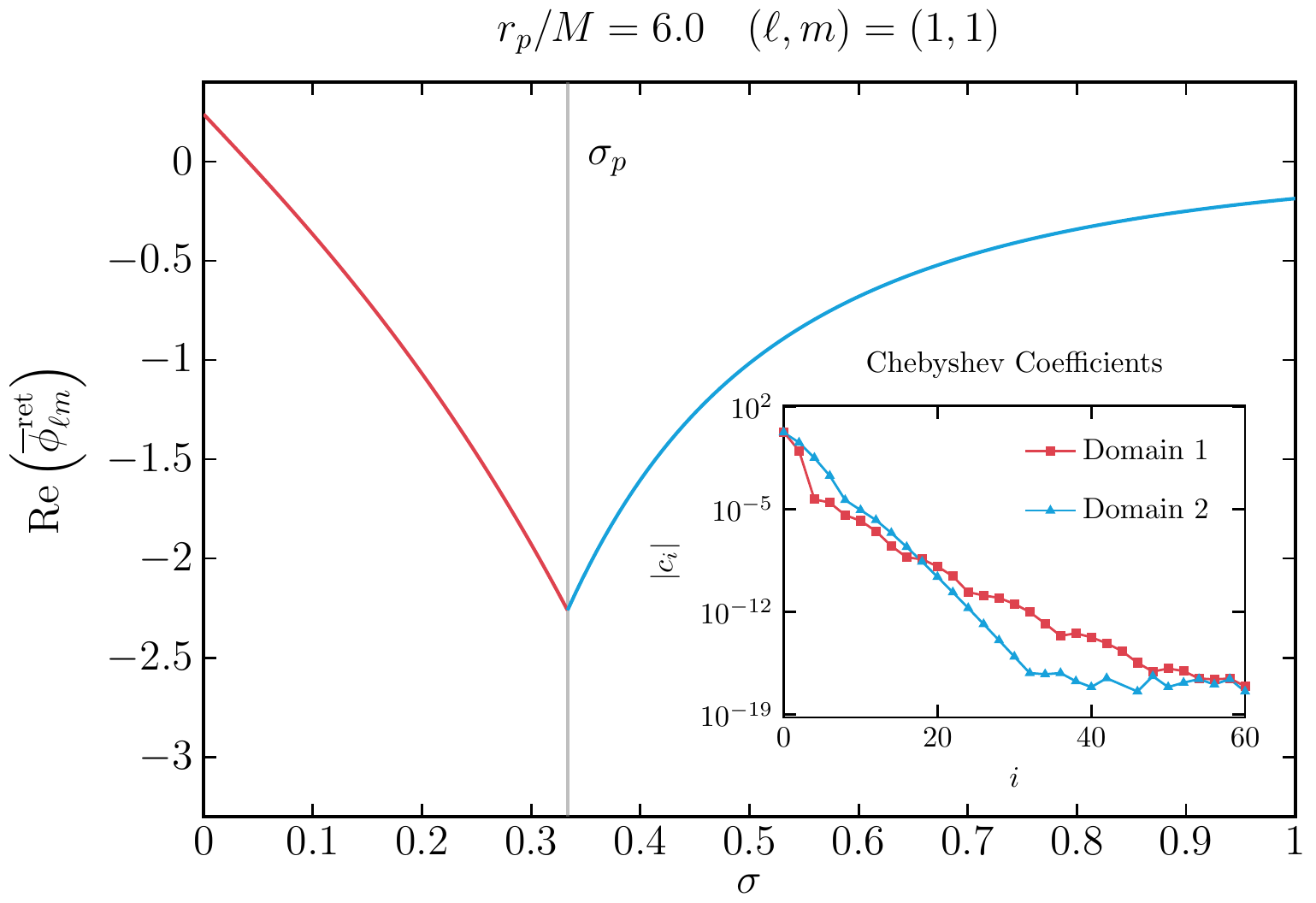}
     }
     \hfill
     \subfloat{
         \centering
         \includegraphics[width=0.48\textwidth]{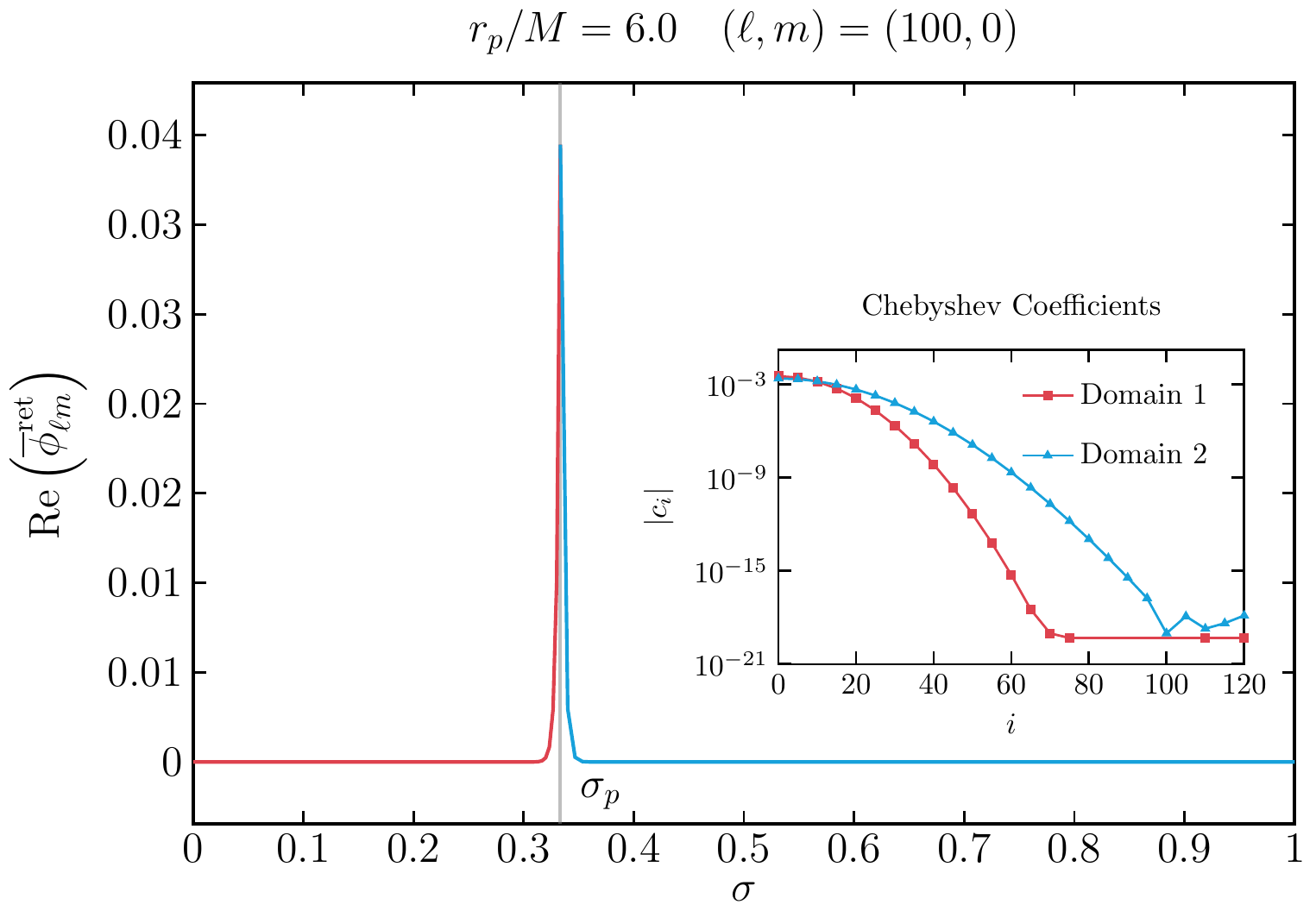}
     }
     \caption{{\em Left panel:} Real parts of hyperboloidal retarded field $\overline{\phi}^{{\rm ret }}_{\ell m}$ for the angular mode $(\ell,m) = (1,1)$ with $r_p=6M$ and $N=60$.  
         The numerical domain extends from future null infinity, $\sigma=0$, to the future event horizon, $\sigma=1$.  
         The inset demonstrates exponential decay of Chebyshev coefficients indicating spectral convergence.  {\em Right panel:} Same fields as in the left panel 
         but for angular mode $(\ell,m) = (100,0)$ and resolution $N=120$. 
         We need higher resolution at high mode numbers because of the steep gradient around the particle.  
         Insets demonstrate slower spectral convergence than in left panel.}
         \label{fig:ModeSum_Distributional}
\end{figure*}

The left panel of Fig.~\ref{fig:ModeSum_Distributional} displays the real part of hyperboloidal retarded field $\overline{\phi}^{{\rm ret }}_{\ell m}$ for the angular mode $(\ell,m) = (1,1)$, where 
the spectral resolution is $N=60$ and the particle is at $r_{p} = 6 M$. 
The retarded field $\overline{\phi}^{{\rm ret }}_{\ell m}$ is continuous with a discontinuity at the first radial derivative. 
Most importantly, the field is accessible in the entire domain, including future null infinity $\sigma = 0$ and the black-hole horizon $\sigma =1$. 
The solutions' accuracy and smoothness are assessed by the behavior of the corresponding Chebyshev coefficients $c_i$\footnote{To simplify the notation, we remove the labels ${({i_{\rm dom}},i_{\rm Fields}})$ used in section \ref{sec:NumMeth}, as this information is available within the plots.}.
The insets show the coefficients' exponential decay up to the round-off saturation of order $10^{-16}$.
In contrast, the right panel of Fig.~\ref{fig:ModeSum_Distributional} explores more extreme regions in the parameter space. 
Similar to the left panel of Fig.~\ref{fig:ModeSum_Distributional} this shows the real part of 
$\overline{\phi}^{{\rm ret }}_{\ell m}$ but with $(\ell,m) = (100,0)$ and $r_p = 6 M$.
The numerical solution requires a higher resolution $N$, especially in domain $2$, due to the steep gradient around the particle.

These computations of the transformed fields $\overline{\phi}^{{\rm ret }}_{\ell m}$ demonstrate the internal consistency of the code in the compact hyperboloidal formulation. 
The field $\phi^{\rm ret }_{\ell m}$ and its derivative are used in the calculation of the self-force and can be reconstructed from $\overline{\phi}^{{\rm ret }}_{\ell m}$ via Eq.~\eqref{eq:HyperFieldReScale}. 
As discussed in Sec.~\ref{sec:NumMeth}, we take a numerical solution with the high resolution $N^{\rm ref}=150$ as reference and evaluate the relative error ${\cal E}_N$ according to Eq.~\eqref{eq:RelError} for the physical retarded field and its $r_p$-derivative at the particle's location. 
We observe spectral convergence, with higher angular modes requiring higher numerical resolution to obtain a given precision. 
The behavior for high angular modes is a consequence of the steep gradients around the particle observed in 
right pane of Fig.~\ref{fig:ModeSum_Distributional}. 
Nevertheless, the required resolution is not prohibitive, as all $\ell-$modes seem to converge similarly. 
The main effect of increasing $\ell$ is an upward shift in the curves, and one obtains accurate solutions for $\ell$-modes as high as $\ell=100$ with moderate resolution $N=100$.

\begin{figure}[ht]
\centering
\includegraphics[width=0.48\textwidth]{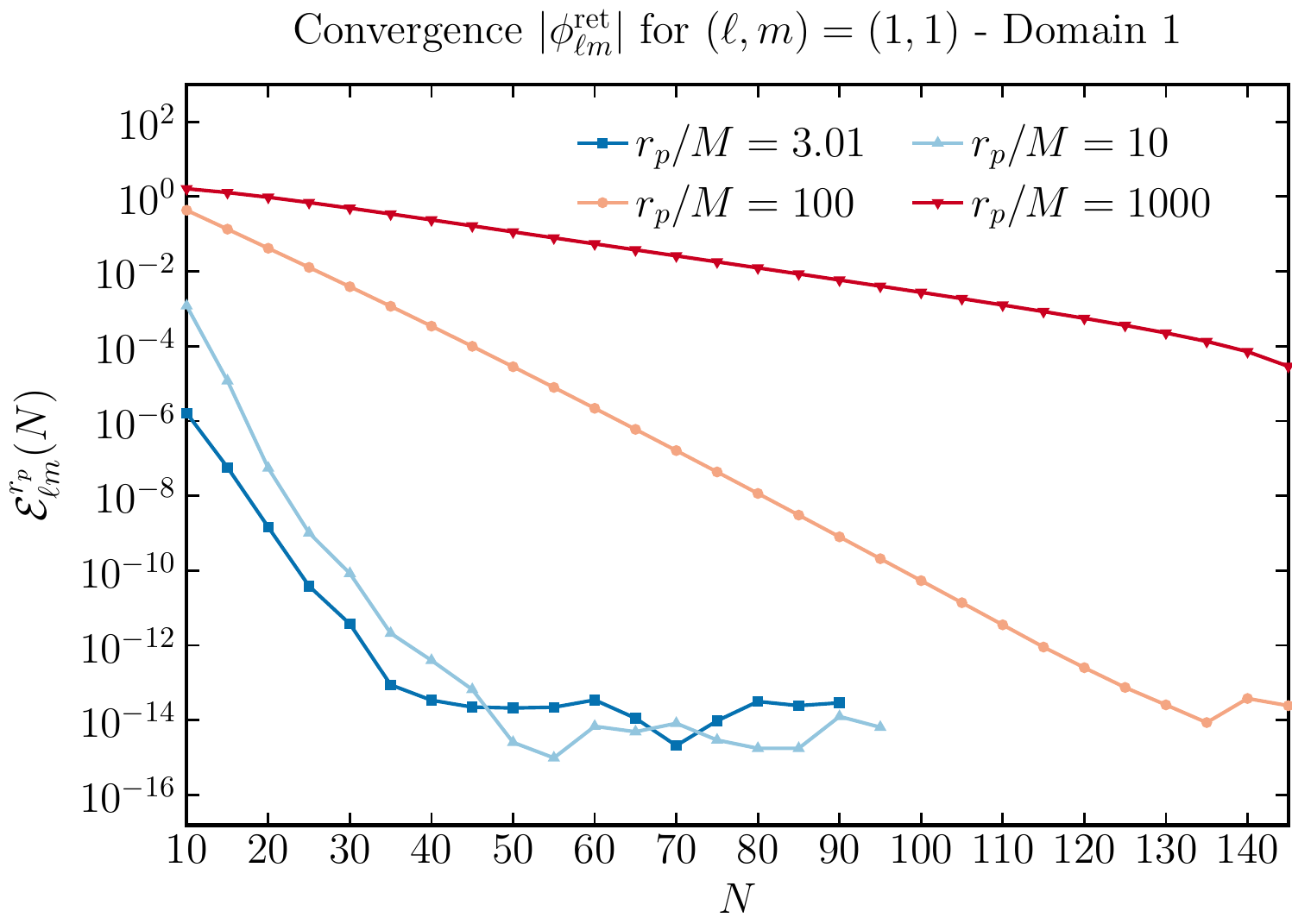}
\caption{Numerical convergence for the retarded field $|{\phi^{{\rm ret }}_{\ell m}}|$ with $(\ell,m)=(1,1)$ as function of particle's orbit $r_p$. Despite the exponential decay of error against a reference solution with $N^{\rm ref}=150$, the decay rate is slower at large orbits $r_p$ due to steep gradients.
}
\label{fig:DoubleDomainConvergence}
\end{figure}

In Fig.~\ref{fig:DoubleDomainConvergence}, we show convergence for a fixed $(\ell, m)=(1, 1)$ but varying $r_p/M=\{ 3.01, 10, 100, 1000\}$. 
Even though we observe exponential convergence regardless of $r_p$, the convergence rate decreases for higher values of $r_p$. 
For instance, resolution $N=150$ yields solutions $|{\phi^{{\rm ret }}_{\ell m}}|$ correct only up to $10^{-4}$ for $r_p=1000M$. 
The reason behind the poor convergence rate is the steep gradients around the particle, so the spectral method loses accuracy for large $r_p$. 
One can, however, solve this problem using analytic mesh refinement (AnMR) as presented in Sec.~\ref{sec:AnMR_results} below.

To further validate our code, we calculate both the energy flux and the local self-force.
The energy flux at infinity, $\dot{E}^{\scri^{+}}$, and the horizon, $\dot{E}^{\mathcal{H}^{+}}$, is computed from the values of the field at $\sigma = 0$ and $\sigma = 1$, respectively, using Eqs.~\eqref{eq:flux_scri} and \eqref{eq:flux_hor}.
In Table \ref{tbl:scalar_flux} we present numerical values for these scalar energy fluxes.
Table \ref{tbl:scalar_flux} also presents a direct comparison with values for the flux at the horizon and spatial infinity computed using the \texttt{Teukolsky} package of the Black Hole Perturbation Toolkit \cite{BHPToolkit}, with relative differences comparable in magnitude to machine precision.

{\renewcommand{\arraystretch}{1.2}
\begin{table*}[hbt!]
\begin{tabular*}{\textwidth}{ @{\extracolsep{\fill}} c S[table-format=3.3] S[table-format=3.3] S[table-format=3.3]  
S[table-format=3.3] S[table-format=3.3]}
\toprule
$r_{p} / M$	
& \multicolumn{1}{c}{$\dot{E}^{\mathcal{H}^{+}} \times \mu(M/q)^{2}$}
& \multicolumn{1}{c}{$\dot{E}^{\scri^{+}} \times \mu(M/q)^{2}$}
& \multicolumn{1}{c}{$F_{t} \times (M/q)^{2}$}	
& \multicolumn{1}{c}{$1 - | F_{t} / u^{t} \dot{E}_{\text{total}} |$}	
& \multicolumn{1}{c}{$1 - \mu | \dot{E}_{\text{{total}}} / \dot{E}
^{\text{BHPT}}_{\text{{total}}} |$} \\
\toprule
$6$	
&		\num{7.85026d-6}		
&		\num{2.47345d-4}
&		\num{3.60907d-4}
&		\num{7.60d-12}
&		\num{-1.04d-10} \\
$7$
&		\num{2.40585d-6}
&		\num{1.31191d-4}
&		\num{1.76732d-4}
&		\num{-6.95d-12}
&		\num{-4.12d-12} \\
$8$
&		\num{8.82307d-7}
&		\num{7.63725d-5}
&		\num{9.77204d-5}
&		\num{6.64d-12}
&		\num{-6.88d-13} \\
$10$
&		\num{1.70076d-10}
&		\num{3.12066d-5}
&		\num{3.75023d-5}
&		\num{3.38d-12}
&		\num{-4.18d-13} \\
$14$
&		\num{1.48586d-8}
&		\num{8.17262d-6}
&		\num{9.23673d-6}
&		\num{4.85d-12}
&		\num{7.92d-13} \\
$20$
&		\num{1.15966d-9}
&		\num{1.98251d-6}
&		\num{2.15159d-6}
&		\num{1.13d-12}
&		\num{6.42d-13} \\
$30$
&		\num{6.53417d-11}
&		\num{3.96179d-7}
&		\num{4.17679d-7}
&		\num{-2.18d-12}
&		\num{1.96d-12} \\
$50$
&		\num{1.77767d-12}
&		\num{5.19670d-8}
&		\num{5.36017d-8}
&		\num{-7.93d-12}
&		\num{1.10d-11} \\
$70$
&		\num{1.66651d-6}
&		\num{1.36106d-8}
&		\num{1.39122d-8}
&		\num{3.39d-12}
&		\num{3.03d-11} \\
$100$
&		\num{1.36047d-14}
&		\num{3.28462d-9}
&		\num{3.33504d-9}
&		\num{4.42d-9}
&		\num{1.35d-10} \\
\botrule
\end{tabular*}
\caption{Sample numerical results for the scalar-field energy flux for a range of numerical values of $r_p$ at exactly $
\mathcal{H}^{+}$ and $\scri^{+}$ in the second and third column respectively. The fourth column displays $t$-component of
the SSF calculated locally using Eq.~\eqref{eq:F_lt}. Column five is an internal consistency check comparing the 
$t$-component of the self-force calculated locally with and using the total energy flux and balance law in 
Eq.~\eqref{eq:work_done_relation}.  Column six presents a comparison of the total energy flux with 
results obtained from the \texttt{Teukolsky} package of the Black Hole Perturbation Toolkit \cite{BHPToolkit}.}
\label{tbl:scalar_flux}
\end{table*}
}

\begin{figure}[ht!]
	\centering
	\includegraphics[width=0.48\textwidth]{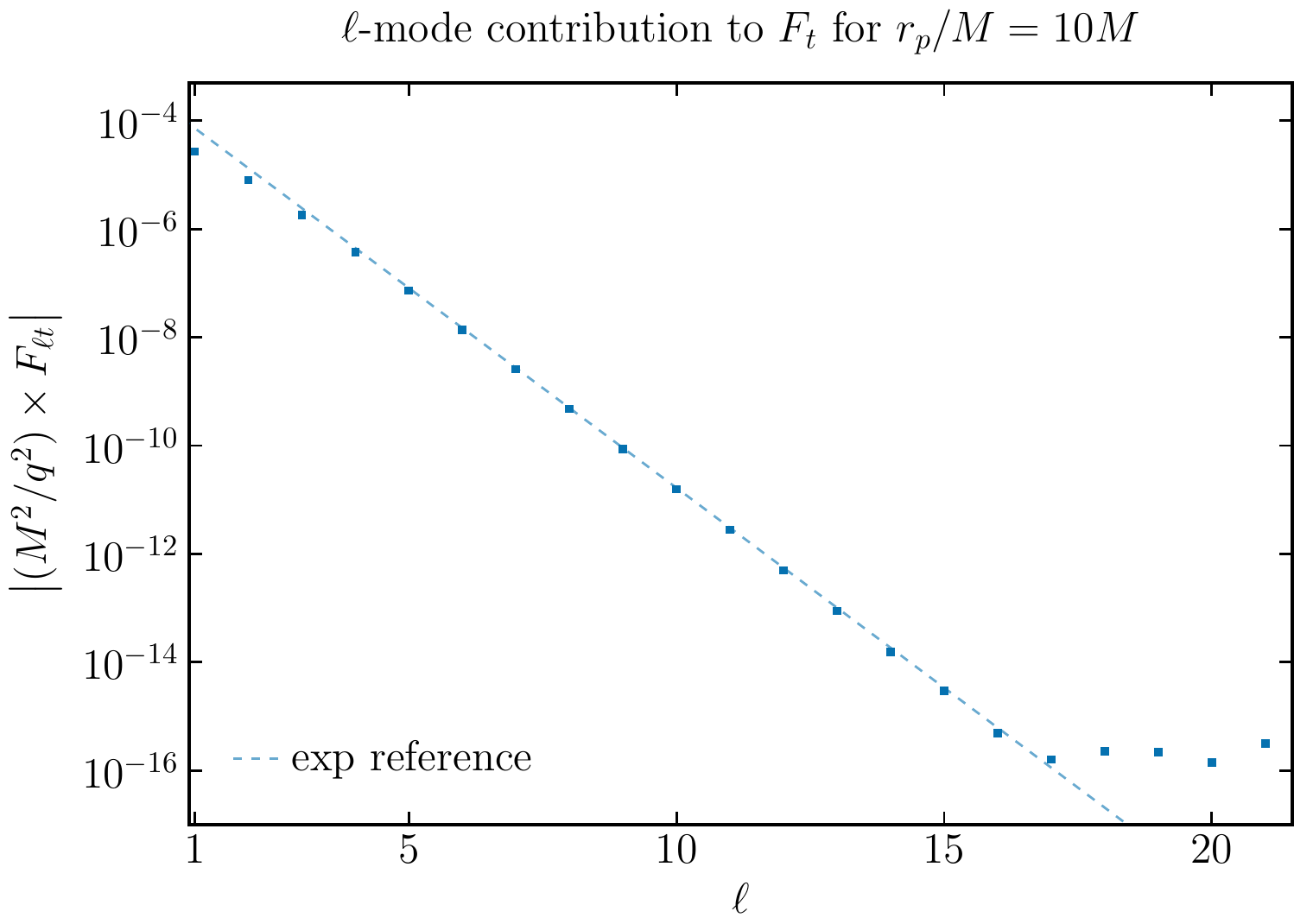}
	\caption{The $\ell-$mode contributions to the $t$-component of SSF, $F_{\ell t}$, for a particle on a circular orbit of radius $r_p=10M$.
	The modes of $F^\ell_t$ converge exponentially until machine precision round-off is encountered near $\ell = 17$.}
	\label{fig:Flt_ModeSum}
\end{figure}

\begin{figure}[ht!] 
	\centering
	\includegraphics[width=0.48\textwidth]{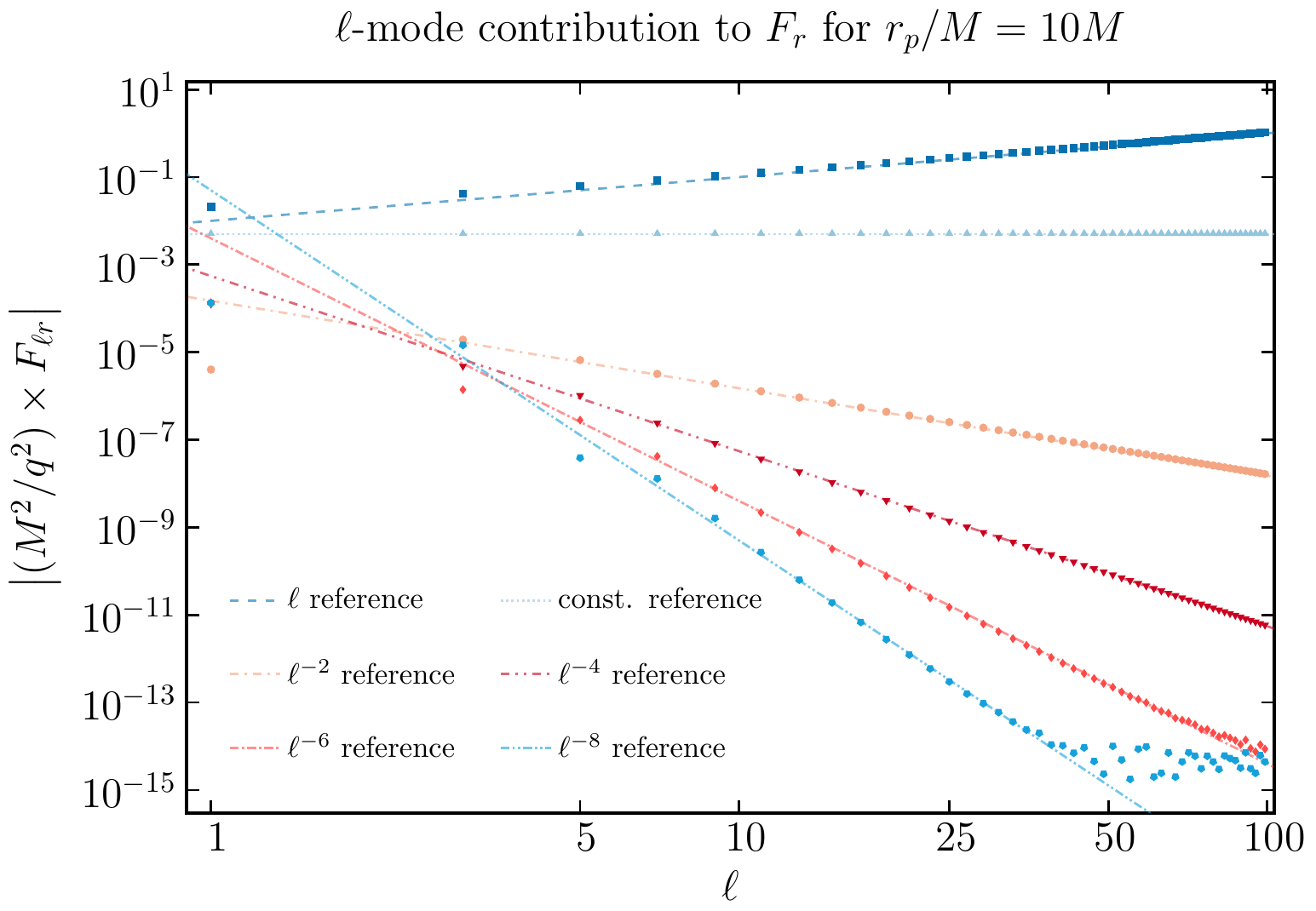}
	\caption{The $\ell-$mode contribution to the $r$-component of the SSF, $F_{\ell r}$, for a particle on a circular orbit of radius $r_p=10M$.
	For large $\ell$ the unregularized modes (blue squares) grow linearly.
	After subtracting the leading regularization parameter, the modes (light blue triangles) tend to a constant for large $\ell$.
	Further subtracting the next regularization parameter, the regular modes (orange circles) fall off as $\ell^{-2}$ for large $\ell$.
	The convergence of the $\ell$-mode sum is then accelerated using higher-order regularization parameters with each additional parameter changing the large $\ell$ behavior by $\ell^{-2}$.
	After all the known regularization parameters are subtracted, the modes quickly reach machine round-off.
	Note that the agreement with the expected large-$\ell$ behavior is excellent out to $\ell=100$ (when the contributions are above machine precision).}
	\label{fig:Flr_ModeSum}
\end{figure}

We also compute the self-force from the values of the derivative of the scalar field at the particle's location.
The $t$-component of the self-force, $F_{\ell t}$, is computed using Eq.~\eqref{eq:F_lt}.
The modes of $F_{\ell t}$ do not require any regularization, and in Fig.~\ref{fig:Flt_ModeSum} we see that the contribution from each $\ell$-mode falls off exponentially to machine round-off.
The $r$-component of the self-force, $F_{\ell r}$, is computed using Eq.~\eqref{eq:F_lr}.
The individual $\ell$-modes of the radial self-force do require regularization, which we perform using Eq.~\eqref{eq:mode-sum}.
In Fig.~\ref{fig:Flr_ModeSum} we show the behavior of both the unregularized and regularized $\ell$-modes of the self-force.
The delicate cancellation between the modes of the retarded field and the regularization parameters is a good test of the 
correctness of our code, and we find that our numerical results are excellent for modes as high as $\ell=100$.

\subsection{Worldtube sources}\label{sec:Results_EffectiveSource}

As an example of a worldtube source, we solve Eq.~\eqref{eq:HypEq_EffecSource} for the hyperboloidal residual field 
$\overline{\phi}^{\mathcal{R}}_{\ell m}$.
The effective-source for this equation has compact support within a region around the particle.
This naturally suggests a four-domain grid for our spectral solver. 
We scale the numerical resolution as $N_1 = 2 N_2 = 2 N_3 = N_4 = N$. 
In the compact radial coordinate $\sigma$, the puncture field regularizing the source takes values in a window around the particle fixed by
\beq
	\sigma_- = \frac{\sigma_p}{2}, \quad \sigma_+ = \frac{1+\sigma_p}{2}.
\eeq
The corresponding physical coordinates $r_\pm(\sigma_\mp)$ read
\beq
	r_+ = 2 r_p, \quad r_- = \dfrac{2 r_p}{1+r_p/(2M)}.
\eeq
This choice halves the region between future null infinity, $\sigma=0$, and the particle, $\sigma=\sigma_p$, as well as between the particle and the horizon $\sigma=1$. 
Thus, the problem is formulated on the four domains
\bit
	\item Domain 1: $\sigma \in [0,\sigma_-]$,
	\item Domain 2: $\sigma \in [\sigma_-,\sigma_p]$,
	\item Domain 3: $\sigma \in [\sigma_p,\sigma_+]$,
	\item Domain 4: $\sigma \in [\sigma_+,1]$.
\eit
\begin{figure}[ht]
	\centering
	\includegraphics[width=0.48\textwidth]{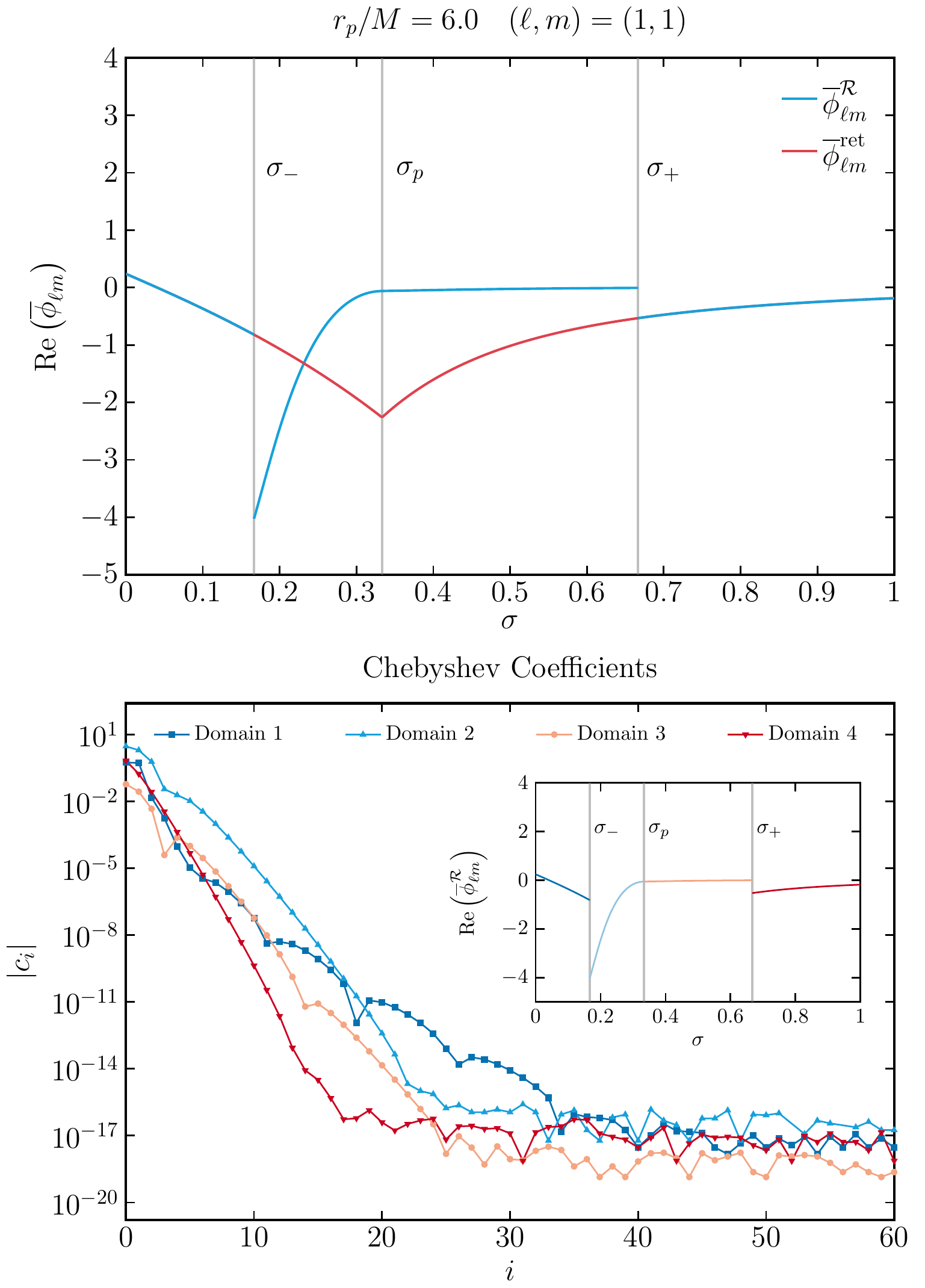}
	\caption{{\em Top panel:} Effective source solution using four-domains with angular mode $(\ell,m)=(1,1)$ and particle location $r_p = 6M$. {\em Bottom panel:} Decay of Chebyshev coefficients in all domains 
	demonstrates spectral convergence. The inset displays the real part of residual field $\overline{\phi}^{\mathcal{R}}_{\ell m}$ with a separate color for each domain.}
	\label{fig:EffectiveSource_X_ModeSum}
\end{figure}
\begin{figure}[ht]
	\includegraphics[width=0.48\textwidth]{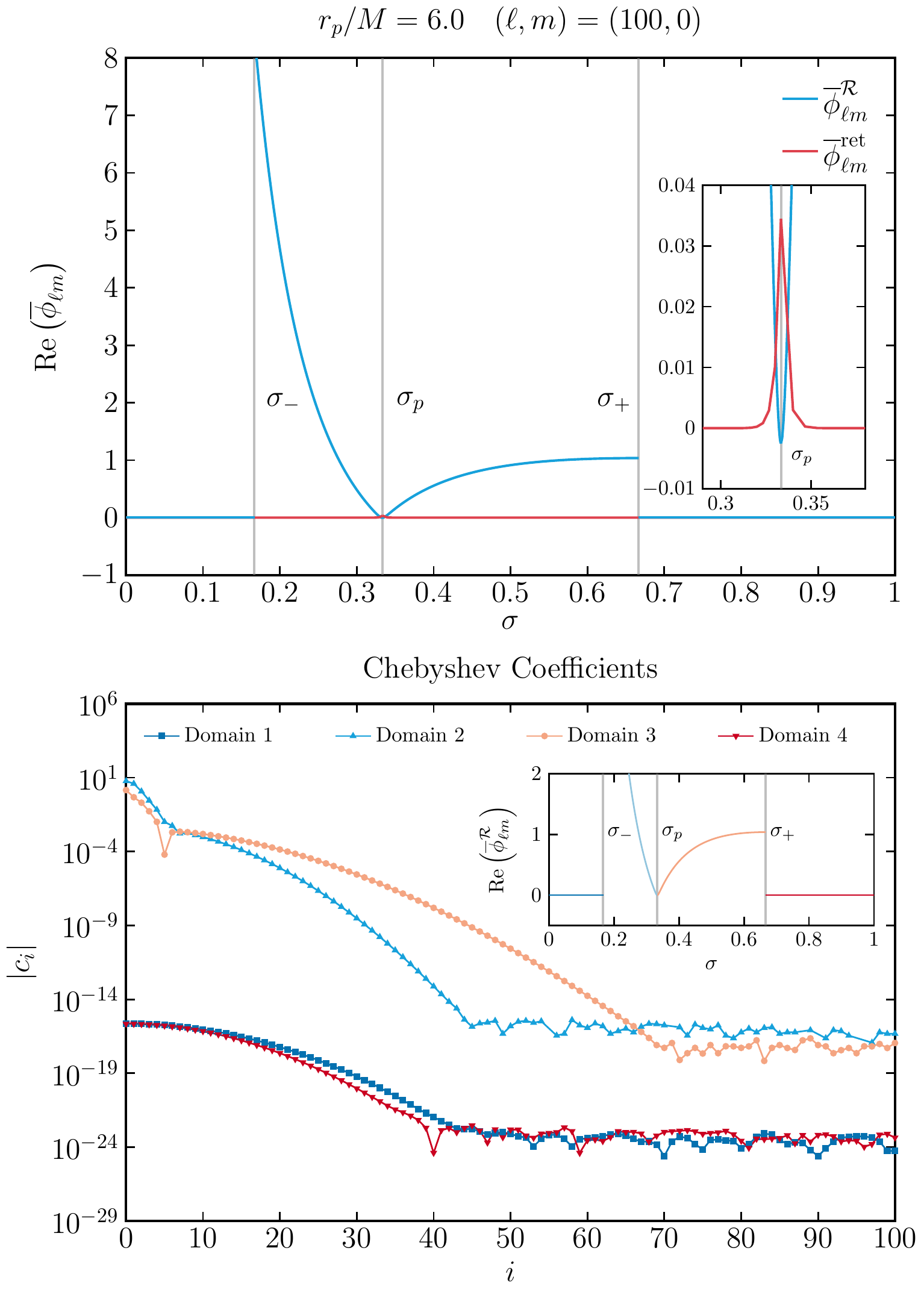}
	\caption{Same setup as in Fig.~\ref{fig:EffectiveSource_X_ModeSum} but for a high angular mode $(\ell,m)=(100,0)$.
	}\label{fig:EffectiveSource_X_ModeSum_SteepGrad}
\end{figure}
We explore the same set of parameters as in the previous section. 
Figure \ref{fig:EffectiveSource_X_ModeSum} displays the results for a fixed angular mode $(\ell,m)=(1,1)$ with the particle located at $r_p = 6M$ and numerical resolution $N=60$. 
Figure \ref{fig:EffectiveSource_X_ModeSum_SteepGrad} shows the results for large angular $(\ell,m) = (100,0)$ and $N=100$.

The hyperboloidal residual field $\overline{\phi}^{\mathcal{R}}_{\ell m}$ (blue) is discontinuous across the window boundaries $\sigma_\mp$, but continuous at the particle's location $\sigma_p$. 
For a consistent comparison, these panels also display (in red) the corresponding retarded field $\overline{\phi}^{\rm ret}_{\ell m} = \overline{\phi}^{\mathcal{R}}_{\ell m} + \overline{\phi}^{\mathcal{P}}_{\ell m}$. 
As the effective-source only has support inside the worldtube we have $\overline{\phi}^{\mathcal{R}}_{\ell m} = \overline{\phi}^{\rm ret}_{\ell m}$ in the domains $1$ and $4$, and $\overline{\phi}^{\mathcal{R}}_{\ell m} = \overline{\phi}^{\rm ret}_{\ell m}-\overline{\phi}^{\mathcal{P}}_{\ell m}$ at domains $2$ and $3$. 
The smoothness of the retarded field across the worldtube boundaries is an important consistency check on the results for $\overline{\phi}^{\mathcal{R}}_{\ell m}$.

The bottom panels on Figs.~\ref{fig:EffectiveSource_X_ModeSum} and \ref{fig:EffectiveSource_X_ModeSum_SteepGrad} display the Chebyshev coefficients within each domain. 
These plots have an inset, where we reproduce the real part of residual field $\overline{\phi}^{\mathcal{R}}_{\ell m}$ with a color code identifying each of the four domains. 
As in the previous section, the coefficients' spectral decay to numerical round-off indicates high accuracy. 
High angular modes as in Fig.~\ref{fig:EffectiveSource_X_ModeSum_SteepGrad} require higher resolution due to steep gradients around the particle.
\begin{figure}[ht]
	\centering
	\includegraphics[width=0.48\textwidth]{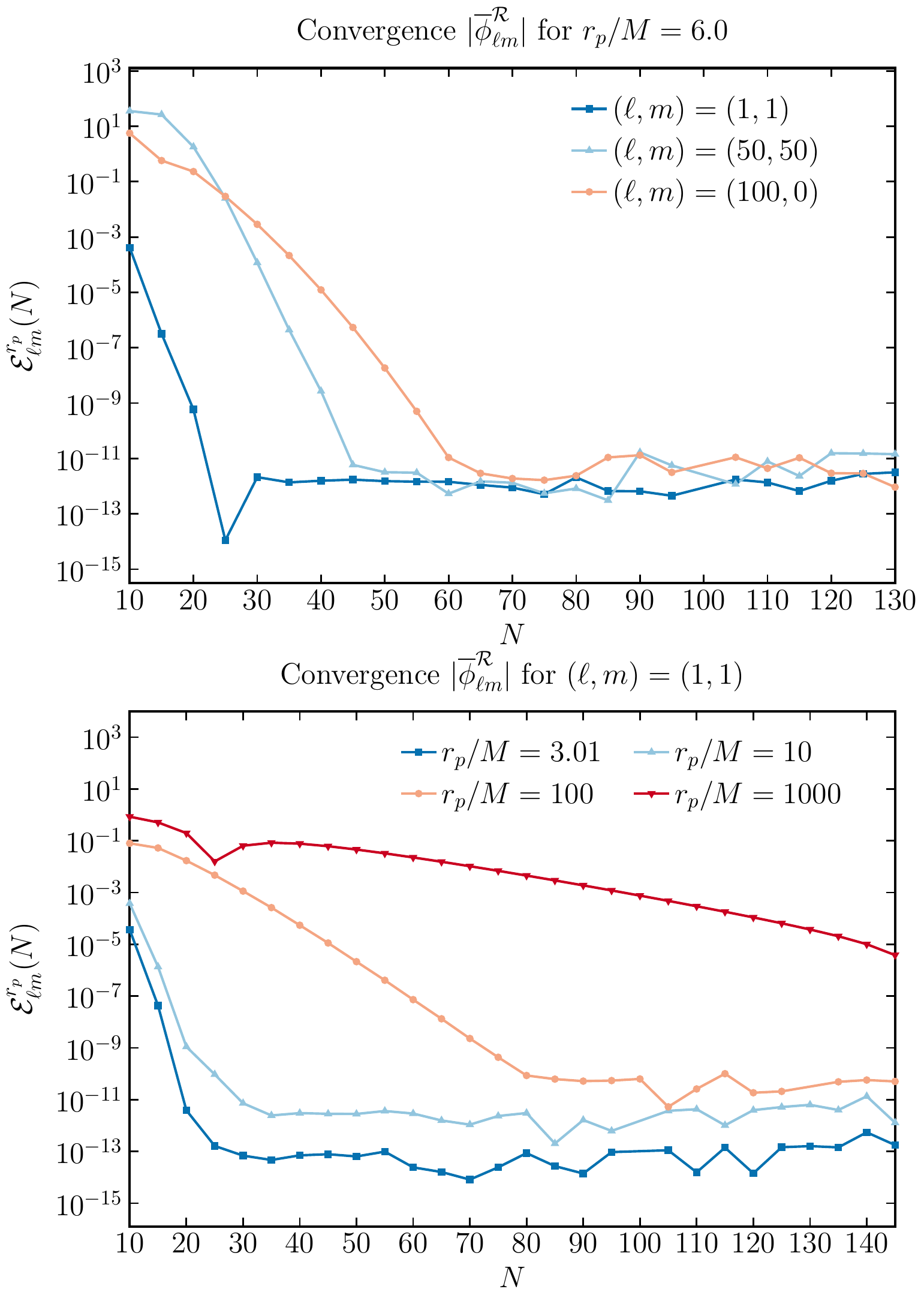}
	\caption{Convergence tests for the residual field $\left|{\phi}^{\mathcal{R}}_{\ell m}\right|$ displaying the error against a reference solution with $N^{\rm ref} = 150$ according to Eq.~\eqref{eq:RelError}. 
	{\em Top panel:} Angular modes $(\ell, m) = (1,1), (50,50)$ and $(100,0)$ with the particle at $r_p = 6M$. Higher angular modes require slightly higher resolution. {\em Bottom panel:} Various particle locations 
	$r_p/M = \{3.01, 10, 100, 1000\}$ with a fixed angular mode $(\ell, m) = (1,1)$. The exponential decay rate is lower for larger orbits.}
	\label{fig:EffectiveSource_Convergence}
\end{figure}

Next, we discuss convergence tests for the residual field $\left|{\phi}^{\mathcal{R}}_{\ell m}\right|$. 
By fixing a reference solution with $N^{\rm ref} = 150$, we calculate the relative error at the particle according to Eq.~\eqref{eq:RelError}. 
The top panel of Fig.~\ref{fig:EffectiveSource_Convergence} compares the code's convergence for the angular modes $(\ell, m) = (1,1), (50,50)$ and $(100,0)$ with the particle at $r_p = 6M$. 
We encounter the expected exponential convergence, with higher angular modes requiring slightly higher resolution. 
The bottom panel compares the convergence for various particle locations $r_p/M = \{3.01, 10, 100, 1000\}$ with a fixed angular mode $(\ell, m) = (1,1)$. 
As in Sec.~\ref{sec:results_distributional_source}, the error decays exponentially in all cases. As before, the decay rate is lower for larger $r_p$ values. 
For instance, when $r_p=1000M$, one only achieves an accuracy of $\sim 10^{-6}$ with $N=150$.

\begin{figure}[ht!] 
	\centering
	\includegraphics[width=0.48\textwidth]{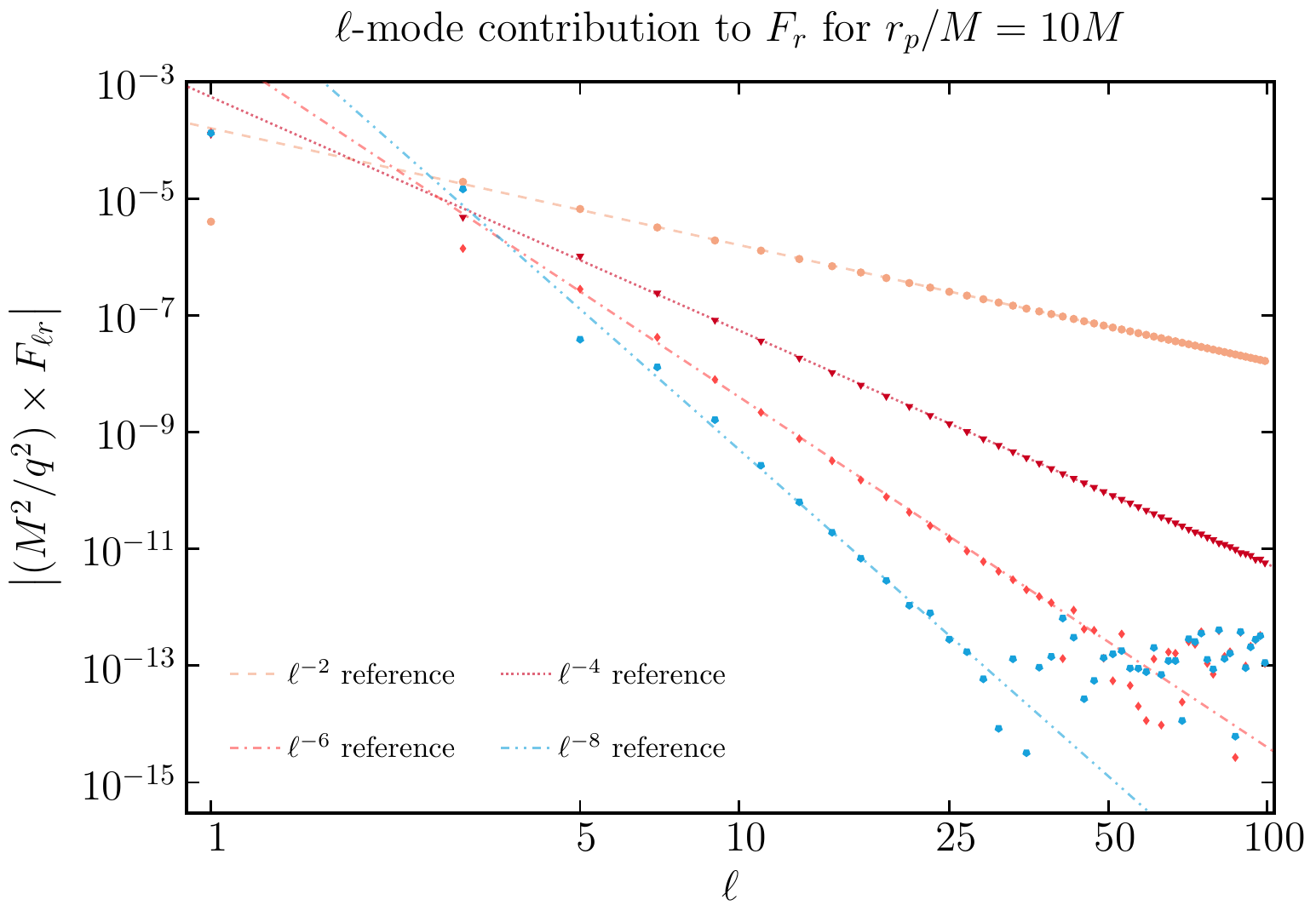}
	\caption{The $\ell$-mode contribution to the $r$-component of self-force, $F_{\ell r}$, computed using the effective-source method.
	The direct output of the hyperboloidal calculation with the effective-source are shown as the (orange) circles which fall off as $\ell^{-2}$.
	We then use higher-order regularization parameters to accelerate the convergence of the series.
	The more rapidly convergent series quickly reaches machine precision round-off.
	The results presented here  for the regularized force are, as expected, the same as the results from the mode-sum approach
	-- see Fig.~\ref{fig:Flr_ModeSum}.
	With our setup the effective-source method is more efficient than the distributional source and mode-sum approach.
	This is because with the distributional source large gradients of the field occur near the particle which necessitates $N=150$ Chebyshev nodes in each domain where the effective-source only requires $N=50$.
	}
	\label{fig:Flr_EffectiveSource}
\end{figure}

We further check our results by computing components of the self-force.
For the $t$-component, $F_{\ell t}$, our results are almost identical to those presented for the distributional source in Fig.~\ref{fig:Flt_ModeSum}.
Using the effective-source approach we directly compute the modes of the residual field, and from their radial derivatives the modes of the radial self-force, $F_{\ell r}$, can be computed using Eq.~\eqref{eq:F_lr}.
With the effective-source in Eq.~\eqref{eq:HyperEffectSource} we expect $\ell^{-2}$ convergence of the $\ell$-modes of the self-force which we observe for modes up to $\ell=100$ -- see Fig.~\ref{fig:Flr_EffectiveSource}.
We then use higher-order regularization parameters to accelerate further the convergence of the $\ell$-mode sum \cite{Heffernan:2012su}.
This faster rate of convergence could also be achieved by using a higher-order puncture which would leave to a smoother effective-source \cite{Warburton:2013lea}.

\subsection{Sources with unbounded support}

\begin{figure}[ht!]
	\includegraphics[width=0.48\textwidth]{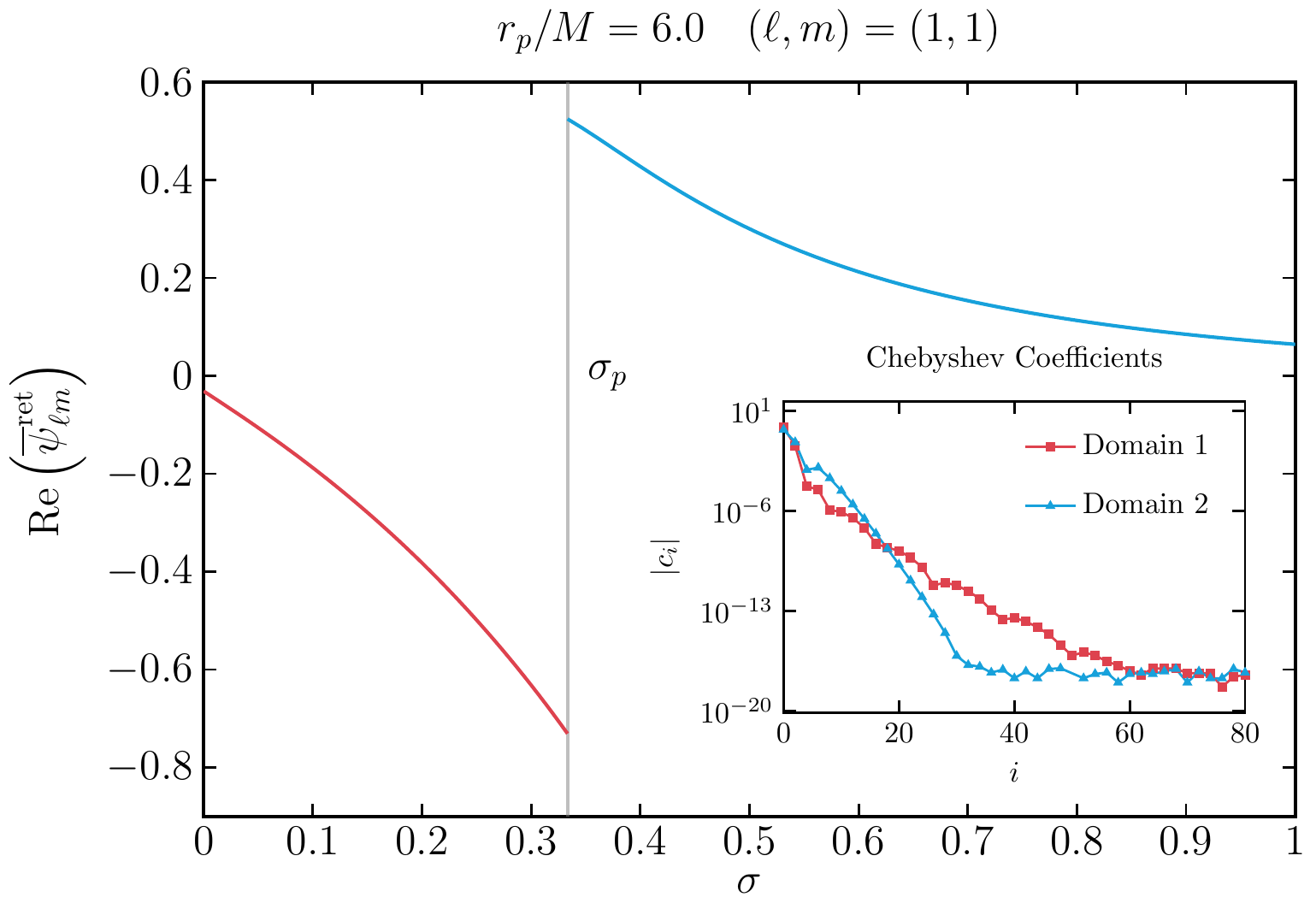}
	\caption{The $r_p$-derivative of the scalar field, $\overline{\psi}^{\rm ret}_{\ell m}$, computed for $r_p = 6M$ and $(l,m) = (1,1)$.
	The source for $\overline{\psi}^{\rm ret}_{\ell m}$ is unbounded but our approach handles it with ease.
	The inset shows the exponential convergence (until machine round-off is reached) for the Chebyshev coefficients in each 
	domain.}
	\label{fig:psi_r6_l1m1}
\end{figure}

\begin{figure}[ht!]
	\includegraphics[width=0.48\textwidth]{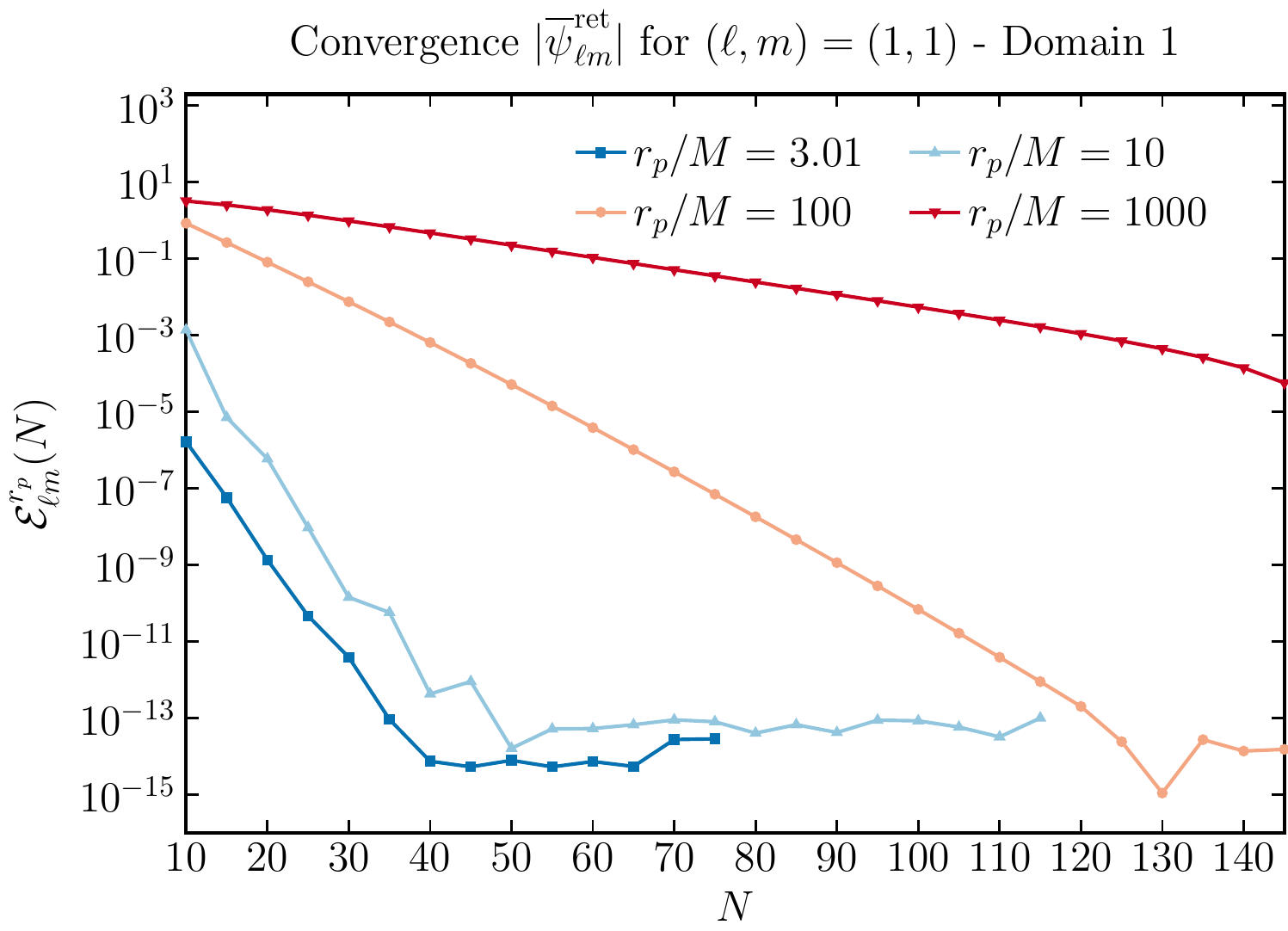}
	\caption{Convergence of $\psi_{\ell m}^{\rm ret}$ with increasing number of Chebyshev nodes, $N$, for different orbital 
	radii.
	In all cases the convergence is exponential but for large radius orbits the convergence can be quite slow.
	The rate of convergence can be improved with Analytic Mesh Refinement -- see Fig.~\ref{fig:AnMR_convergence}.}
	\label{fig:psi_convergence}
\end{figure}

As an example of a problem with an unbounded support source, we compute, $\overline{\psi}^{\rm ret}_{\ell m} = \overline{\phi}^{\rm ret}_{\ell m,r_p}$ which satisfies the field equation \eqref{eq:ExtendedSourceOperator_Hyp}.
The source for Eq.~\eqref{eq:ExtendedSourceOperator_Hyp} contains $\overline{\phi}^{\rm ret}_{\ell m}$ and so we solve for both fields simultaneously.
These problems are not well suited to the variations of parameter approach as explained in the introduction, but we find our hyperboloidal spectral approach handles them with ease.
In Fig.~\ref{fig:psi_r6_l1m1} we show the calculation of the $(1,1)$-mode of the $\psi_{\ell m}^{\rm ret}$ for a particle orbiting at $r_p=6M$.
As with the compact sources, the decay of the Chebyshev coefficients in the two domains demonstrate spectral convergence.
We see similar convergence properties for other orbital radii -- see Fig.~\ref{fig:psi_convergence}.
Again, the convergence is slower for large radius orbits with compact sources.

To check our results further, we compute the $r_p$-derivative of the energy flux radiated through the event horizon and to infinity.
For reference values to compare against, we use the \texttt{Teukolsky} package from the BHPToolkit to compute the numerical $r_{p}$-derivative of the fluxes.  
This is achieved by fitting a Taylor series centred around the $r_{p}$ value of interest using a densely populated grid of fluxes around $r_{p}$.  
It suffices for our expansion to be truncated at $\mathcal{O}(r_{p} - r)^{5}$ for a grid of 50 points equally spaced over the range $[ r_{p} - 0.05,\, r_{p} + 0.05 ]$.
This approach is very slow as we must solve the scalar wave equation many times for each $r_p$ value at which we wish to compute the $r_p$-derivative of the fluxes. 
We compare our hyperboloidal data to the numerically compute the $r_{p}$-derivative in Table \ref{table:dE/drp} and find excellent agreement.

{\renewcommand{\arraystretch}{1.2}
\begin{table*}[hbt!]\label{table:dE/drp}
\begin{tabular*}{\textwidth}{ @{\extracolsep{\fill}} c S[table-format=3.3] S[table-format=3.3] S[table-format=3.3]  
S[table-format=3.3] S[table-format=3.3]}
\toprule
$r_{p} / M$	
& \multicolumn{1}{c}{$\partial_{r_{p}}\dot{E}^{\mathcal{H}^{+}}_{\ell = 1} \times \mu(M/q)^{2}$}
& \multicolumn{1}{c}{$\partial_{r_{p}}\dot{E}^{\scri^{+}}_{\ell = 1} \times \mu(M/q)^{2}$}
& \multicolumn{1}{c}{$\mathcal{D}_{r_{p}}F_{1 t} \times (M/q)^{2}$}
& \multicolumn{1}{c}{$1 - | \mathcal{D}_{r_{p}}F_{1 t} / \partial_{r_{p}}(u^{t} \dot{E}^{\ell = 1}_{\text{total}}) |$}	
& \multicolumn{1}{c}{$1 - \mu | \mathcal{D}_{r_{p}}\dot{E}^{\ell = 1}_{\text{total}} / 
\partial_{r_{p}}\dot{E}^{\text{BHPT}}_{\text{{total}}} |$}	\\
\toprule
$6$	
&		\num{-9.37367d-6}		
&		\num{-7.51347d-5}
&		\num{-8.45083d-5}
&		\num{-2.32d-11}
&		\num{2.35d-13} \\
$7$
&		\num{-2.48526d-6}
&		\num{-3.96139d-5}
&		\num{-4.20992d-5}
&		\num{9.04d-11}
&		\num{-5.17d-13}\\
$8$
&		\num{-7.97541d-7}
&		\num{-1.11122d-5}
&		\num{-2.30219d-5}
&		\num{5.32d-11}
&		\num{-5.82d-13}\\
$10$
&		\num{-1.22413d-7}
&		\num{-8.20090d-6}
&		\num{-8.32331d-6}
&		\num{5.03d-11}
&		\num{-2.67d-13}\\
$14$
&		\num{-7.57877d-9}
&		\num{-1.73916d-6}
&		\num{-1.74674d-6}
&		\num{1.22d-10}
&		\num{8.92d-14}\\
$20$
&		\num{-4.11511d-10}
&		\num{-3.22739d-7}
&		\num{-3.23151d-7}
&		\num{3.63d-10}
&		\num{-2.71d-13}\\
$30$
&		\num{-1.53865d-11}
&		\num{-4.60066d-8}
&		\num{-4.60220d-8}
&		\num{7.24d-10}
&		\num{-9.41d-14}\\
$50$
&		\num{-2.50251d-13}
&		\num{-3.82238d-9}
&		\num{-3.82263d-9}
&		\num{2.23d-12}
&		\num{-5.50d-13}\\
$70$
&		\num{-1.67314d-14}
&		\num{-7.32062d-10}
&		\num{-7.32079d-10}
&		\num{2.44d-9}
&		\num{-1.41d-13}\\
$100$
&		\num{-9.54997d-16}
&		\num{-1.25886d-10}
&		\num{-1.25887d-10}
&		\num{3.65d-10}
&		\num{-4.03d-13}\\
\botrule
\end{tabular*}
\caption{Sample numerical results for the $r_{p}$-derivative of the scalar-field energy flux for the $\ell = 1$ mode for a 
range of numerical values of $r_{p}$ at exactly $\mathcal{H}^{+}$ and $\scri^{+}$ in the second and third column respectively.  
The fourth column presents the $\ell = 1$ mode of the $r_{p}$-derivative of the $t$-component of the SSF calculated using 
Eq.~\eqref{eq:dr0_F_lt}. Column five is an internal consistency check comparing the $t$-component of the self-force 
calculated locally and using the total energy flux and balance law in Eq.~\eqref{eq:work_done_relation}.  Column six presents 
a comparison of the $r_{p}$-derivative of the total energy flux with results obtained via numerically differentiating 
solutions from the \texttt{Teukolsky} package of the Black Hole Perturbation Toolkit (BHPToolkit) \cite{BHPToolkit} as 
described in the main text.}
\end{table*}
}

\subsection{Large radius orbits}\label{sec:AnMR_results}
\begin{figure}[ht!]
	\includegraphics[width=0.48\textwidth]{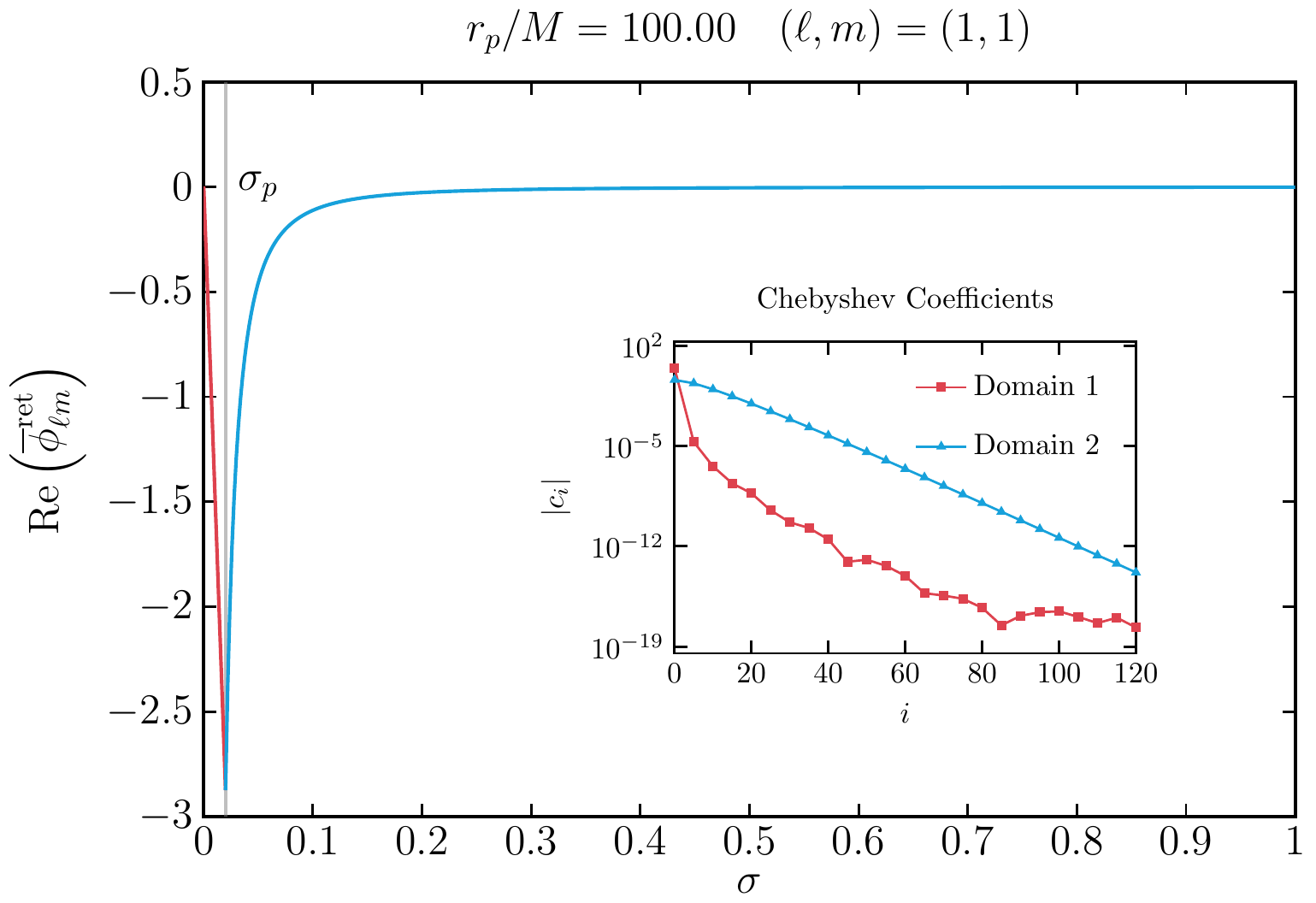}
	\caption{In self-force calculations for large orbits, strong gradients form in the compact coordinate $\sigma$ around the particle. 
}
\label{fig:ModeSum_ExtendedSource_Large_rp}
\end{figure}

We see in Figs.~\ref{fig:DoubleDomainConvergence} and \ref{fig:EffectiveSource_Convergence} that the convergence of the solution slows down for large orbits.
This slow convergence is due to the fixed mapping of the unbounded domain to a compact domain. 
In the compact radial coordinate $\sigma$, the region between null infinity at $\sigma=0$ and the particle at $\sigma=\sigma_p$ becomes very small as it scales as $\sim r_p^{-1}$ while the domain between the particle and the horizon becomes comparatively large. 
Strong gradients form because the main contribution to $\overline{\phi}_{\ell m}^{\rm ret}$ (or 
$\overline{\phi}_{\ell m}^{\mathcal{R}}$) comes from the region around the particle. 
These strong gradients are already visible for $r_p=100M$ depicted in Fig.~\ref{fig:ModeSum_ExtendedSource_Large_rp}.

One way to improve the accuracy of our results for large $r_p$ is to increase the number of sub-domains. 
We observe faster convergence at $r_p=1000M$ in the four-domain code (solving for $\overline{\phi}_{\ell m}^{\cal R}$ in Fig.~\ref{fig:EffectiveSource_Convergence}) than in the two-domain code (solving for $\overline{\phi}_{\ell m}^{\rm ret}$ in the bottom panel of Fig.~\ref{fig:DoubleDomainConvergence}).
These codes solve for different fields and the comparison between the errors are only valid at a qualitative level, but we can still confirm that the better convergence for the four-domain computation is a direct consequence of having more domains.

However, increasing the number of subdomains can quickly become prohibitive with the current algorithm. 
The ODE solver used in these computations employs an lower-upper (LU) decomposition scheme with a computational scaling as $n_{\rm total}^3$. 
Assuming that all $n_{\rm dom}$ have a numerical resolution of order $N$, one obtains $n_{\rm total} \sim n_{\rm dom} N$ and thus a scaling $n_{\rm dom}^3 N^3$. 
More subdomains require significantly more computational resources. 
It is evident from Fig.~\ref{fig:ModeSum_ExtendedSource_Large_rp} that the solution on much of the computational domain does not show any features that need to be resolved. 
Shifting the existing resources towards the steep gradients seems the appropriate solution. 
Therefore, instead of increasing the subdomains, we employ analytic mesh refinement (AnMR) described in Sec.~\ref{sec:AnMR} to achieve the desired high accuracy for large orbits while keeping the computational requirements low.
To demonstrate the effects of AnMR we will focus on problems with two domains below, but also present results for four domains using an effective source in Appendix \ref{apdx:AnMR_with_effective_source}.

\subsubsection{Analytic mesh refinement with two domains}

In this section we present large orbit calculations in our two-domain code, i.e., solving for the retarded field $\overline{\phi}_{\ell m}^{\rm ret}$ and its $r_p$-derivative $\overline{\psi}_{\ell m}^{\rm ret}$. 
We first concentrate on domain $2$, $\sigma \in [\sigma_p,1]$. 
Because the particle is located at the domain's left boundary, the mapping \eqref{eq:AnMR} is employed with $x_{\rm B} = -1$. 
\begin{figure}[ht]
	\centering
	\includegraphics[width=0.48\textwidth]{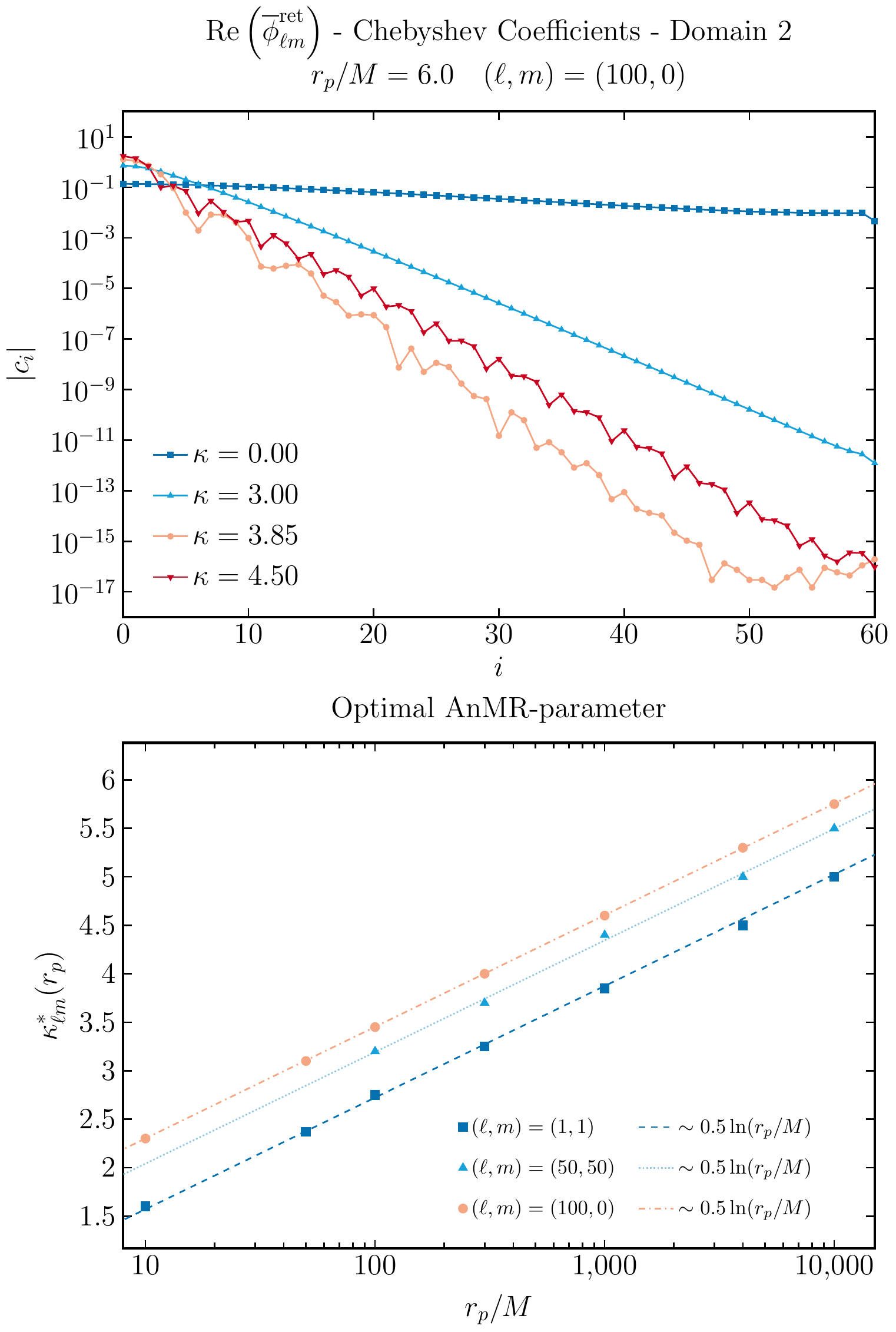}
	\caption{{\em Top panel:} Chebyshev coefficients of ${\rm Re}(\overline{\phi}_{\ell m}^{\rm ret})$ for $r_p=1000M$ and 
	$(\ell,m)=(1,1)$ in domain $2$ between the particle and the horizon. The optimal value for the AnMR parameter 
	is $\kappa=3.85$. {\em Bottom panel:} Optimal AnMR parameters plotted against $r_p/M$ with the fit Eq.~\eqref{eq:kappa_dom2} 
	for different angular modes. 
}
\label{fig:cheb_rp1000_ret_phi_AnMR_dom2}
\end{figure}
The top panel of Fig.~\ref{fig:cheb_rp1000_ret_phi_AnMR_dom2} displays the Chebyshev coefficients of ${\rm Re}(\overline{\phi}_{\ell m}^{\rm ret})$ for several AnMR-parameters $\kappa$ when $r_p=1000M$ and $(\ell,m)=(1,1)$. 
We observe slow convergence without AnMR ($\kappa = 0$). Increasing $\kappa$ increases the grid point density around the left boundary, and, as a consequence, the function becomes better represented by its spectral approximation, which improves the convergence rate. 
For instance, at $\kappa=0$, the coefficients assume values only of order $\sim10^{-2}$, while $\kappa=3$ yields coefficients down to order $\sim10^{-10}$.
For each combination of parameters $r_p$ and $(\ell,m)$, there exits an optimal value $\kappa^*_{\ell m}(r_p)$ leading to the fastest convergence. 
In Fig.~\ref{fig:cheb_rp1000_ret_phi_AnMR_dom2}, optimal decay is achieved at $\kappa^*_{11}(10^3M) \approx 3.85$, where the $c_i$'s reach the numerical round-off saturation at around $N\approx 50$. 
As we further increase $\kappa$, the coefficients' decay rate decreases once again.
We empirically find the optimal value for $\kappa^*_{\ell m}(r_p)$ on domain $2$ at several radii $r_p/M = \{10, 50, 100, 300, 1000, 4000, 10000\}$. 
The bottom panel of Fig.~\ref{fig:cheb_rp1000_ret_phi_AnMR_dom2} shows $\kappa^*_{\ell m}(r_p)$ for the angular modes used as example: $(\ell,m)=(1,1), (50,50),$ and $(100,0)$. We find that the fit
\beq
	\label{eq:kappa_dom2}
	\kappa^*_{\ell m}(r_p) \approx A_{\ell m} + 0.5 \ln\left(\dfrac{r_p}{M}\right),
\eeq
captures the $r_p$-dependence for the optimal $\kappa^*_{\ell m}$. 
Interestingly, the log-dependence is independent of the $(\ell, m)$-mode, and the only effect of the angular parameters is to shift the curve upwards. 
For instance, we have $A_{1,1} \approx 0.42$, $A_{50,50}\approx 0.89$, and $A_{100,0}\approx 1.15$.
One can also exploit the AnMR to increase the accuracy in the domain extending up to future null infinity (domain $1$). 
However, a systematic pattern for the optimal $\kappa^*_{\ell m}(r_p)$ in domain $1$ [similar to Eq.~\eqref{eq:kappa_dom2} in domain $2$] is absent. 
Appendix~\ref{app:AnMR} discusses this possibility and it brings an explicit example for the configuration $(\ell,m)=(1,1)$. 
Since the experiments with AnMR on domain $1$ demonstrate marginal accuracy improvements, we employ AnMR only on the domain extending to the black-hole horizon.
\begin{figure*}[ht!]
	 \centering
     \subfloat{
         \centering
         \includegraphics[width=0.48\textwidth]{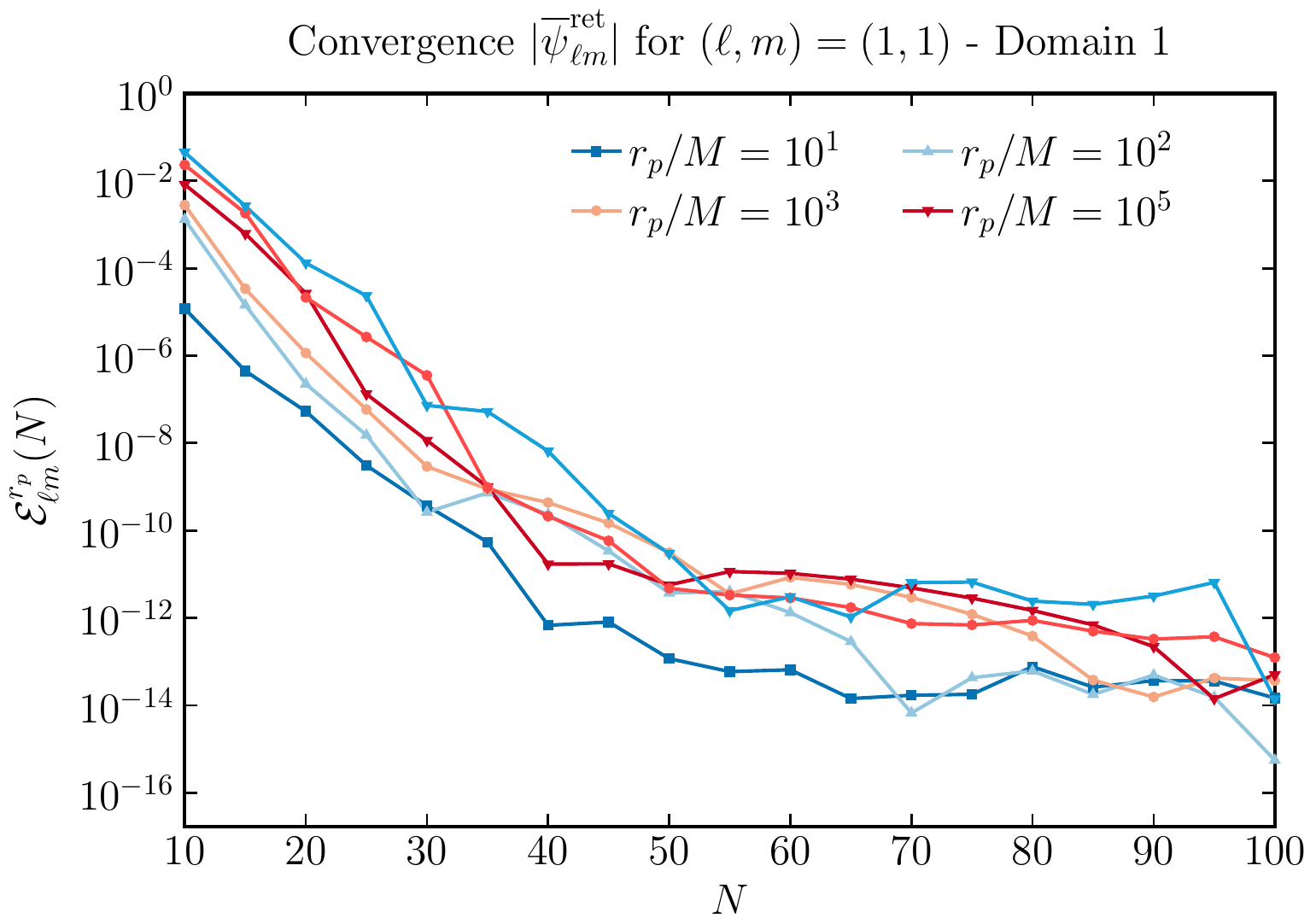}
     }
     \hfill
     \subfloat{
         \centering
         \includegraphics[width=0.48\textwidth]{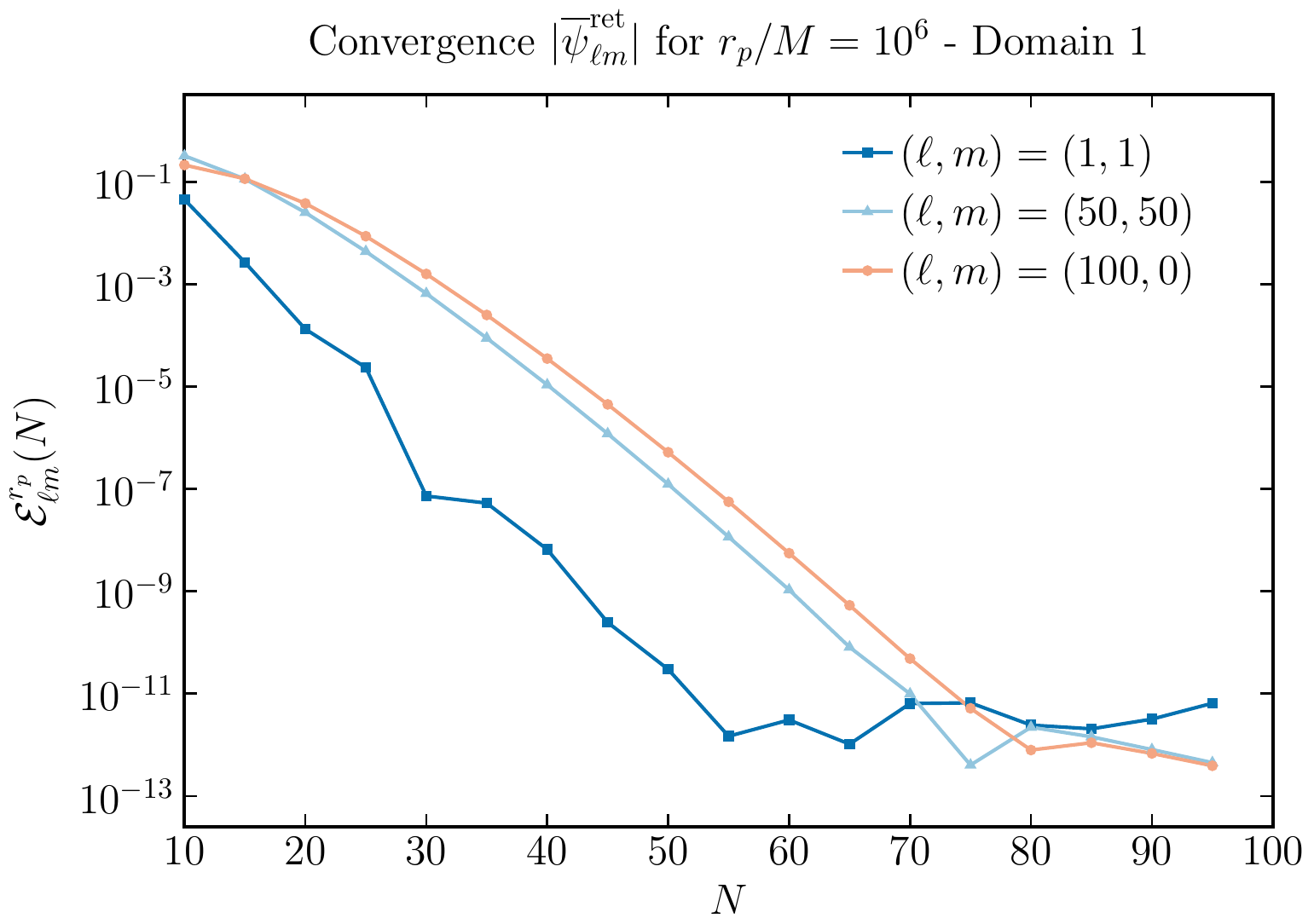}
     }
    \caption{{\em Left panel:} Exponential error decay for large orbits with optimal AnMR.
	Comparison with Fig.~\ref{fig:psi_convergence} demonstrates the power of AnMR in calculating accurate self-force results for large radius orbits.
	{\em Right panel:} Convergence for different spherical harmonic modes with $r_p/M=10^{6}$ including a high-$\ell$ mode where machine precision is reached around $N=80$. 
	This computation would be prohibitively resource-intensive in standard self-force calculations.}
    \label{fig:AnMR_convergence}
\end{figure*}
Using the value $\kappa^*_{\ell m}(r_p)$ from Eq.~\eqref{eq:kappa_dom2}, we can compute accurate solutions for any $r_p$ with a relatively low numerical resolution. 
In Fig.~\ref{fig:AnMR_convergence} we display convergence tests similar to the bottom panel of Fig.~\ref{fig:EffectiveSource_Convergence}. 
The numerical resolution is set as $N_1 = N_2 = N$, with $N^{\rm ref} =100$ for reference solution in Eq.~\eqref{eq:RelError}. 
Convergence is spectral with saturation at machine precision around $N\sim 70$, regardless of $r_p$. 
Similarly, the right panel of Fig.~\ref{fig:AnMR_convergence} shows the equivalent results for a fixed $r_p = 10^6 M$, but comparing the different angular modes $(\ell, m) = (1,1)$, $(50,50)$ and $(100,0)$. 
The exponential decay saturates at $N\sim 70$ even for high-$\ell$ modes. 
Fig.~\ref{fig:AnMR_convergence} clearly demonstrates the significant gain offered by the AnMR combined with compactification for large orbits.

\subsubsection{Post-Newtonian comparison for large radius orbits}\label{sec:EnergyFlux}
To demonstrate the significant improvement the analytic mesh refinement provides for large radius orbits, we compute the $t$-component of the self-force, $F_t$ and its $r_{p}$-derivative and compare it against a post-Newtonian series in the weak field.
For a scalar particle in a Schwarzschild background, a weak-field expression for $F_t$ was derived to high PN order in Ref.~\cite{Hikida:2004hs}, with the terms up to 4PN terms given explicitly.  
After summation over $\ell$-modes, the 4PN expression is given by
\begin{multline}\label{eq:Ft_PN}
    	F_{t}(r_{p} \gg M) = \frac{q^2 V^{4}}{3 \, r_{p}^{2}} \bigg[ 1 - \frac{1}{2} V^{2} + 2\pi V^{3}\\
	- \frac{77}{8} V^{4} + \frac{27\pi}{5} V^{5} +\mathcal{O}(V^6) \bigg],
\end{multline}
where $V = \sqrt{M/r_{p}}$.  
Note that our definition of the scalar field differs from that of \cite{Hikida:2004hs} by a factor of $4\pi$, which leads to the same difference in $F_t$.  
The $t$-component represents the energy lost due to the SSF and hence the expression begins at 1.5PN order since this is due to dipole radiation.
The PN-expression for the $r_{p}$-derivative of the $t$-component of the SSF, after some simplification, is
\begin{multline}\label{eq:drp_Ft_PN}
    \mathcal{D}_{r_{p}}F_{t}(r_{p} \gg M) = -\frac{4\, q^2 V^{4}}{3\,r_{p}^{3}} \bigg[ 1 - \frac{5}{8} V^{2}
    + \frac{11\pi}{4} V^{3}\\
	- \frac{231}{16} V^{4} 
	+ \frac{351\pi}{40} V^{5} +\mathcal{O}(V^6) \bigg].
\end{multline}

\begin{figure*}[ht!]
	 \centering
     \subfloat{
         \centering
         \includegraphics[width=0.48\textwidth]{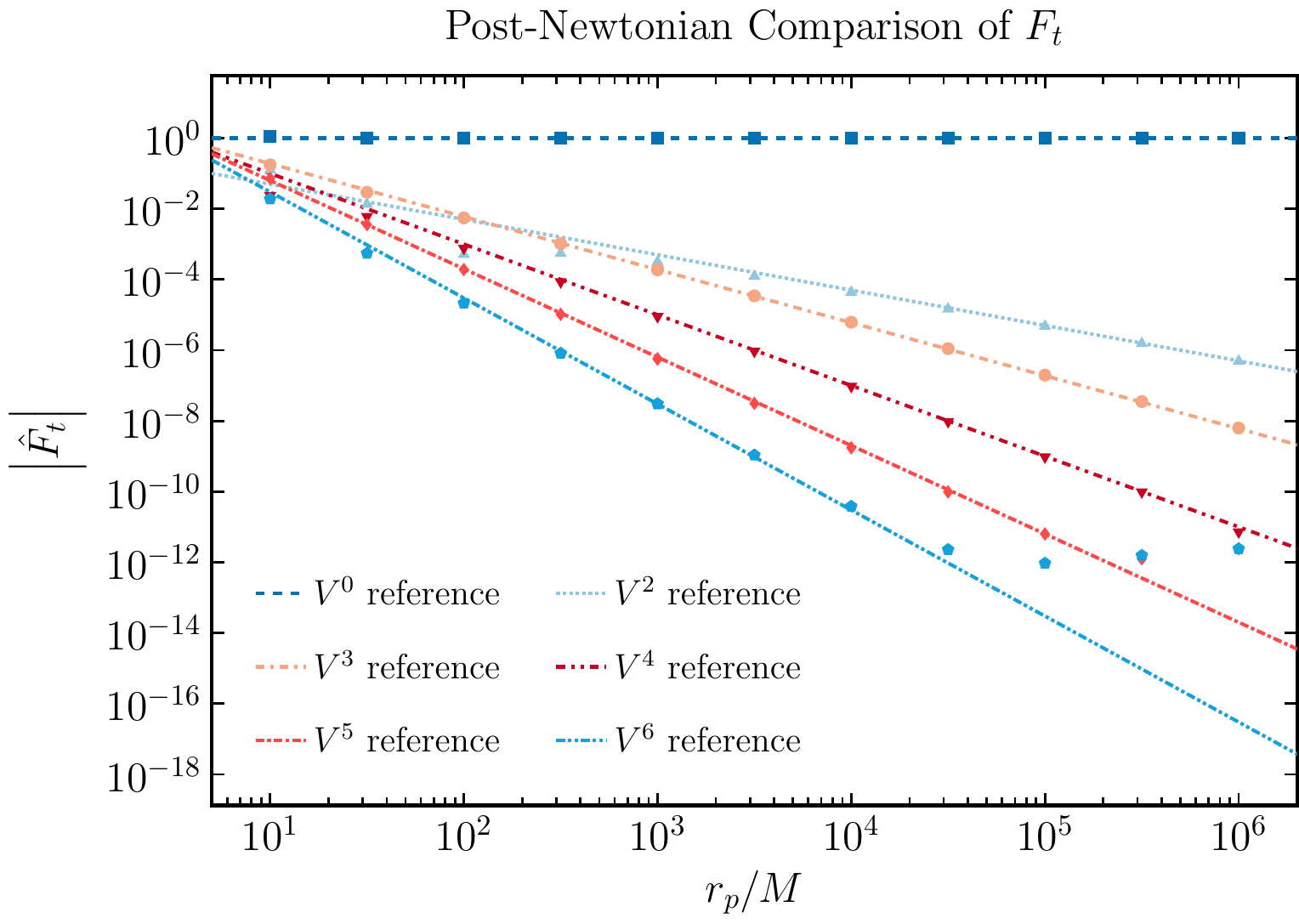}
     }
     \hfill
     \subfloat{
         \centering
         \includegraphics[width=0.48\textwidth]{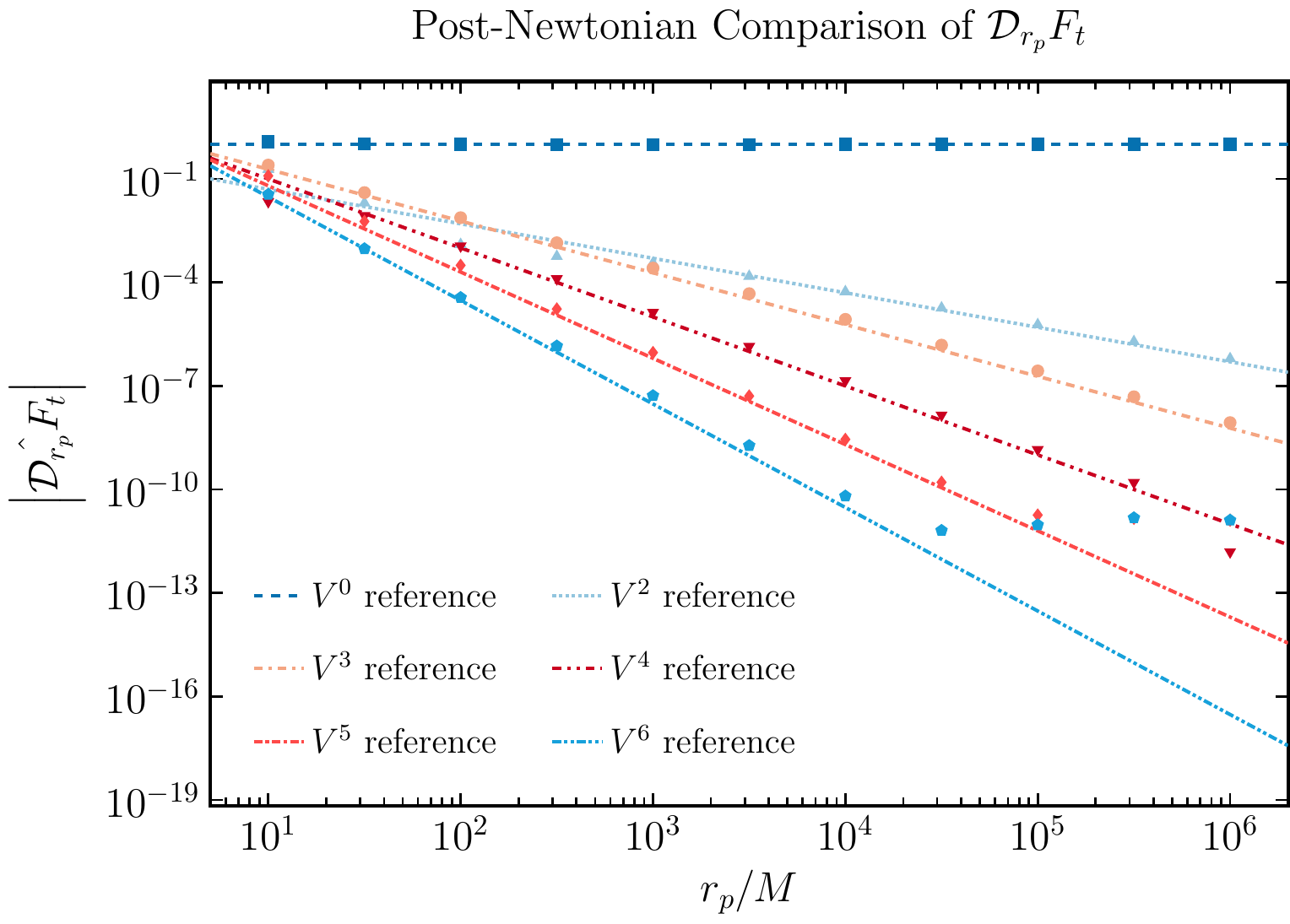}
     }
     \caption{{\em Left panel:} Comparison of the Newtonian-normalised $t$-component of the self-force, $\hat{F_t}$, with its 4.5PN expansion.  
     The (dark blue) squares show our numerical results for $\hat{F_t}$.  
     This data approaches the leading (normalized) PN result for large radius orbits.  
     When we subtract the leading PN term from the numerical data we get the (light blue) triangles.  
     For large $r_p$ this data approaches a $V^2$ reference curve as expected from the PN series in Eq.~\eqref{eq:Ft_PN}.  
     When we subtract the first subdominant term in the PN series and see that the residual (orange circles) falls off as $V^3$, as expected.  
     We repeat this procedure with the remaining terms in the PN series to compute the other data and find agreement with the PN series (until machine round-off is reached).  
     This shows our numerical results are accurate even for extreme large radius orbits with $r_p/M\sim10^6$.  
     {\em Right panel:} Comparison of the $r_{p}$-derivative of the Newtonian-normalised $t$-component of the self-force obtained from flux-balance laws with the 4.5PN expression.  
     This figure constructed in the same way as the left panel except we subtract terms from the PN series in Eq.~\eqref{eq:drp_Ft_PN}.  
     Again we see that our numerical results are accurate even for extreme large radius orbits with $r_p/M\sim10^6$.  
     Accurate results for large radius orbits for self-force problem with unbounded sources are very difficult to achieve with the standard variation of parameters approach.  
     Our hyperboloidal method can thus be instrumental in future precision comparisons with PN theory for self-force problems with unbounded sources, e.g., second-order self-force calculations.}
     \label{fig:Ft_PN}
\end{figure*}

We compare the numerical results of our code to the above two PN series in Fig.~\ref{fig:Ft_PN}.
In both panels we plot the force or its $r_p$-derivative normalized by the leading term in the relevant PN series, i.e., the coefficient in front of 
the square brackets in Eqs.~\eqref{eq:Ft_PN} or \eqref{eq:drp_Ft_PN}, respectively.
We denote these normalized quantities with an overhat.

When we subtract the leading (normalized) PN term (i.e. $1$) from $\hat{F_{t}}$ we observe that the residual scales as $V^{2}$, as expected.  
The order of the scaling increases by $\mathcal{O}(V)$ for the subtraction of each subsequent sub-leading PN term up to 4.5PN order.
We find excellent agreement with the PN series for orbits as large as $r_p = 10^6 M$ even up to $\mathcal{O}(V^{6})$.
For the calculation of $F_t$ this is a significant improvement of what is usually possible with the numerical integration method that 
relies on boundary condition expansions evaluated in the wave zone (though note the MST method works well for large orbits with distributional sources).
For $\partial_{r_{p}}F_{t}$, neither the numerical integration nor MST methods work well for large radius orbits, but the results from our hyperboloidal 
approach agree very well with the PN series. We find the same scaling arguments as previously and agreement up for $r_{p} = 10^{6}M$ up to $\mathcal{O}{(V^{6})}$ for the residuals.

\section{Conclusion}
This work presents the hyperboloidal approach to self-force calculations in the frequency domain.
This approach works well for the three classes of sources typically found in self-force calculations: distributional, worldtube, and unbounded support.
The latter, in particular, is challenging for current techniques but crucial for emerging second-order (in the mass ratio) calculations \cite{Pound:2019lzj,Warburton:2021kwk,Wardell:2021fyy}. Another challenging problem for current self-force techniques, present for all three classes of sources, is the comparison to post-Newtonian results for large orbital radii. Compactification along hyperboloidal surfaces combined with analytic mesh refinement is an elegant solution to these challenging problems.

Our approach relies on two essential ingredients.
On the theoretical side, we employ scri-fixing hyperboloidal coordinates for the background black-hole spacetime in minimal gauge \cite{Zenginoglu:2007jw, Zenginoglu:2011jz, Ansorg:2016ztf,PanossoMacedo:2018hab, PanossoMacedo:2019npm}.
On the numerical side, we solve the self-force equations with a spectral ODE solver, enhanced with analytic mesh refinement to resolve functions with steep gradients~\cite{Ansorg:2003br,Ansorg:2006gd,Meinel:2008kpy,Ansorg13}.
The combination of these theoretical and numerical frameworks provides us with a powerful novel scheme to address the current limitations of the numerical techniques in the self-force program. 

We emphasize various advantages of hyperboloidal slices relevant to the self-force problem, as demonstrated in this work. First, the boundary conditions at the black hole and the wave zone become trivial.
Specifically, the geometric construction of hyperboloidal slices ensures the absence of incoming characteristics as the radial coordinate approaches the horizon or extends towards the wave zone.
Consequently, the treatment of the boundary conditions is behavioral and not numerical. The outgoing behavior of solutions near the boundaries follows directly from the regularity of solutions as discussed with Eq.~\eqref{eq:BC_hyp}. This simplification of boundary treatment is both a conceptual and a practical advantage because one does not need to impose boundary conditions by hand to ensure the uniqueness of the solution, and one does not need to compute lengthy and tedious approximations at finite radii for each type of perturbation or source.  

Second, radiation extraction becomes a trivial evaluation at the outer boundary, whereas current calculations extrapolate fluxes from finite radii up to infinity. Such extrapolations are particularly difficult to perform for unbounded support sources and introduces additional systematic errors that must be controlled. In contrast, we evaluate fluxes directly from the hyperboloidal solutions at the spacetime boundaries as discussed in Sec.~\ref{sec:EnergyFlux}. The extraction of fluxes are as accurate as the numerical solution of the equations without additional systematic errors.

Third, hyperboloidal slices improve the numerical efficiency of ODE solvers. Typically, the accuracy of frequency domain calculations is limited by the number of grid points per wavelength. Hyperboloidal transformations flatten the waves and reduce the number of spatial oscillations along the time slice, thereby enabling a highly efficient numerical solver.
The hyperboloidal solution is smooth and non-oscillatory throughout the domain except for discontinuities at the particle location or worldtube boundaries. Therefore, multi-domain spectral methods are ideally adapted to generate highly accurate solutions for little computational cost. 
Such spectral methods have been successfully employed both for hyperboloidal formulations~\cite{Schinkel:2013zm,Schinkel:2013tka,PanossoMacedo:2014dnr} and self-force calculations~\cite{Canizares:2010yx}.
Our code, based on Refs.~\cite{Ansorg:2016ztf, PanossoMacedo:2018hab}, brings these two applications together in a multi-domain spectral code for hyperboloidal self-force calculations where the compactified exterior black-hole region is divided into subdomains to properly treat the singular behaviors and discontinuities at the particle location while efficiently resolving the non-oscillatory solution with spectral accuracy away from the discontinuities.

Fourth, hyperboloidal compactification efficiently solves the problem of unbounded support sources with support extending across the entire exterior black-hole region. Such sources provide a significant numerical challenge in second-order self-force calculations. 
Present implementations compute the second-order source on a finite radial domain and expend significant effort making the source fall off more rapidly to make the integrals in the variation of parameters approach converge more rapidly.
Our approach avoids these issues entirely and handles the case of unbounded support sources with ease.
There is also an additional advantage of using of hyperboloidal slicing in second-order calculations as it improves the behavior of the source near the boundaries \cite{Pound:2015wva,Miller:2020bft}.

Fifth, compactification allows us to compute self-force for orbits with very large radii, e.g., $r_p\sim10^6 M$. 
The large radius regime is important for connecting self-force results to post-Newtonian theory \cite{Blanchet:2009sd, Blanchet:2010zd, Dolan:2013roa, Dolan:2014pja}.
This regime is challenging for current numerical integration methods because they place the outer boundary far into the wave zone for convergence of the boundary series.
For weak-field orbits, the wave zone moves out into the very weak field, requiring many steps for the numerical integrator to reach the particle's radius.
While this problem can be overcome with the Mano-Suzuki-Takasugi method \cite{Mino:1996nk} for distributional or worldtube sources, the computational cost of this approach prohibits the application of the method to unbounded support sources.
Hyperboloidal compactification maps the entire exterior domain onto the finite numerical grid and therefore includes automatically any large radii in the domain. 
We resolve the steep gradients around the particle that form due to compactification by using analytic mesh-refinement~\cite{Meinel:2008kpy,PanossoMacedo:2014dnr,Ammon:2016szz,Pynn:2016mtw,Kalisch2016}. 
We demonstrate that our approach works exceptionally well for these cases, as well as for distributional and worldtube sources in Sec.~\ref{sec:AnMR_results} and Appendix \ref{apdx:AnMR_with_effective_source}.

Sixth, we compute solutions with high $\ell$ modes very accurately.
This is essential for studies of the behavior of the self-force and related gauge-invariant quantities near the light-ring \cite{Akcay:2012ea,Dolan:2014pja}.

Given the geometric elegance of the hyperboloidal framework and the strong evidence for its advantages, we conclude that future studies in black-hole perturbation theory will make heavy use of hyperboloidal foliations. We note that the benefits we list arise not so much from the hyperboloidal nature of the coordinates but from the regularity of the foliation in the entire exterior domain. This regularity allows us to include the black hole horizon and future null infinity on our numerical grid. One would expect similar advantages from a double-null foliation with compactification. The main reason we prefer the hyperboloidal framework is its flexibility. It is straightforward to extend hyperboloidal coordinates from Schwarzschild to Kerr spacetimes \cite{Zenginoglu:2007jw, PanossoMacedo:2019npm}, whereas it is highly nontrivial to do the same for double-null coordinates. 

Presently, the results discussed in this work are restricted to the first-order scalar-self force for a particle on a circular orbit around a Schwarzschild black hole. 
There are many steps to take on the path to second-order, gravitational self-force for a particle on a general orbit in a Kerr spacetime. 
We expect that our approach will readily extend to, e.g., the Lorenz-gauge gravitational case \cite{Akcay:2010dx,Akcay:2013wfa,Wardell:2015ada} and to Kerr spacetime using the Teukolsky formalism \cite{Teukolsky:1973ha}, both of which are commonly used in frequency domain self-force calculations (see Appendix \ref{app:operators_spin} for the operators in the Regge-Wheeler-Zerilli and Bardeen-Press-Teukolsky formalisms using the minimal gauge).
 
\section{Acknowledgments}
RPM acknowledges financial support provided by the STFC grant number ST/V000551/1,  COST Action CA16104 via the Short Term Scientific Mission grant, and European Research Council Grant ERC-2014-StG 639022-NewNGR ``New frontiers in numerical general relativity".
NW acknowledges support from a Royal Society - Science Foundation Ireland University Research Fellowship via grants UF160093 and RGF\textbackslash R1\textbackslash180022.
This work makes use of the Black Hole Perturbation Toolkit \cite{BHPToolkit}.

\appendix

\section{Black-hole perturbation theory}\label{app:operators_spin}
Black-hole perturbation theory on spherically symmetric BH spacetime is commonly formulated either in the Regge-Wheeler-Zerilli (RWZ) or the Bardeen-Press-Teukolsky (BPT) formalism. 
Both describes perturbative field characterised by their spin-weight $p=0, \pm 1, \pm 2$. The RWZ approach considers specific combinations of the perturbed metric as the propagating field on the Schwarzschild background, 
whereas the BPT formulation has scalar fields ($p=0$), and the propagating degrees of freedom for the Faraday-Maxwell ($p=\pm1$) and Weyl tensors ($p=\pm2$) as perturbative fields. We denote $\phi_{p,\ell m}$ and $u_{p,\ell m}$ 
fields with spin $p$ within the BPT and RWZ formalism respectively. For scalar fields $p=0$, they are trivially related by
\beq
\label{eq:BPT_RWZ}
\phi_{0,\ell m} = \dfrac{u_{0,\ell m}}{r}.
\eeq

Hyperboloidal formulations of the RWZ and BPT equations were first implemented in time domain using constant mean curvature time surfaces \cite{zenginouglu2009gravitational, zenginouglu2010asymptotics}. In this appendix, we present the frequency domain expressions for the left-hand side of Eq.~\eqref{eq:FreqDomain} for a field with spin $p$ using the minimal gauge. We also discuss the corresponding factors $Z$ and ${\cal F}$ involved in the frequency-domain hyperboloidal transformations via eqs.~\eqref{eq:HyperFieldReScale} and \eqref{eq:HyperOperatorReScale}, respectively, as well as the hyperboloidal operator ${\boldsymbol A}$.
\subsection{Bardeen-Press-Teukolsky formalism}
With the BPT formalism, the left-hand side of Eq.~\eqref{eq:FreqDomain} reads $\Delta^{\rm BPT}_{\ell m} \phi_{p, \ell m}$, with
\bea
\label{eq:BPT}
&&\Delta^{\rm BPT}_{\ell m} = \dfrac{d^2}{dr^2 } + 2(1+p)\dfrac{r-M}{r^2 f} \dfrac{d}{dr}  \\
&& - \dfrac{1}{f}\Bigg( \dfrac{ \ell(\ell +1) - p(p+1)}{r^2} + 2 i p \dfrac{\omega}{r^2} \left( \dfrac{M}{f} - r \right) - \dfrac{\omega^2}{f} \Bigg). \nn
\eea
The above operator differs from the usual format for the BPT equation by an overall factor $r^2 f$ and it reduces to Eq.~\eqref{eq:LaplaceOperator} when $p=0$. 
With hyperboloidal transformation, the regularisation factor in Eq.~\eqref{eq:HyperFieldReScale} reads~\cite{Zenginoglu:2011jz, PanossoMacedo:2019npm}
\beq
Z = \Omega^{1+2p}\left( r^2 f \right)^{-p}e^{s H},
\eeq
while ${\cal F}$ is still given by Eq.~\eqref{eq:F}.
Finally, the coefficients $\alpha_2$, $\alpha_1$ and $\alpha_0$ for the operator ${\boldsymbol A}^{\rm BPT}_{\ell m}$ in Eq.~\eqref{eq:Hyperboloidal_Laplace_operator} read~\cite{Ansorg:2016ztf}
\bea
\alpha_2 &=& \sigma^2(1-\sigma), \\
\alpha_1 &=& s(1-2\sigma^2) + \sigma\bigg(2-3\sigma + p(2-\sigma) \bigg), \\
\alpha_0 &=& - \bigg( s^2(1+\sigma) + s\left[ 2\sigma - p(1-\sigma) \right] \nn \\
&&+ \ell(\ell+1)+(1+p)(\sigma-p)\bigg) .
\eea
\subsection{The Regge-Wheeler-Zerilli formalism}
With the RWZ formalism, the left-hand side of Eq.~\eqref{eq:FreqDomain} reads $\Delta^{\rm RWZ}_{\ell m} \, u_{p, \ell m}$, with
\beq
\Delta^{\rm RWZ}_{\ell m} = \dfrac{d^2}{dr_*^2} - (P^{\rm RW,Z}_{\ell m}-\omega^2).
\eeq
The potential $P^{\rm RW,Z}_{\ell m}$ depends on the type of perturbation. The potential for polar perturbations (RW) are (with $\ell\geq |p|$)
\beq
P^{\rm RW}_{\ell m} = \dfrac{f}{r^2}\Bigg( \ell(1+\ell) + (1-p^2)\dfrac{M}{r} \Bigg),
\eeq
whereas the potential for axial perturbations (Z) reads (with $n=(\ell-1)(\ell+2)/2$ and $\ell\geq2$)
\beq
P^{\rm Z}_{\ell m} = \dfrac{f}{r^2}\Bigg( \dfrac{2n^2(n+1)r^3 + 6n^2Mr^2 + 18nM^2r +18M^3}{r(nr+3M)^2} \Bigg).
\eeq
With hyperboloidal transformation, the regularisation factors in Eqs.~\eqref{eq:HyperFieldReScale} and \eqref{eq:HyperOperatorReScale} read
\beq
Z =\dfrac{2M}{\lambda} e^{s H}, \quad {\cal F} = \dfrac{Z f}{r^2}.
\eeq
Finally, the coefficients $\alpha_2$, $\alpha_1$ and $\alpha_0$ for the operator ${\boldsymbol A}^{\rm RWZ}_{\ell m}$ in Eq.~\eqref{eq:Hyperboloidal_Laplace_operator} read~\cite{jaramillo2021pseudospectrum}
\bea
\alpha_2 &=&  \sigma^2(1-\sigma) \\
\alpha_1 &=&   2\sigma(1-3\sigma^2) - s  (1-2\sigma^2), \\
\alpha_0 &=& -  \bigg( s^2 \, (1+\sigma) +  s\, \sigma^2   + V_{\ell m}^{\rm RW,Z}  \bigg),
\eea
with $V_{\ell m}^{\rm RW,Z} = \dfrac{r^2}{f}  P_{\ell m}^{\rm RW,Z}$.

\section{Evaluation of the stress-energy tensor in hyperboloidal coordinates}
\label{app:hyperboloidal_flux}
In order to calculate the flux towards future null infinity and the horizon we
need to consider the limits of the integrand towards $\sigma \rightarrow 0$
and $\sigma \rightarrow 1$ respectively.  The relevant
components of the stress energy tensor
are $\tensor{T}{_{\tau\tau}}$ and $\tensor{T}{_{\tau\sigma}}$.  Under the conformal
rescaling,
\beq
	g_{\alpha\beta} = \Omega^{-2}\tilde{g}_{\alpha\beta}, \quad
	g^{\alpha\beta} = \Omega^{2}\tilde{g}^{\alpha\beta},
\eeq
and $\nabla_{\alpha}\Phi = \partial_{\alpha}\Phi$.  Hence, the stress-energy tensor becomes
\beq
	\tensor{T}{_{\alpha\beta}} = \frac{1}{4\pi} \left(\nabla_{\alpha}\Phi
	\nabla_{\beta}\Phi - \frac{1}{2}\tilde{g}_{\alpha\beta}
	\tilde{g}^{\mu\nu} \nabla_{\mu}\Phi
	\nabla_{\nu}\Phi \right).
	\label{eq:conformal_stress_energy_tensor}
\eeq
\begin{widetext}
The components $\tensor{T}{_{\tau\tau}}$ and $\tensor{T}{_{\tau\sigma}}$ can
then be expressed as
\begin{align}
	\tensor{T}{_{\tau\tau}} &=
	\frac{1}{4\pi}\bigg[ \Omega^{2}(\partial_{\tau}\overline{\Phi})^{2}
	+ \frac{1}{2}\tilde{g}_{\tau\tau} \tilde{g}^{\mu\nu} \nabla_{\mu}\Phi
	\nabla_{\nu}\Phi \bigg]
	= \frac{1}{4\pi}\bigg[ \Omega^{2}(\partial_{\tau}\overline{\Phi})^{2}
	+ \frac{1}{2} \sigma^{2}(\sigma - 1)
	\tilde{g}^{\mu\nu}\nabla_{\mu}\Phi \nabla_{\nu}\Phi \bigg],\\
	\tensor{T}{_{\tau\sigma}} &=
	\frac{1}{4\pi}\bigg[
	\Omega^{2}\partial_{\tau}\overline{\Phi}\partial_{\sigma}\overline{\Phi} 
	+ \frac{1}{2}\tilde{g}_{\tau\sigma} \tilde{g}^{\mu\nu}
	\nabla_{\mu}\Phi \nabla_{\nu}\Phi \bigg]
	= \frac{1}{4\pi}\bigg[
	\Omega^{2}\partial_{\tau}\overline{\Phi}\partial_{\sigma}\overline{\Phi}
	+ \frac{1}{4}(1 - 2\sigma^{2}) \tilde{g}^{\mu\nu} \nabla_{\mu}\Phi
	\nabla_{\nu}\Phi \bigg].
\end{align}
The second term in these expression can be written as 
\beq
	\frac{1}{2}\tilde{g}_{\alpha\beta} \tilde{g}^{\mu\nu} \nabla_{\mu}\Phi
	\nabla_{\nu}\Phi =
	\frac{1}{2}\tilde{g}_{\alpha\beta} \left[ \tilde{g}^{\tau\tau}
	(\partial_{\tau}\Phi)^{2} + \tilde{g}^{\sigma\sigma}
	(\partial_{\sigma}\Phi)^{2} + 2 \tilde{g}^{\tau\sigma}
	\partial_{\tau}\Phi \partial_{\sigma}\Phi +
	\tilde{g}^{\theta\theta} (\partial_{\theta}\Phi)^{2} +
	\tilde{g}^{\varphi\varphi} (\partial_{\varphi}\Phi)^{2} \right].
	\label{eq:stress_energy_components}
\eeq
But since our scalar-field scales as $\Phi = \Omega \overline{\Phi}$ then
\begin{align}
	\frac{1}{2}\tilde{g}_{\alpha\beta} \tilde{g}^{\mu\nu} \nabla_{\mu}\Phi
	\nabla_{\nu}\Phi =
	\frac{1}{2}\tilde{g}_{\alpha\beta} \big[ \Omega^{2}	\tilde{g}^{\tau\tau}
	( \partial_{\tau}\overline{\Phi} )^{2} + \tilde{g}^{\sigma\sigma}
	( \overline{\Phi} + \Omega\, \partial_{\sigma}\overline{\Phi} )^{2}
	&+ 2\tilde{g}^{\tau\sigma} \Omega\, (\partial_{\tau}\overline{\Phi})
	( \overline{\Phi} + \Omega\, \partial_{\sigma}\overline{\Phi} )\nonumber \\
	&+ \tilde{g}^{\theta\theta} \Omega^{2}\, (\partial_{\theta}\overline{\Phi})^{2}
	+ \tilde{g}^{\varphi\varphi} \Omega^{2}\, (\partial_{\varphi}\overline{\Phi})^{2}
	\big].
\end{align}
Inserting the components of the conformal metric we are left with
\begin{align}
	\frac{1}{2}\tilde{g}_{\alpha\beta} \tilde{g}^{\mu\nu} \nabla_{\mu}\Phi
	\nabla_{\nu}\Phi =
	\frac{1}{2}\tilde{g}_{\alpha\beta} \big[ -4\Omega^{2}(1 + \sigma)	
	( \partial_{\tau}\overline{\Phi} )^{2} + 4\sigma^{2}(1 - \sigma)
	( \overline{\Phi} + \Omega\, &\partial_{\sigma}\overline{\Phi} )^{2}
	+ 4(1 - 2\sigma^{2}) \Omega\, (\partial_{\tau}\overline{\Phi})
	( \overline{\Phi} + \Omega\, \partial_{\sigma}\overline{\Phi} )\nonumber \\
	&+ 4 \Omega^{2}\, (\partial_{\theta}\overline{\Phi})^{2}
	+ 4\,\csc^{2}\theta\, \Omega^{2}\,
	(\partial_{\varphi}\overline{\Phi})^{2}
	\big].
	\label{eq:stress_energy_second_term}
\end{align}
Taking our results from Eqs.~\eqref{eq:stress_energy_components} and
(\ref{eq:stress_energy_second_term}) we find 
\begin{align}
	\tensor{T}{_{\tau\tau}} &= \frac{\sigma^{2}}{4\pi\lambda^{3}}	 \bigg[
	\lambda (\partial_{\tau}\overline{\Phi})^{2} - (1 - \sigma)\sigma \big( (2 -
	4\sigma^{2}) (\partial_{\tau}\overline{\Phi}) (\overline{\Phi} + \lambda
	\partial_{\sigma}\overline{\Phi} )^{2} + 2 \lambda \sigma
	(\partial_{\theta}\overline{\Phi})^{2} - 2\lambda(\sigma - 1)\sigma 
	( \overline{\Phi} + \lambda \partial_{\sigma}\overline{\Phi} )^{2}\nonumber \\
	&- 2\lambda\sigma(\sigma + 1) (\partial_{\tau}\overline{\Phi})^{2} + 2\lambda
	\sigma \csc^{2}\theta\, \partial_{\varphi}\overline{\Phi} \big) \bigg],\\
	\tensor{T}{_{\tau\sigma}} &= \frac{\sigma}{8\pi\lambda^{3}}	 \bigg[
	2\lambda\sigma (\partial_{\sigma}\overline{\Phi}) (\partial_{\tau}\overline{\Phi}) +
	(1 - 2\sigma^{2})\big( (2 - 4\sigma^{2})(\partial_{\tau}\overline{\Phi})
	\big( \lambda ( \overline{\Phi} + \lambda \partial_{\sigma}\overline{\Phi} )^{2} +
	2\lambda \sigma (\partial_{\theta}\overline{\Phi})^{2} - 2\lambda\sigma(\sigma -
	1) ( \lambda ( \overline{\Phi} + \lambda \partial_{\sigma}\overline{\Phi} )^{2} 
	\nonumber \\
	&- 2\lambda\sigma (\sigma + 1) (\partial_{\tau}\overline{\Phi})^{2} 
	+ 2\lambda\sigma \csc^{2}\theta\, \partial_{\varphi}\overline{\Phi} \big)
	\bigg].
\end{align}
\end{widetext}

\section{Analytic Mesh-Refinement}\label{apdx:AnMR_with_effective_source}
This appendix complements Sec. \ref{sec:AnMR_results} and discusses the effects of the AnMR in two cases: the accuracy on domain $1$ extending between future null infinity and the particle's orbit for distributional sources (with qualitatively similar results in the case of sources with unbounded support); and scenarios with worldtube sources, whose results follow from a code with 4-domains. 
\subsection{Treatment at $\scri^+$}
\label{app:AnMR}
\begin{figure}[h!]
\centering
\includegraphics[width=0.48\textwidth]{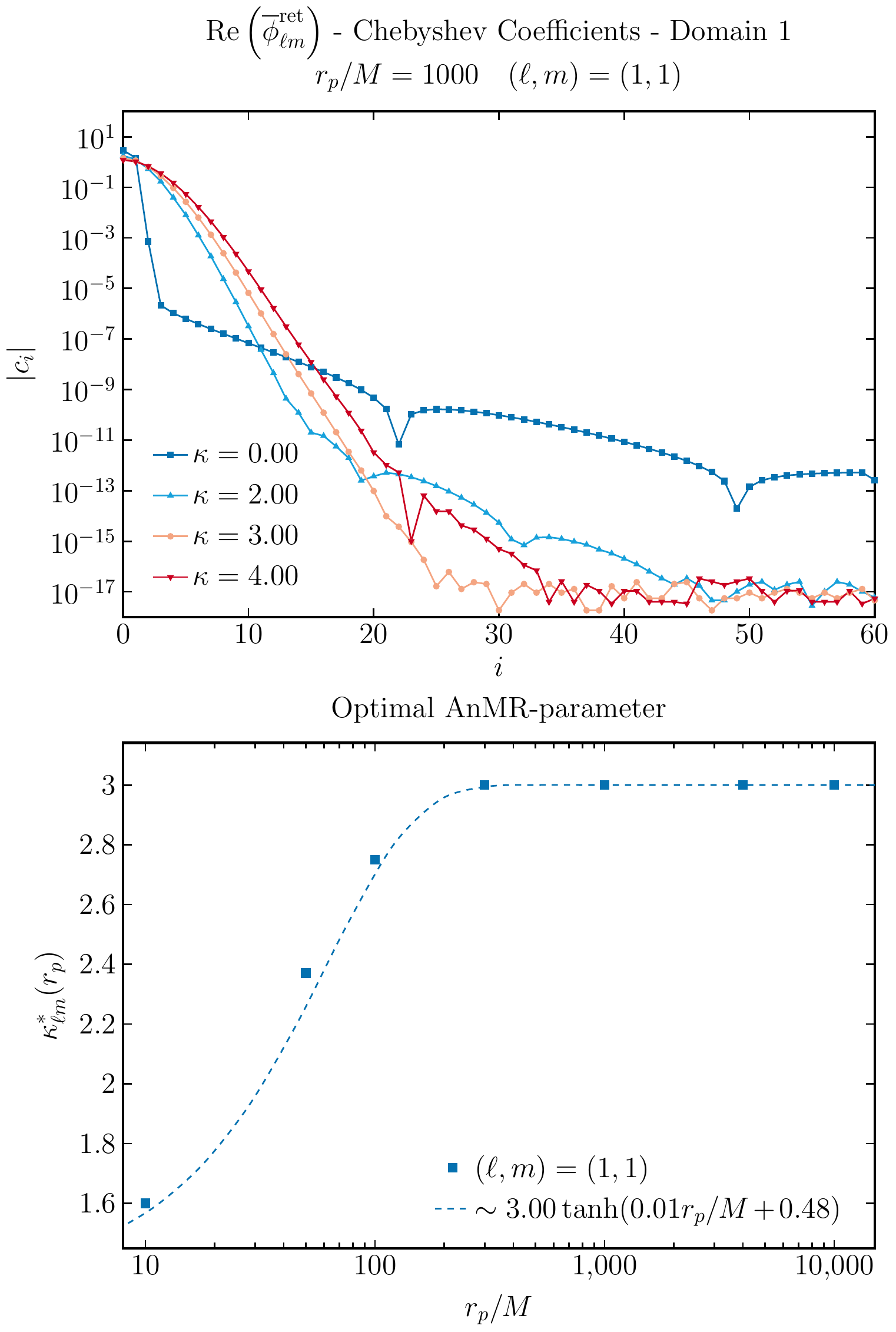}
\caption{AnMR effect over $\overline{\phi}^{\rm ret}_{\ell m}$ on domain $1$ (between future null infinity and particle's orbit). {\em Top panel:} Chebyshev coefficients of ${\rm Re}(\overline{\phi}_{\ell m}^{\rm ret})$ with $r_p=1000M$ and $(\ell, m)=(1,1)$ in domain $1$ between the particle and future null infinity. The improvement with AnMR is not as compelling as in domain $2$ presented in Fig.~\ref{fig:cheb_rp1000_ret_phi_AnMR_dom2}. {\em Bottom panel:} Optimal AnMR parameters plotted against $r_p/M$ with the fit Eq.~\eqref{eq:kappa_dom1}.
}
\label{fig:cheb_rp1000_ret_phi_AnMR_dom1}
\end{figure}

Contrary to the systematic pattern observed by Eq.~\eqref{eq:kappa_dom2}, the method to optimise the solutions' accuracy in domain $1$ is very sensitive to the particular $(\ell,m)$-mode, and the improvement in accuracy is not so significant. As example, we consider the solution for $\overline{\phi}^{\rm ret}_{\ell m}$ with $(\ell,m)=(1,1)$. One needs to employ the AnMR with $x_{\rm B}=-1$, i.e., the map \eqref{eq:AnMR} populates the grid points around future null infinity $\sigma=0$. Fig.~\ref{fig:cheb_rp1000_ret_phi_AnMR_dom1} demonstrates this effect. 
The top panel in Fig.~\ref{fig:cheb_rp1000_ret_phi_AnMR_dom1} shows the Chebyshev coefficients of ${\rm Re}(\overline{\phi}_{\ell m}^{\rm ret})$ with $r_p=1000M$ and $(\ell, m)=(1,1)$ for several AnMR-parameters $\kappa$ for domain $1$. The coefficients decay moderately fast for $\kappa=0$, but one can improve the decay rate by varying $\kappa$ (e.g.~$\kappa=2,3$ and $4$). 
As explained in Sec. \ref{sec:AnMR_results}, one typically encounters and optimal value $\kappa^*_{\ell m}(r_p)$ on domain $1$, for which the decay is the fastest. 
In this example, the optimal value $\kappa^*_{\ell m}(r_p)$ corresponds to $\kappa^*_{1,1}(1000M)=3$.
The bottom panel displays the fit of $\kappa^*_{1,1}(r_p)$ against $r_p$. Contrary to the log-dependence of Eq.~\eqref{eq:kappa_dom2}, we observe that $\kappa^*_{1,1}(r_p)$ quickly saturates around $\sim 3$, according to
\beq\label{eq:kappa_dom1}
\kappa^*_1 = 3.00 \tanh\left( 0.01 \dfrac{r_p}{M} + 0.48 \right).
\eeq
On the other hand, we observe that the coefficients on domain $1$ for modes $(\ell,m) = (50,50)$ and $(\ell,m) = (100,0)$ are optimised with a map \eqref{eq:AnMR} with $x_{\rm B}=1$, i.e., with an increase of grid points around the particle. We observe a slight improvement on the coefficients decay rate for $(\ell,m) = (50,50)$ and $(100,0)$, respectively when $\kappa = 1$ or $\kappa=1.5$, regardless of the particle location. Because the effects on the accuracy is marginal and highly dependent on the angular mode $(\ell,m)$, we refrain form using the AnMR technique on domain $1$.

\subsection{Worldtube sources}\label{apdx:AnMR_with_effective_source}
\begin{figure}[ht!]	
	\centering
	\includegraphics[width=0.48\textwidth]{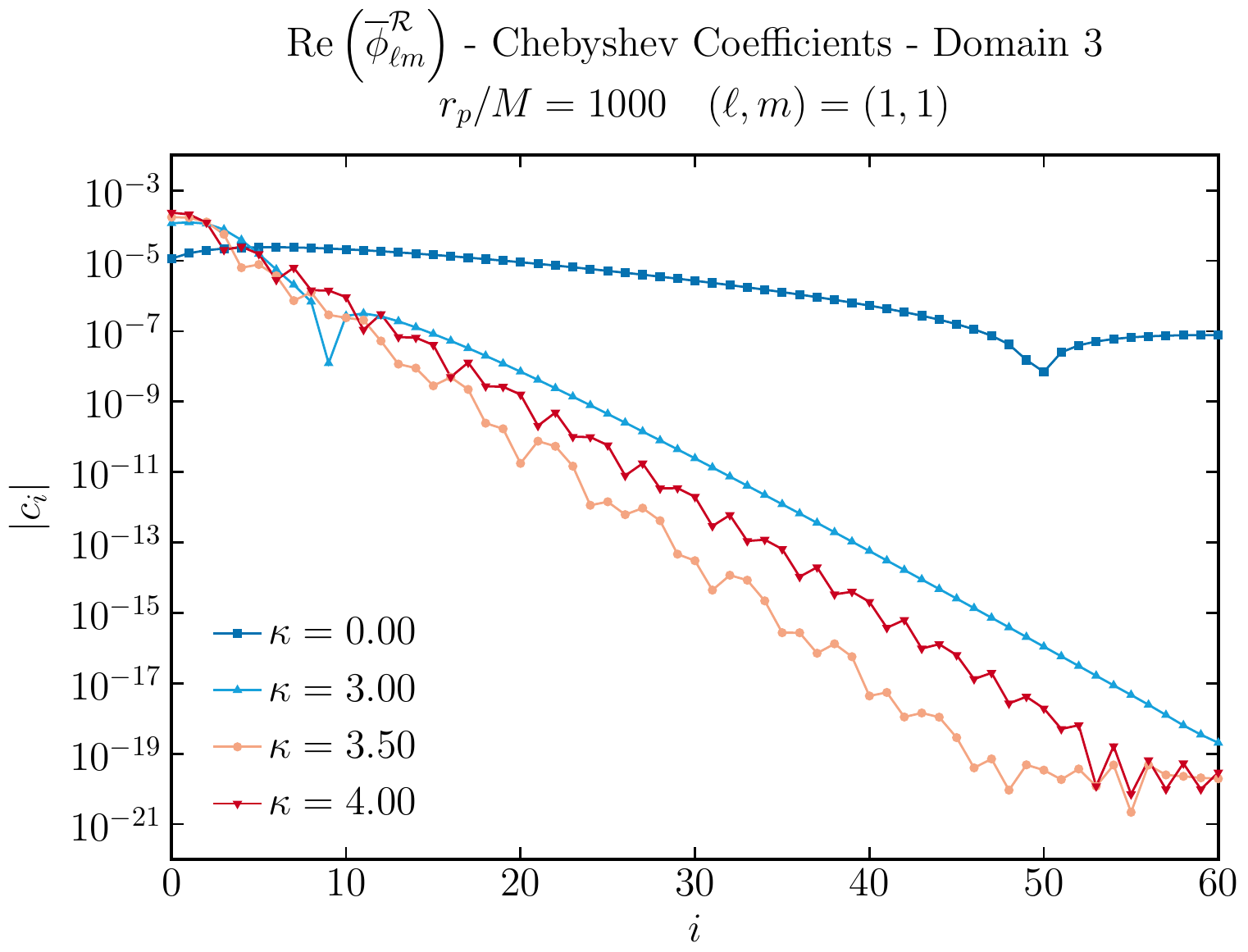}  
	\caption{AnMR for scenarios with worldtube sources, requiring a code with 4 domains. Chebyshev coefficients of ${\rm Re}(\overline{\phi}_{\ell m}^{\mathcal{R}})$ for $r_p=1000M$ and $(\ell,m)=(1,1)$ in domain $3$ (between the particle $\sigma_p$ and the worldtube boundary $\sigma_+$). Here, the optimal value for the AnMR parameter is $\kappa=3.5$ and the overall tendency on $r_p$ and $(\ell,m)$ follows the tendency observed in Eq.~\eqref{eq:kappa_dom2}.
}
\label{fig:cheb_rp1000_eff_phi_AnMR_dom1}
\end{figure}

We apply the AnMR also to effective-source computations. 
As discussed in Sec.~\ref{sec:Results_EffectiveSource}, this problem requires a four-domain code. 
The accuracy loss for large orbits arise from domain $3$ where $\sigma\in[\sigma_p, \sigma_+]$. 
Therefore, we use the AnMR ($x_{\rm B}=-1$) on domain $3$. The top panel of Fig.~\ref{fig:cheb_rp1000_eff_phi_AnMR_dom1} displays the Chebyshev coefficients 
for the residual field ${\rm Re}(\overline{\Phi}^{\mathcal{R}}_{\ell m})$ with $r_p=1000M$ and $(\ell, m)=(1,1)$. 
As previously, we observe the coefficients' slow decay rate when $\kappa=0$ and significant improvement for $\kappa>0$. 
In this example, the best decay rate is achieved for the value $\kappa\sim 3.5$. 
The calibration for the optimal $k^*_{\ell m}(r_p)$ is the same as in Eq.~\eqref{eq:kappa_dom2}. 
In particular, for $(\ell, m)= (1,1)$, $(50,50)$ and $(100,0)$, the offsets $A_{\ell, m}$ are $A_{1,1} \approx - 0.02$, $A_{50,50} \approx 0.59$ and $A_{100,0} \approx 0.85$.
\begin{figure}[ht!]
\centering
\includegraphics[width=0.48\textwidth]{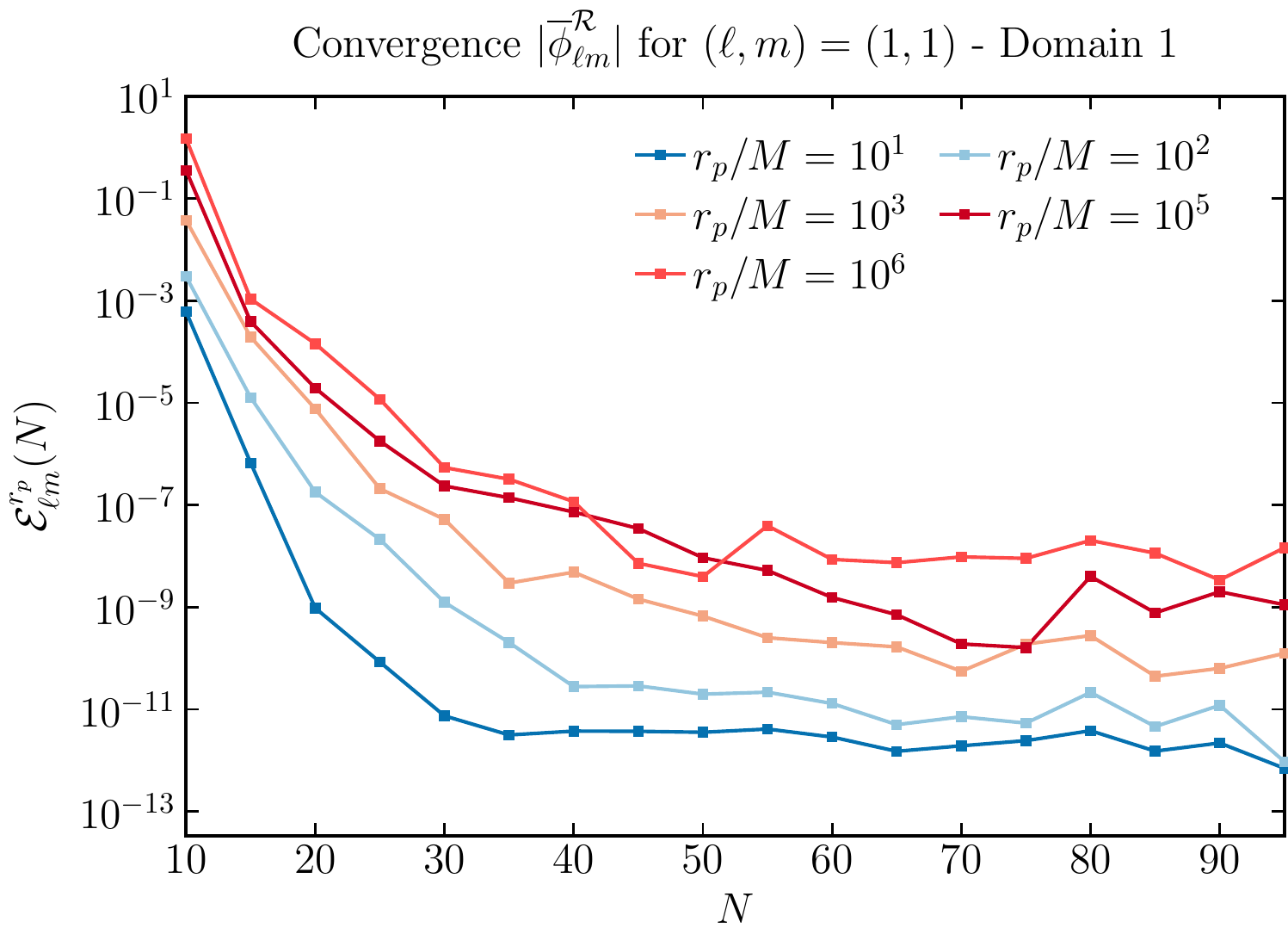}
	\caption{Exponential error decay for large orbits with optimal AnMR for worldtube sources. It demonstrates the challenging computation of $r_p/M=10^6$ and $(\ell,m)=(1,1)$ where machine precision is reached around $N=50$. In this particular example, the error saturates at $\sim 10^{-6}$ because the residual field $\overline{\Phi}^{\mathcal{R}}_{\ell,m}$ assumes values $\left|\Phi^{\mathcal{R}}_{1,1}\right| \sim 10^{-13}$, i.e. the around roundoff limits of double float operations.}
	\label{fig:EffectiveSource_AnMR_convergence_ell_m}
\end{figure}
We show in Fig.~\ref{fig:EffectiveSource_AnMR_convergence_ell_m} the relative error for a fixed angular mode $(\ell,m) = (1,1)$ for several values of $\log_{10}(r_p/M)=\{2\cdots 6\}$. 
We find exponential convergence, but the saturation happens at larger errors as one increases the particle's location. 
Note that for $r_p=10^6M$, the solution already approaches machine precision for double float operations. 
One has $\left|\Phi^{\mathcal{R}}_{1,1}\right| \sim 10^{-13}$, $\left|\Phi^{\mathcal{R}}_{50,50}\right| \sim 10^{-14}$, and $\left|\Phi^{\mathcal{R}}_{100,0}\right| \sim 10^{-15}$. 
The high saturation error in Fig.~\ref{fig:EffectiveSource_AnMR_convergence_ell_m} reflects limitations with respect to precision. A more accurate calculation would require higher internal precision.

\bibliography{bibitems}

\begin{thebibliography}{97}%
\makeatletter
\providecommand \@ifxundefined [1]{%
 \@ifx{#1\undefined}
}%
\providecommand \@ifnum [1]{%
 \ifnum #1\expandafter \@firstoftwo
 \else \expandafter \@secondoftwo
 \fi
}%
\providecommand \@ifx [1]{%
 \ifx #1\expandafter \@firstoftwo
 \else \expandafter \@secondoftwo
 \fi
}%
\providecommand \natexlab [1]{#1}%
\providecommand \enquote  [1]{``#1''}%
\providecommand \bibnamefont  [1]{#1}%
\providecommand \bibfnamefont [1]{#1}%
\providecommand \citenamefont [1]{#1}%
\providecommand \href@noop [0]{\@secondoftwo}%
\providecommand \href [0]{\begingroup \@sanitize@url \@href}%
\providecommand \@href[1]{\@@startlink{#1}\@@href}%
\providecommand \@@href[1]{\endgroup#1\@@endlink}%
\providecommand \@sanitize@url [0]{\catcode `\\12\catcode `\$12\catcode
  `\&12\catcode `\#12\catcode `\^12\catcode `\_12\catcode `\%12\relax}%
\providecommand \@@startlink[1]{}%
\providecommand \@@endlink[0]{}%
\providecommand \url  [0]{\begingroup\@sanitize@url \@url }%
\providecommand \@url [1]{\endgroup\@href {#1}{\urlprefix }}%
\providecommand \urlprefix  [0]{URL }%
\providecommand \Eprint [0]{\href }%
\providecommand \doibase [0]{https://doi.org/}%
\providecommand \selectlanguage [0]{\@gobble}%
\providecommand \bibinfo  [0]{\@secondoftwo}%
\providecommand \bibfield  [0]{\@secondoftwo}%
\providecommand \translation [1]{[#1]}%
\providecommand \BibitemOpen [0]{}%
\providecommand \bibitemStop [0]{}%
\providecommand \bibitemNoStop [0]{.\EOS\space}%
\providecommand \EOS [0]{\spacefactor3000\relax}%
\providecommand \BibitemShut  [1]{\csname bibitem#1\endcsname}%
\let\auto@bib@innerbib\@empty
\bibitem [{\citenamefont {Abbott}\ \emph
  {et~al.}(2021{\natexlab{a}})\citenamefont {Abbott} \emph
  {et~al.}}]{LIGOScientific:2020kqk}%
  \BibitemOpen
  \bibfield  {author} {\bibinfo {author} {\bibfnamefont {R.}~\bibnamefont
  {Abbott}} \emph {et~al.} (\bibinfo {collaboration} {LIGO Scientific,
  Virgo}),\ }\bibfield  {title} {\bibinfo {title} {{Population Properties of
  Compact Objects from the Second LIGO-Virgo Gravitational-Wave Transient
  Catalog}},\ }\href {https://doi.org/10.3847/2041-8213/abe949} {\bibfield
  {journal} {\bibinfo  {journal} {Astrophys. J. Lett.}\ }\textbf {\bibinfo
  {volume} {913}},\ \bibinfo {pages} {L7} (\bibinfo {year}
  {2021}{\natexlab{a}})},\ \Eprint {https://arxiv.org/abs/2010.14533}
  {arXiv:2010.14533 [astro-ph.HE]} \BibitemShut {NoStop}%
\bibitem [{\citenamefont {Abbott}\ \emph
  {et~al.}(2021{\natexlab{b}})\citenamefont {Abbott} \emph
  {et~al.}}]{LIGOScientific:2020tif}%
  \BibitemOpen
  \bibfield  {author} {\bibinfo {author} {\bibfnamefont {R.}~\bibnamefont
  {Abbott}} \emph {et~al.} (\bibinfo {collaboration} {LIGO Scientific,
  Virgo}),\ }\bibfield  {title} {\bibinfo {title} {{Tests of general relativity
  with binary black holes from the second LIGO-Virgo gravitational-wave
  transient catalog}},\ }\href {https://doi.org/10.1103/PhysRevD.103.122002}
  {\bibfield  {journal} {\bibinfo  {journal} {Phys. Rev. D}\ }\textbf {\bibinfo
  {volume} {103}},\ \bibinfo {pages} {122002} (\bibinfo {year}
  {2021}{\natexlab{b}})},\ \Eprint {https://arxiv.org/abs/2010.14529}
  {arXiv:2010.14529 [gr-qc]} \BibitemShut {NoStop}%
\bibitem [{\citenamefont {Amaro-Seoane}\ \emph {et~al.}(2017)\citenamefont
  {Amaro-Seoane}, \citenamefont {Audley}, \citenamefont {Babak}, \citenamefont
  {Baker}, \citenamefont {Barausse}, \citenamefont {Bender}, \citenamefont
  {Berti}, \citenamefont {Binetruy}, \citenamefont {Born}, \citenamefont
  {Bortoluzzi}, \citenamefont {Camp}, \citenamefont {Caprini}, \citenamefont
  {Cardoso}, \citenamefont {Colpi}, \citenamefont {Conklin}, \citenamefont
  {Cornish}, \citenamefont {Cutler}, \citenamefont {Danzmann}, \citenamefont
  {Dolesi}, \citenamefont {Ferraioli}, \citenamefont {Ferroni}, \citenamefont
  {Fitzsimons}, \citenamefont {Gair}, \citenamefont {Bote}, \citenamefont
  {Giardini}, \citenamefont {Gibert}, \citenamefont {Grimani}, \citenamefont
  {Halloin}, \citenamefont {Heinzel}, \citenamefont {Hertog}, \citenamefont
  {Hewitson}, \citenamefont {Holley-Bockelmann}, \citenamefont {Hollington},
  \citenamefont {Hueller}, \citenamefont {Inchauspe}, \citenamefont {Jetzer},
  \citenamefont {Karnesis}, \citenamefont {Killow}, \citenamefont {Klein},
  \citenamefont {Klipstein}, \citenamefont {Korsakova}, \citenamefont {Larson},
  \citenamefont {Livas}, \citenamefont {Lloro}, \citenamefont {Man},
  \citenamefont {Mance}, \citenamefont {Martino}, \citenamefont {Mateos},
  \citenamefont {McKenzie}, \citenamefont {McWilliams}, \citenamefont {Miller},
  \citenamefont {Mueller}, \citenamefont {Nardini}, \citenamefont {Nelemans},
  \citenamefont {Nofrarias}, \citenamefont {Petiteau}, \citenamefont {Pivato},
  \citenamefont {Plagnol}, \citenamefont {Porter}, \citenamefont {Reiche},
  \citenamefont {Robertson}, \citenamefont {Robertson}, \citenamefont {Rossi},
  \citenamefont {Russano}, \citenamefont {Schutz}, \citenamefont {Sesana},
  \citenamefont {Shoemaker}, \citenamefont {Slutsky}, \citenamefont {Sopuerta},
  \citenamefont {Sumner}, \citenamefont {Tamanini}, \citenamefont {Thorpe},
  \citenamefont {Troebs}, \citenamefont {Vallisneri}, \citenamefont {Vecchio},
  \citenamefont {Vetrugno}, \citenamefont {Vitale}, \citenamefont {Volonteri},
  \citenamefont {Wanner}, \citenamefont {Ward}, \citenamefont {Wass},
  \citenamefont {Weber}, \citenamefont {Ziemer},\ and\ \citenamefont
  {Zweifel}}]{amaroseoane2017laser}%
  \BibitemOpen
  \bibfield  {author} {\bibinfo {author} {\bibfnamefont {P.}~\bibnamefont
  {Amaro-Seoane}}, \bibinfo {author} {\bibfnamefont {H.}~\bibnamefont
  {Audley}}, \bibinfo {author} {\bibfnamefont {S.}~\bibnamefont {Babak}},
  \bibinfo {author} {\bibfnamefont {J.}~\bibnamefont {Baker}}, \bibinfo
  {author} {\bibfnamefont {E.}~\bibnamefont {Barausse}}, \bibinfo {author}
  {\bibfnamefont {P.}~\bibnamefont {Bender}}, \bibinfo {author} {\bibfnamefont
  {E.}~\bibnamefont {Berti}}, \bibinfo {author} {\bibfnamefont
  {P.}~\bibnamefont {Binetruy}}, \bibinfo {author} {\bibfnamefont
  {M.}~\bibnamefont {Born}}, \bibinfo {author} {\bibfnamefont {D.}~\bibnamefont
  {Bortoluzzi}}, \bibinfo {author} {\bibfnamefont {J.}~\bibnamefont {Camp}},
  \bibinfo {author} {\bibfnamefont {C.}~\bibnamefont {Caprini}}, \bibinfo
  {author} {\bibfnamefont {V.}~\bibnamefont {Cardoso}}, \bibinfo {author}
  {\bibfnamefont {M.}~\bibnamefont {Colpi}}, \bibinfo {author} {\bibfnamefont
  {J.}~\bibnamefont {Conklin}}, \bibinfo {author} {\bibfnamefont
  {N.}~\bibnamefont {Cornish}}, \bibinfo {author} {\bibfnamefont
  {C.}~\bibnamefont {Cutler}}, \bibinfo {author} {\bibfnamefont
  {K.}~\bibnamefont {Danzmann}}, \bibinfo {author} {\bibfnamefont
  {R.}~\bibnamefont {Dolesi}}, \bibinfo {author} {\bibfnamefont
  {L.}~\bibnamefont {Ferraioli}}, \bibinfo {author} {\bibfnamefont
  {V.}~\bibnamefont {Ferroni}}, \bibinfo {author} {\bibfnamefont
  {E.}~\bibnamefont {Fitzsimons}}, \bibinfo {author} {\bibfnamefont
  {J.}~\bibnamefont {Gair}}, \bibinfo {author} {\bibfnamefont {L.~G.}\
  \bibnamefont {Bote}}, \bibinfo {author} {\bibfnamefont {D.}~\bibnamefont
  {Giardini}}, \bibinfo {author} {\bibfnamefont {F.}~\bibnamefont {Gibert}},
  \bibinfo {author} {\bibfnamefont {C.}~\bibnamefont {Grimani}}, \bibinfo
  {author} {\bibfnamefont {H.}~\bibnamefont {Halloin}}, \bibinfo {author}
  {\bibfnamefont {G.}~\bibnamefont {Heinzel}}, \bibinfo {author} {\bibfnamefont
  {T.}~\bibnamefont {Hertog}}, \bibinfo {author} {\bibfnamefont
  {M.}~\bibnamefont {Hewitson}}, \bibinfo {author} {\bibfnamefont
  {K.}~\bibnamefont {Holley-Bockelmann}}, \bibinfo {author} {\bibfnamefont
  {D.}~\bibnamefont {Hollington}}, \bibinfo {author} {\bibfnamefont
  {M.}~\bibnamefont {Hueller}}, \bibinfo {author} {\bibfnamefont
  {H.}~\bibnamefont {Inchauspe}}, \bibinfo {author} {\bibfnamefont
  {P.}~\bibnamefont {Jetzer}}, \bibinfo {author} {\bibfnamefont
  {N.}~\bibnamefont {Karnesis}}, \bibinfo {author} {\bibfnamefont
  {C.}~\bibnamefont {Killow}}, \bibinfo {author} {\bibfnamefont
  {A.}~\bibnamefont {Klein}}, \bibinfo {author} {\bibfnamefont
  {B.}~\bibnamefont {Klipstein}}, \bibinfo {author} {\bibfnamefont
  {N.}~\bibnamefont {Korsakova}}, \bibinfo {author} {\bibfnamefont {S.~L.}\
  \bibnamefont {Larson}}, \bibinfo {author} {\bibfnamefont {J.}~\bibnamefont
  {Livas}}, \bibinfo {author} {\bibfnamefont {I.}~\bibnamefont {Lloro}},
  \bibinfo {author} {\bibfnamefont {N.}~\bibnamefont {Man}}, \bibinfo {author}
  {\bibfnamefont {D.}~\bibnamefont {Mance}}, \bibinfo {author} {\bibfnamefont
  {J.}~\bibnamefont {Martino}}, \bibinfo {author} {\bibfnamefont
  {I.}~\bibnamefont {Mateos}}, \bibinfo {author} {\bibfnamefont
  {K.}~\bibnamefont {McKenzie}}, \bibinfo {author} {\bibfnamefont {S.~T.}\
  \bibnamefont {McWilliams}}, \bibinfo {author} {\bibfnamefont
  {C.}~\bibnamefont {Miller}}, \bibinfo {author} {\bibfnamefont
  {G.}~\bibnamefont {Mueller}}, \bibinfo {author} {\bibfnamefont
  {G.}~\bibnamefont {Nardini}}, \bibinfo {author} {\bibfnamefont
  {G.}~\bibnamefont {Nelemans}}, \bibinfo {author} {\bibfnamefont
  {M.}~\bibnamefont {Nofrarias}}, \bibinfo {author} {\bibfnamefont
  {A.}~\bibnamefont {Petiteau}}, \bibinfo {author} {\bibfnamefont
  {P.}~\bibnamefont {Pivato}}, \bibinfo {author} {\bibfnamefont
  {E.}~\bibnamefont {Plagnol}}, \bibinfo {author} {\bibfnamefont
  {E.}~\bibnamefont {Porter}}, \bibinfo {author} {\bibfnamefont
  {J.}~\bibnamefont {Reiche}}, \bibinfo {author} {\bibfnamefont
  {D.}~\bibnamefont {Robertson}}, \bibinfo {author} {\bibfnamefont
  {N.}~\bibnamefont {Robertson}}, \bibinfo {author} {\bibfnamefont
  {E.}~\bibnamefont {Rossi}}, \bibinfo {author} {\bibfnamefont
  {G.}~\bibnamefont {Russano}}, \bibinfo {author} {\bibfnamefont
  {B.}~\bibnamefont {Schutz}}, \bibinfo {author} {\bibfnamefont
  {A.}~\bibnamefont {Sesana}}, \bibinfo {author} {\bibfnamefont
  {D.}~\bibnamefont {Shoemaker}}, \bibinfo {author} {\bibfnamefont
  {J.}~\bibnamefont {Slutsky}}, \bibinfo {author} {\bibfnamefont {C.~F.}\
  \bibnamefont {Sopuerta}}, \bibinfo {author} {\bibfnamefont {T.}~\bibnamefont
  {Sumner}}, \bibinfo {author} {\bibfnamefont {N.}~\bibnamefont {Tamanini}},
  \bibinfo {author} {\bibfnamefont {I.}~\bibnamefont {Thorpe}}, \bibinfo
  {author} {\bibfnamefont {M.}~\bibnamefont {Troebs}}, \bibinfo {author}
  {\bibfnamefont {M.}~\bibnamefont {Vallisneri}}, \bibinfo {author}
  {\bibfnamefont {A.}~\bibnamefont {Vecchio}}, \bibinfo {author} {\bibfnamefont
  {D.}~\bibnamefont {Vetrugno}}, \bibinfo {author} {\bibfnamefont
  {S.}~\bibnamefont {Vitale}}, \bibinfo {author} {\bibfnamefont
  {M.}~\bibnamefont {Volonteri}}, \bibinfo {author} {\bibfnamefont
  {G.}~\bibnamefont {Wanner}}, \bibinfo {author} {\bibfnamefont
  {H.}~\bibnamefont {Ward}}, \bibinfo {author} {\bibfnamefont {P.}~\bibnamefont
  {Wass}}, \bibinfo {author} {\bibfnamefont {W.}~\bibnamefont {Weber}},
  \bibinfo {author} {\bibfnamefont {J.}~\bibnamefont {Ziemer}},\ and\ \bibinfo
  {author} {\bibfnamefont {P.}~\bibnamefont {Zweifel}},\ }\href@noop {}
  {\bibinfo {title} {Laser interferometer space antenna}} (\bibinfo {year}
  {2017}),\ \Eprint {https://arxiv.org/abs/1702.00786} {arXiv:1702.00786
  [astro-ph.IM]} \BibitemShut {NoStop}%
\bibitem [{\citenamefont {Amaro-Seoane}(2018)}]{Amaro-Seoane:2018gbb}%
  \BibitemOpen
  \bibfield  {author} {\bibinfo {author} {\bibfnamefont {P.}~\bibnamefont
  {Amaro-Seoane}},\ }\bibfield  {title} {\bibinfo {title} {{Detecting
  Intermediate-Mass Ratio Inspirals From The Ground And Space}},\ }\href
  {https://doi.org/10.1103/PhysRevD.98.063018} {\bibfield  {journal} {\bibinfo
  {journal} {Phys. Rev. D}\ }\textbf {\bibinfo {volume} {98}},\ \bibinfo
  {pages} {063018} (\bibinfo {year} {2018})},\ \Eprint
  {https://arxiv.org/abs/1807.03824} {arXiv:1807.03824 [astro-ph.HE]}
  \BibitemShut {NoStop}%
\bibitem [{\citenamefont {Poisson}\ \emph {et~al.}(2011)\citenamefont
  {Poisson}, \citenamefont {Pound},\ and\ \citenamefont
  {Vega}}]{Poisson:2011nh}%
  \BibitemOpen
  \bibfield  {author} {\bibinfo {author} {\bibfnamefont {E.}~\bibnamefont
  {Poisson}}, \bibinfo {author} {\bibfnamefont {A.}~\bibnamefont {Pound}},\
  and\ \bibinfo {author} {\bibfnamefont {I.}~\bibnamefont {Vega}},\ }\bibfield
  {title} {\bibinfo {title} {{The Motion of point particles in curved
  spacetime}},\ }\href {https://doi.org/10.12942/lrr-2011-7} {\bibfield
  {journal} {\bibinfo  {journal} {Living Rev. Rel.}\ }\textbf {\bibinfo
  {volume} {14}},\ \bibinfo {pages} {7} (\bibinfo {year} {2011})},\ \Eprint
  {https://arxiv.org/abs/1102.0529} {arXiv:1102.0529 [gr-qc]} \BibitemShut
  {NoStop}%
\bibitem [{\citenamefont {Barack}\ and\ \citenamefont
  {Pound}(2019)}]{Barack:2018yvs}%
  \BibitemOpen
  \bibfield  {author} {\bibinfo {author} {\bibfnamefont {L.}~\bibnamefont
  {Barack}}\ and\ \bibinfo {author} {\bibfnamefont {A.}~\bibnamefont {Pound}},\
  }\bibfield  {title} {\bibinfo {title} {{Self-force and radiation reaction in
  general relativity}},\ }\href {https://doi.org/10.1088/1361-6633/aae552}
  {\bibfield  {journal} {\bibinfo  {journal} {Rept. Prog. Phys.}\ }\textbf
  {\bibinfo {volume} {82}},\ \bibinfo {pages} {016904} (\bibinfo {year}
  {2019})},\ \Eprint {https://arxiv.org/abs/1805.10385} {arXiv:1805.10385
  [gr-qc]} \BibitemShut {NoStop}%
\bibitem [{\citenamefont {Pound}\ and\ \citenamefont
  {Wardell}(2021)}]{Pound:2021qin}%
  \BibitemOpen
  \bibfield  {author} {\bibinfo {author} {\bibfnamefont {A.}~\bibnamefont
  {Pound}}\ and\ \bibinfo {author} {\bibfnamefont {B.}~\bibnamefont
  {Wardell}},\ }\bibfield  {title} {\bibinfo {title} {{Black hole perturbation
  theory and gravitational self-force}},\ }\href@noop {} {\  (\bibinfo {year}
  {2021})},\ \Eprint {https://arxiv.org/abs/2101.04592} {arXiv:2101.04592
  [gr-qc]} \BibitemShut {NoStop}%
\bibitem [{\citenamefont {Hinderer}\ and\ \citenamefont
  {Flanagan}(2008)}]{Hinderer:2008dm}%
  \BibitemOpen
  \bibfield  {author} {\bibinfo {author} {\bibfnamefont {T.}~\bibnamefont
  {Hinderer}}\ and\ \bibinfo {author} {\bibfnamefont {E.~E.}\ \bibnamefont
  {Flanagan}},\ }\bibfield  {title} {\bibinfo {title} {{Two timescale analysis
  of extreme mass ratio inspirals in Kerr. I. Orbital Motion}},\ }\href
  {https://doi.org/10.1103/PhysRevD.78.064028} {\bibfield  {journal} {\bibinfo
  {journal} {Phys. Rev. D}\ }\textbf {\bibinfo {volume} {78}},\ \bibinfo
  {pages} {064028} (\bibinfo {year} {2008})},\ \Eprint
  {https://arxiv.org/abs/0805.3337} {arXiv:0805.3337 [gr-qc]} \BibitemShut
  {NoStop}%
\bibitem [{\citenamefont {Wardell}\ \emph {et~al.}(2021)\citenamefont
  {Wardell}, \citenamefont {Pound}, \citenamefont {Warburton}, \citenamefont
  {Miller}, \citenamefont {Durkan},\ and\ \citenamefont
  {Le~Tiec}}]{Wardell:2021fyy}%
  \BibitemOpen
  \bibfield  {author} {\bibinfo {author} {\bibfnamefont {B.}~\bibnamefont
  {Wardell}}, \bibinfo {author} {\bibfnamefont {A.}~\bibnamefont {Pound}},
  \bibinfo {author} {\bibfnamefont {N.}~\bibnamefont {Warburton}}, \bibinfo
  {author} {\bibfnamefont {J.}~\bibnamefont {Miller}}, \bibinfo {author}
  {\bibfnamefont {L.}~\bibnamefont {Durkan}},\ and\ \bibinfo {author}
  {\bibfnamefont {A.}~\bibnamefont {Le~Tiec}},\ }\bibfield  {title} {\bibinfo
  {title} {{Gravitational waveforms for compact binaries from second-order
  self-force theory}},\ }\href@noop {} {\  (\bibinfo {year} {2021})},\ \Eprint
  {https://arxiv.org/abs/2112.12265} {arXiv:2112.12265 [gr-qc]} \BibitemShut
  {NoStop}%
\bibitem [{\citenamefont {Pound}(2012)}]{Pound:2012nt}%
  \BibitemOpen
  \bibfield  {author} {\bibinfo {author} {\bibfnamefont {A.}~\bibnamefont
  {Pound}},\ }\bibfield  {title} {\bibinfo {title} {{Second-order gravitational
  self-force}},\ }\href {https://doi.org/10.1103/PhysRevLett.109.051101}
  {\bibfield  {journal} {\bibinfo  {journal} {Phys. Rev. Lett.}\ }\textbf
  {\bibinfo {volume} {109}},\ \bibinfo {pages} {051101} (\bibinfo {year}
  {2012})},\ \Eprint {https://arxiv.org/abs/1201.5089} {arXiv:1201.5089
  [gr-qc]} \BibitemShut {NoStop}%
\bibitem [{\citenamefont {Gralla}(2012)}]{Gralla:2012db}%
  \BibitemOpen
  \bibfield  {author} {\bibinfo {author} {\bibfnamefont {S.~E.}\ \bibnamefont
  {Gralla}},\ }\bibfield  {title} {\bibinfo {title} {{Second Order
  Gravitational Self Force}},\ }\href
  {https://doi.org/10.1103/PhysRevD.85.124011} {\bibfield  {journal} {\bibinfo
  {journal} {Phys. Rev. D}\ }\textbf {\bibinfo {volume} {85}},\ \bibinfo
  {pages} {124011} (\bibinfo {year} {2012})},\ \Eprint
  {https://arxiv.org/abs/1203.3189} {arXiv:1203.3189 [gr-qc]} \BibitemShut
  {NoStop}%
\bibitem [{\citenamefont {Diaz-Rivera}\ \emph {et~al.}(2004)\citenamefont
  {Diaz-Rivera}, \citenamefont {Messaritaki}, \citenamefont {Whiting},\ and\
  \citenamefont {Detweiler}}]{Diaz-Rivera:2004nim}%
  \BibitemOpen
  \bibfield  {author} {\bibinfo {author} {\bibfnamefont {L.~M.}\ \bibnamefont
  {Diaz-Rivera}}, \bibinfo {author} {\bibfnamefont {E.}~\bibnamefont
  {Messaritaki}}, \bibinfo {author} {\bibfnamefont {B.~F.}\ \bibnamefont
  {Whiting}},\ and\ \bibinfo {author} {\bibfnamefont {S.~L.}\ \bibnamefont
  {Detweiler}},\ }\bibfield  {title} {\bibinfo {title} {{Scalar field
  self-force effects on orbits about a Schwarzschild black hole}},\ }\href
  {https://doi.org/10.1103/PhysRevD.70.124018} {\bibfield  {journal} {\bibinfo
  {journal} {Phys. Rev. D}\ }\textbf {\bibinfo {volume} {70}},\ \bibinfo
  {pages} {124018} (\bibinfo {year} {2004})},\ \Eprint
  {https://arxiv.org/abs/gr-qc/0410011} {arXiv:gr-qc/0410011} \BibitemShut
  {NoStop}%
\bibitem [{\citenamefont {Warburton}\ and\ \citenamefont
  {Barack}(2011)}]{Warburton:2011hp}%
  \BibitemOpen
  \bibfield  {author} {\bibinfo {author} {\bibfnamefont {N.}~\bibnamefont
  {Warburton}}\ and\ \bibinfo {author} {\bibfnamefont {L.}~\bibnamefont
  {Barack}},\ }\bibfield  {title} {\bibinfo {title} {{Self force on a scalar
  charge in Kerr spacetime: eccentric equatorial orbits}},\ }\href
  {https://doi.org/10.1103/PhysRevD.83.124038} {\bibfield  {journal} {\bibinfo
  {journal} {Phys. Rev. D}\ }\textbf {\bibinfo {volume} {83}},\ \bibinfo
  {pages} {124038} (\bibinfo {year} {2011})},\ \Eprint
  {https://arxiv.org/abs/1103.0287} {arXiv:1103.0287 [gr-qc]} \BibitemShut
  {NoStop}%
\bibitem [{\citenamefont {Akcay}\ \emph {et~al.}(2013)\citenamefont {Akcay},
  \citenamefont {Warburton},\ and\ \citenamefont {Barack}}]{Akcay:2013wfa}%
  \BibitemOpen
  \bibfield  {author} {\bibinfo {author} {\bibfnamefont {S.}~\bibnamefont
  {Akcay}}, \bibinfo {author} {\bibfnamefont {N.}~\bibnamefont {Warburton}},\
  and\ \bibinfo {author} {\bibfnamefont {L.}~\bibnamefont {Barack}},\
  }\bibfield  {title} {\bibinfo {title} {{Frequency-domain algorithm for the
  Lorenz-gauge gravitational self-force}},\ }\href
  {https://doi.org/10.1103/PhysRevD.88.104009} {\bibfield  {journal} {\bibinfo
  {journal} {Phys. Rev. D}\ }\textbf {\bibinfo {volume} {88}},\ \bibinfo
  {pages} {104009} (\bibinfo {year} {2013})},\ \Eprint
  {https://arxiv.org/abs/1308.5223} {arXiv:1308.5223 [gr-qc]} \BibitemShut
  {NoStop}%
\bibitem [{\citenamefont {Akcay}(2011)}]{Akcay:2010dx}%
  \BibitemOpen
  \bibfield  {author} {\bibinfo {author} {\bibfnamefont {S.}~\bibnamefont
  {Akcay}},\ }\bibfield  {title} {\bibinfo {title} {{A Fast Frequency-Domain
  Algorithm for Gravitational Self-Force: I. Circular Orbits in Schwarzschild
  Spacetime}},\ }\href {https://doi.org/10.1103/PhysRevD.83.124026} {\bibfield
  {journal} {\bibinfo  {journal} {Phys. Rev. D}\ }\textbf {\bibinfo {volume}
  {83}},\ \bibinfo {pages} {124026} (\bibinfo {year} {2011})},\ \Eprint
  {https://arxiv.org/abs/1012.5860} {arXiv:1012.5860 [gr-qc]} \BibitemShut
  {NoStop}%
\bibitem [{\citenamefont {Merlin}\ and\ \citenamefont
  {Shah}(2015)}]{Merlin:2014qda}%
  \BibitemOpen
  \bibfield  {author} {\bibinfo {author} {\bibfnamefont {C.}~\bibnamefont
  {Merlin}}\ and\ \bibinfo {author} {\bibfnamefont {A.~G.}\ \bibnamefont
  {Shah}},\ }\bibfield  {title} {\bibinfo {title} {{Self-force from
  reconstructed metric perturbations: numerical implementation in Schwarzschild
  spacetime}},\ }\href {https://doi.org/10.1103/PhysRevD.91.024005} {\bibfield
  {journal} {\bibinfo  {journal} {Phys. Rev. D}\ }\textbf {\bibinfo {volume}
  {91}},\ \bibinfo {pages} {024005} (\bibinfo {year} {2015})},\ \Eprint
  {https://arxiv.org/abs/1410.2998} {arXiv:1410.2998 [gr-qc]} \BibitemShut
  {NoStop}%
\bibitem [{\citenamefont {van~de Meent}(2016)}]{vandeMeent:2016pee}%
  \BibitemOpen
  \bibfield  {author} {\bibinfo {author} {\bibfnamefont {M.}~\bibnamefont
  {van~de Meent}},\ }\bibfield  {title} {\bibinfo {title} {{Gravitational
  self-force on eccentric equatorial orbits around a Kerr black hole}},\ }\href
  {https://doi.org/10.1103/PhysRevD.94.044034} {\bibfield  {journal} {\bibinfo
  {journal} {Phys. Rev. D}\ }\textbf {\bibinfo {volume} {94}},\ \bibinfo
  {pages} {044034} (\bibinfo {year} {2016})},\ \Eprint
  {https://arxiv.org/abs/1606.06297} {arXiv:1606.06297 [gr-qc]} \BibitemShut
  {NoStop}%
\bibitem [{\citenamefont {van~de Meent}(2018)}]{vandeMeent:2017bcc}%
  \BibitemOpen
  \bibfield  {author} {\bibinfo {author} {\bibfnamefont {M.}~\bibnamefont
  {van~de Meent}},\ }\bibfield  {title} {\bibinfo {title} {{Gravitational
  self-force on generic bound geodesics in Kerr spacetime}},\ }\href
  {https://doi.org/10.1103/PhysRevD.97.104033} {\bibfield  {journal} {\bibinfo
  {journal} {Phys. Rev. D}\ }\textbf {\bibinfo {volume} {97}},\ \bibinfo
  {pages} {104033} (\bibinfo {year} {2018})},\ \Eprint
  {https://arxiv.org/abs/1711.09607} {arXiv:1711.09607 [gr-qc]} \BibitemShut
  {NoStop}%
\bibitem [{\citenamefont {Warburton}\ and\ \citenamefont
  {Barack}(2010)}]{Warburton:2010eq}%
  \BibitemOpen
  \bibfield  {author} {\bibinfo {author} {\bibfnamefont {N.}~\bibnamefont
  {Warburton}}\ and\ \bibinfo {author} {\bibfnamefont {L.}~\bibnamefont
  {Barack}},\ }\bibfield  {title} {\bibinfo {title} {{Self force on a scalar
  charge in Kerr spacetime: circular equatorial orbits}},\ }\href
  {https://doi.org/10.1103/PhysRevD.81.084039} {\bibfield  {journal} {\bibinfo
  {journal} {Phys. Rev. D}\ }\textbf {\bibinfo {volume} {81}},\ \bibinfo
  {pages} {084039} (\bibinfo {year} {2010})},\ \Eprint
  {https://arxiv.org/abs/1003.1860} {arXiv:1003.1860 [gr-qc]} \BibitemShut
  {NoStop}%
\bibitem [{\citenamefont {Warburton}\ and\ \citenamefont
  {Wardell}(2014)}]{Warburton:2013lea}%
  \BibitemOpen
  \bibfield  {author} {\bibinfo {author} {\bibfnamefont {N.}~\bibnamefont
  {Warburton}}\ and\ \bibinfo {author} {\bibfnamefont {B.}~\bibnamefont
  {Wardell}},\ }\bibfield  {title} {\bibinfo {title} {{Applying the
  effective-source approach to frequency-domain self-force calculations}},\
  }\href {https://doi.org/10.1103/PhysRevD.89.044046} {\bibfield  {journal}
  {\bibinfo  {journal} {Phys. Rev. D}\ }\textbf {\bibinfo {volume} {89}},\
  \bibinfo {pages} {044046} (\bibinfo {year} {2014})},\ \Eprint
  {https://arxiv.org/abs/1311.3104} {arXiv:1311.3104 [gr-qc]} \BibitemShut
  {NoStop}%
\bibitem [{\citenamefont {Wardell}\ and\ \citenamefont
  {Warburton}(2015)}]{Wardell:2015ada}%
  \BibitemOpen
  \bibfield  {author} {\bibinfo {author} {\bibfnamefont {B.}~\bibnamefont
  {Wardell}}\ and\ \bibinfo {author} {\bibfnamefont {N.}~\bibnamefont
  {Warburton}},\ }\bibfield  {title} {\bibinfo {title} {{Applying the
  effective-source approach to frequency-domain self-force calculations:
  Lorenz-gauge gravitational perturbations}},\ }\href
  {https://doi.org/10.1103/PhysRevD.92.084019} {\bibfield  {journal} {\bibinfo
  {journal} {Phys. Rev. D}\ }\textbf {\bibinfo {volume} {92}},\ \bibinfo
  {pages} {084019} (\bibinfo {year} {2015})},\ \Eprint
  {https://arxiv.org/abs/1505.07841} {arXiv:1505.07841 [gr-qc]} \BibitemShut
  {NoStop}%
\bibitem [{\citenamefont {Miller}\ and\ \citenamefont
  {Pound}(2021)}]{Miller:2020bft}%
  \BibitemOpen
  \bibfield  {author} {\bibinfo {author} {\bibfnamefont {J.}~\bibnamefont
  {Miller}}\ and\ \bibinfo {author} {\bibfnamefont {A.}~\bibnamefont {Pound}},\
  }\bibfield  {title} {\bibinfo {title} {{Two-timescale evolution of
  extreme-mass-ratio inspirals: waveform generation scheme for quasicircular
  orbits in Schwarzschild spacetime}},\ }\href
  {https://doi.org/10.1103/PhysRevD.103.064048} {\bibfield  {journal} {\bibinfo
   {journal} {Phys. Rev. D}\ }\textbf {\bibinfo {volume} {103}},\ \bibinfo
  {pages} {064048} (\bibinfo {year} {2021})},\ \Eprint
  {https://arxiv.org/abs/2006.11263} {arXiv:2006.11263 [gr-qc]} \BibitemShut
  {NoStop}%
\bibitem [{\citenamefont {Sasaki}\ and\ \citenamefont
  {Tagoshi}(2003)}]{Sasaki:2003xr}%
  \BibitemOpen
  \bibfield  {author} {\bibinfo {author} {\bibfnamefont {M.}~\bibnamefont
  {Sasaki}}\ and\ \bibinfo {author} {\bibfnamefont {H.}~\bibnamefont
  {Tagoshi}},\ }\bibfield  {title} {\bibinfo {title} {{Analytic black hole
  perturbation approach to gravitational radiation}},\ }\href
  {https://doi.org/10.12942/lrr-2003-6} {\bibfield  {journal} {\bibinfo
  {journal} {Living Rev. Rel.}\ }\textbf {\bibinfo {volume} {6}},\ \bibinfo
  {pages} {6} (\bibinfo {year} {2003})},\ \Eprint
  {https://arxiv.org/abs/gr-qc/0306120} {arXiv:gr-qc/0306120} \BibitemShut
  {NoStop}%
\bibitem [{\citenamefont {Osburn}\ \emph {et~al.}(2014)\citenamefont {Osburn},
  \citenamefont {Forseth}, \citenamefont {Evans},\ and\ \citenamefont
  {Hopper}}]{Osburn:2014hoa}%
  \BibitemOpen
  \bibfield  {author} {\bibinfo {author} {\bibfnamefont {T.}~\bibnamefont
  {Osburn}}, \bibinfo {author} {\bibfnamefont {E.}~\bibnamefont {Forseth}},
  \bibinfo {author} {\bibfnamefont {C.~R.}\ \bibnamefont {Evans}},\ and\
  \bibinfo {author} {\bibfnamefont {S.}~\bibnamefont {Hopper}},\ }\bibfield
  {title} {\bibinfo {title} {{Lorenz gauge gravitational self-force
  calculations of eccentric binaries using a frequency domain procedure}},\
  }\href {https://doi.org/10.1103/PhysRevD.90.104031} {\bibfield  {journal}
  {\bibinfo  {journal} {Phys. Rev. D}\ }\textbf {\bibinfo {volume} {90}},\
  \bibinfo {pages} {104031} (\bibinfo {year} {2014})},\ \Eprint
  {https://arxiv.org/abs/1409.4419} {arXiv:1409.4419 [gr-qc]} \BibitemShut
  {NoStop}%
\bibitem [{\citenamefont {Shah}\ \emph {et~al.}(2014)\citenamefont {Shah},
  \citenamefont {Friedman},\ and\ \citenamefont {Whiting}}]{Shah:2013uya}%
  \BibitemOpen
  \bibfield  {author} {\bibinfo {author} {\bibfnamefont {A.~G.}\ \bibnamefont
  {Shah}}, \bibinfo {author} {\bibfnamefont {J.~L.}\ \bibnamefont {Friedman}},\
  and\ \bibinfo {author} {\bibfnamefont {B.~F.}\ \bibnamefont {Whiting}},\
  }\bibfield  {title} {\bibinfo {title} {{Finding high-order analytic
  post-Newtonian parameters from a high-precision numerical self-force
  calculation}},\ }\href {https://doi.org/10.1103/PhysRevD.89.064042}
  {\bibfield  {journal} {\bibinfo  {journal} {Phys. Rev. D}\ }\textbf {\bibinfo
  {volume} {89}},\ \bibinfo {pages} {064042} (\bibinfo {year} {2014})},\
  \Eprint {https://arxiv.org/abs/1312.1952} {arXiv:1312.1952 [gr-qc]}
  \BibitemShut {NoStop}%
\bibitem [{\citenamefont {Bini}\ and\ \citenamefont
  {Damour}(2013)}]{Bini:2013zaa}%
  \BibitemOpen
  \bibfield  {author} {\bibinfo {author} {\bibfnamefont {D.}~\bibnamefont
  {Bini}}\ and\ \bibinfo {author} {\bibfnamefont {T.}~\bibnamefont {Damour}},\
  }\bibfield  {title} {\bibinfo {title} {{Analytical determination of the
  two-body gravitational interaction potential at the fourth post-Newtonian
  approximation}},\ }\href {https://doi.org/10.1103/PhysRevD.87.121501}
  {\bibfield  {journal} {\bibinfo  {journal} {Phys. Rev. D}\ }\textbf {\bibinfo
  {volume} {87}},\ \bibinfo {pages} {121501} (\bibinfo {year} {2013})},\
  \Eprint {https://arxiv.org/abs/1305.4884} {arXiv:1305.4884 [gr-qc]}
  \BibitemShut {NoStop}%
\bibitem [{\citenamefont {Kavanagh}\ \emph {et~al.}(2015)\citenamefont
  {Kavanagh}, \citenamefont {Ottewill},\ and\ \citenamefont
  {Wardell}}]{Kavanagh:2015lva}%
  \BibitemOpen
  \bibfield  {author} {\bibinfo {author} {\bibfnamefont {C.}~\bibnamefont
  {Kavanagh}}, \bibinfo {author} {\bibfnamefont {A.~C.}\ \bibnamefont
  {Ottewill}},\ and\ \bibinfo {author} {\bibfnamefont {B.}~\bibnamefont
  {Wardell}},\ }\bibfield  {title} {\bibinfo {title} {{Analytical high-order
  post-Newtonian expansions for extreme mass ratio binaries}},\ }\href
  {https://doi.org/10.1103/PhysRevD.92.084025} {\bibfield  {journal} {\bibinfo
  {journal} {Phys. Rev. D}\ }\textbf {\bibinfo {volume} {92}},\ \bibinfo
  {pages} {084025} (\bibinfo {year} {2015})},\ \Eprint
  {https://arxiv.org/abs/1503.02334} {arXiv:1503.02334 [gr-qc]} \BibitemShut
  {NoStop}%
\bibitem [{\citenamefont {Munna}(2020)}]{Munna:2020iju}%
  \BibitemOpen
  \bibfield  {author} {\bibinfo {author} {\bibfnamefont {C.}~\bibnamefont
  {Munna}},\ }\bibfield  {title} {\bibinfo {title} {{Analytic post-Newtonian
  expansion of the energy and angular momentum radiated to infinity by
  eccentric-orbit nonspinning extreme-mass-ratio inspirals to the 19th
  order}},\ }\href {https://doi.org/10.1103/PhysRevD.102.124001} {\bibfield
  {journal} {\bibinfo  {journal} {Phys. Rev. D}\ }\textbf {\bibinfo {volume}
  {102}},\ \bibinfo {pages} {124001} (\bibinfo {year} {2020})},\ \Eprint
  {https://arxiv.org/abs/2008.10622} {arXiv:2008.10622 [gr-qc]} \BibitemShut
  {NoStop}%
\bibitem [{\citenamefont {Throwe}()}]{ThroweThesis}%
  \BibitemOpen
  \bibfield  {author} {\bibinfo {author} {\bibfnamefont {W.}~\bibnamefont
  {Throwe}},\ }\emph {\bibinfo {title}
  {\href{https://dspace.mit.edu/handle/1721.1/61270}{"High precision
  calculation of generic extreme mass ratio inspirals"}}},\ \href@noop {}
  {\bibinfo {type} {{MIT} undergraduate thesis}}\BibitemShut {NoStop}%
\bibitem [{\citenamefont {Zenginoglu}(2008)}]{Zenginoglu:2007jw}%
  \BibitemOpen
  \bibfield  {author} {\bibinfo {author} {\bibfnamefont {A.}~\bibnamefont
  {Zenginoglu}},\ }\bibfield  {title} {\bibinfo {title} {{Hyperboloidal
  foliations and scri-fixing}},\ }\href
  {https://doi.org/10.1088/0264-9381/25/14/145002} {\bibfield  {journal}
  {\bibinfo  {journal} {Class. Quant. Grav.}\ }\textbf {\bibinfo {volume}
  {25}},\ \bibinfo {pages} {145002} (\bibinfo {year} {2008})},\ \Eprint
  {https://arxiv.org/abs/0712.4333} {arXiv:0712.4333 [gr-qc]} \BibitemShut
  {NoStop}%
\bibitem [{\citenamefont {Zenginoglu}(2011)}]{Zenginoglu:2011jz}%
  \BibitemOpen
  \bibfield  {author} {\bibinfo {author} {\bibfnamefont {A.}~\bibnamefont
  {Zenginoglu}},\ }\bibfield  {title} {\bibinfo {title} {{A Geometric framework
  for black hole perturbations}},\ }\href
  {https://doi.org/10.1103/PhysRevD.83.127502} {\bibfield  {journal} {\bibinfo
  {journal} {Phys. Rev.}\ }\textbf {\bibinfo {volume} {D83}},\ \bibinfo {pages}
  {127502} (\bibinfo {year} {2011})},\ \Eprint
  {https://arxiv.org/abs/1102.2451} {arXiv:1102.2451 [gr-qc]} \BibitemShut
  {NoStop}%
\bibitem [{\citenamefont {Panosso~Macedo}(2020)}]{PanossoMacedo:2019npm}%
  \BibitemOpen
  \bibfield  {author} {\bibinfo {author} {\bibfnamefont {R.}~\bibnamefont
  {Panosso~Macedo}},\ }\bibfield  {title} {\bibinfo {title} {{Hyperboloidal
  framework for the Kerr spacetime}},\ }\href
  {https://doi.org/10.1088/1361-6382/ab6e3e} {\bibfield  {journal} {\bibinfo
  {journal} {Class. Quant. Grav.}\ }\textbf {\bibinfo {volume} {37}},\ \bibinfo
  {pages} {065019} (\bibinfo {year} {2020})},\ \Eprint
  {https://arxiv.org/abs/1910.13452} {arXiv:1910.13452 [gr-qc]} \BibitemShut
  {NoStop}%
\bibitem [{\citenamefont
  {Zengino{\u{g}}lu}(2008)}]{zenginouglu2008hyperboloidal}%
  \BibitemOpen
  \bibfield  {author} {\bibinfo {author} {\bibfnamefont {A.}~\bibnamefont
  {Zengino{\u{g}}lu}},\ }\bibfield  {title} {\bibinfo {title} {A hyperboloidal
  study of tail decay rates for scalar and yang--mills fields},\ }\href@noop {}
  {\bibfield  {journal} {\bibinfo  {journal} {Classical and quantum gravity}\
  }\textbf {\bibinfo {volume} {25}},\ \bibinfo {pages} {175013} (\bibinfo
  {year} {2008})}\BibitemShut {NoStop}%
\bibitem [{\citenamefont {Zengino{\u{g}}lu}\ \emph {et~al.}(2009)\citenamefont
  {Zengino{\u{g}}lu}, \citenamefont {Nunez},\ and\ \citenamefont
  {Husa}}]{zenginouglu2009gravitational}%
  \BibitemOpen
  \bibfield  {author} {\bibinfo {author} {\bibfnamefont {A.}~\bibnamefont
  {Zengino{\u{g}}lu}}, \bibinfo {author} {\bibfnamefont {D.}~\bibnamefont
  {Nunez}},\ and\ \bibinfo {author} {\bibfnamefont {S.}~\bibnamefont {Husa}},\
  }\bibfield  {title} {\bibinfo {title} {Gravitational perturbations of
  schwarzschild spacetime at null infinity and the hyperboloidal initial value
  problem},\ }\href@noop {} {\bibfield  {journal} {\bibinfo  {journal}
  {Classical and Quantum Gravity}\ }\textbf {\bibinfo {volume} {26}},\ \bibinfo
  {pages} {035009} (\bibinfo {year} {2009})}\BibitemShut {NoStop}%
\bibitem [{\citenamefont {Bizo{\'n}}\ \emph {et~al.}(2010)\citenamefont
  {Bizo{\'n}}, \citenamefont {Rostworowski},\ and\ \citenamefont
  {Zengino{\u{g}}lu}}]{bizon2010saddle}%
  \BibitemOpen
  \bibfield  {author} {\bibinfo {author} {\bibfnamefont {P.}~\bibnamefont
  {Bizo{\'n}}}, \bibinfo {author} {\bibfnamefont {A.}~\bibnamefont
  {Rostworowski}},\ and\ \bibinfo {author} {\bibfnamefont {A.}~\bibnamefont
  {Zengino{\u{g}}lu}},\ }\bibfield  {title} {\bibinfo {title} {Saddle-point
  dynamics of a yang--mills field on the exterior schwarzschild spacetime},\
  }\href@noop {} {\bibfield  {journal} {\bibinfo  {journal} {Classical and
  Quantum Gravity}\ }\textbf {\bibinfo {volume} {27}},\ \bibinfo {pages}
  {175003} (\bibinfo {year} {2010})}\BibitemShut {NoStop}%
\bibitem [{\citenamefont {Zenginoglu}\ and\ \citenamefont
  {Khanna}(2011)}]{Zenginoglu:2011zz}%
  \BibitemOpen
  \bibfield  {author} {\bibinfo {author} {\bibfnamefont {A.}~\bibnamefont
  {Zenginoglu}}\ and\ \bibinfo {author} {\bibfnamefont {G.}~\bibnamefont
  {Khanna}},\ }\bibfield  {title} {\bibinfo {title} {{Null infinity waveforms
  from extreme-mass-ratio inspirals in Kerr spacetime}},\ }\href
  {https://doi.org/10.1103/PhysRevX.1.021017} {\bibfield  {journal} {\bibinfo
  {journal} {Phys. Rev. X}\ }\textbf {\bibinfo {volume} {1}},\ \bibinfo {pages}
  {021017} (\bibinfo {year} {2011})},\ \Eprint
  {https://arxiv.org/abs/1108.1816} {arXiv:1108.1816 [gr-qc]} \BibitemShut
  {NoStop}%
\bibitem [{\citenamefont {R{\'a}cz}\ and\ \citenamefont
  {T{\'o}th}(2011)}]{racz2011numerical}%
  \BibitemOpen
  \bibfield  {author} {\bibinfo {author} {\bibfnamefont {I.}~\bibnamefont
  {R{\'a}cz}}\ and\ \bibinfo {author} {\bibfnamefont {G.~Z.}\ \bibnamefont
  {T{\'o}th}},\ }\bibfield  {title} {\bibinfo {title} {Numerical investigation
  of the late-time kerr tails},\ }\href@noop {} {\bibfield  {journal} {\bibinfo
   {journal} {Classical and Quantum Gravity}\ }\textbf {\bibinfo {volume}
  {28}},\ \bibinfo {pages} {195003} (\bibinfo {year} {2011})}\BibitemShut
  {NoStop}%
\bibitem [{\citenamefont {Zengino{\u{g}}lu}\ and\ \citenamefont
  {Galley}(2012)}]{zenginouglu2012caustic}%
  \BibitemOpen
  \bibfield  {author} {\bibinfo {author} {\bibfnamefont {A.}~\bibnamefont
  {Zengino{\u{g}}lu}}\ and\ \bibinfo {author} {\bibfnamefont {C.~R.}\
  \bibnamefont {Galley}},\ }\bibfield  {title} {\bibinfo {title} {Caustic
  echoes from a schwarzschild black hole},\ }\href@noop {} {\bibfield
  {journal} {\bibinfo  {journal} {Physical Review D}\ }\textbf {\bibinfo
  {volume} {86}},\ \bibinfo {pages} {064030} (\bibinfo {year}
  {2012})}\BibitemShut {NoStop}%
\bibitem [{\citenamefont {Vega}\ \emph {et~al.}(2013)\citenamefont {Vega},
  \citenamefont {Wardell}, \citenamefont {Diener}, \citenamefont {Cupp},\ and\
  \citenamefont {Haas}}]{vega2013scalar}%
  \BibitemOpen
  \bibfield  {author} {\bibinfo {author} {\bibfnamefont {I.}~\bibnamefont
  {Vega}}, \bibinfo {author} {\bibfnamefont {B.}~\bibnamefont {Wardell}},
  \bibinfo {author} {\bibfnamefont {P.}~\bibnamefont {Diener}}, \bibinfo
  {author} {\bibfnamefont {S.}~\bibnamefont {Cupp}},\ and\ \bibinfo {author}
  {\bibfnamefont {R.}~\bibnamefont {Haas}},\ }\bibfield  {title} {\bibinfo
  {title} {Scalar self-force for eccentric orbits around a schwarzschild black
  hole},\ }\href@noop {} {\bibfield  {journal} {\bibinfo  {journal} {Physical
  Review D}\ }\textbf {\bibinfo {volume} {88}},\ \bibinfo {pages} {084021}
  (\bibinfo {year} {2013})}\BibitemShut {NoStop}%
\bibitem [{\citenamefont {Harms}\ \emph {et~al.}(2014)\citenamefont {Harms},
  \citenamefont {Bernuzzi}, \citenamefont {Nagar},\ and\ \citenamefont
  {Zengino{\u{g}}lu}}]{harms2014new}%
  \BibitemOpen
  \bibfield  {author} {\bibinfo {author} {\bibfnamefont {E.}~\bibnamefont
  {Harms}}, \bibinfo {author} {\bibfnamefont {S.}~\bibnamefont {Bernuzzi}},
  \bibinfo {author} {\bibfnamefont {A.}~\bibnamefont {Nagar}},\ and\ \bibinfo
  {author} {\bibfnamefont {A.}~\bibnamefont {Zengino{\u{g}}lu}},\ }\bibfield
  {title} {\bibinfo {title} {A new gravitational wave generation algorithm for
  particle perturbations of the kerr spacetime},\ }\href@noop {} {\bibfield
  {journal} {\bibinfo  {journal} {Classical and Quantum Gravity}\ }\textbf
  {\bibinfo {volume} {31}},\ \bibinfo {pages} {245004} (\bibinfo {year}
  {2014})}\BibitemShut {NoStop}%
\bibitem [{\citenamefont {Thornburg}\ and\ \citenamefont
  {Wardell}(2017)}]{thornburg2017scalar}%
  \BibitemOpen
  \bibfield  {author} {\bibinfo {author} {\bibfnamefont {J.}~\bibnamefont
  {Thornburg}}\ and\ \bibinfo {author} {\bibfnamefont {B.}~\bibnamefont
  {Wardell}},\ }\bibfield  {title} {\bibinfo {title} {Scalar self-force for
  highly eccentric equatorial orbits in kerr spacetime},\ }\href@noop {}
  {\bibfield  {journal} {\bibinfo  {journal} {Physical Review D}\ }\textbf
  {\bibinfo {volume} {95}},\ \bibinfo {pages} {084043} (\bibinfo {year}
  {2017})}\BibitemShut {NoStop}%
\bibitem [{\citenamefont {Zhang}\ \emph {et~al.}(2020)\citenamefont {Zhang},
  \citenamefont {Wang}, \citenamefont {Wang},\ and\ \citenamefont
  {Saavedra}}]{zhang2020object}%
  \BibitemOpen
  \bibfield  {author} {\bibinfo {author} {\bibfnamefont {S.-J.}\ \bibnamefont
  {Zhang}}, \bibinfo {author} {\bibfnamefont {B.}~\bibnamefont {Wang}},
  \bibinfo {author} {\bibfnamefont {A.}~\bibnamefont {Wang}},\ and\ \bibinfo
  {author} {\bibfnamefont {J.~F.}\ \bibnamefont {Saavedra}},\ }\bibfield
  {title} {\bibinfo {title} {Object picture of scalar field perturbation on
  kerr black hole in scalar-einstein-gauss-bonnet theory},\ }\href@noop {}
  {\bibfield  {journal} {\bibinfo  {journal} {Physical Review D}\ }\textbf
  {\bibinfo {volume} {102}},\ \bibinfo {pages} {124056} (\bibinfo {year}
  {2020})}\BibitemShut {NoStop}%
\bibitem [{\citenamefont {Ripley}\ \emph {et~al.}(2021)\citenamefont {Ripley},
  \citenamefont {Loutrel}, \citenamefont {Giorgi},\ and\ \citenamefont
  {Pretorius}}]{ripley2021numerical}%
  \BibitemOpen
  \bibfield  {author} {\bibinfo {author} {\bibfnamefont {J.~L.}\ \bibnamefont
  {Ripley}}, \bibinfo {author} {\bibfnamefont {N.}~\bibnamefont {Loutrel}},
  \bibinfo {author} {\bibfnamefont {E.}~\bibnamefont {Giorgi}},\ and\ \bibinfo
  {author} {\bibfnamefont {F.}~\bibnamefont {Pretorius}},\ }\bibfield  {title}
  {\bibinfo {title} {Numerical computation of second-order vacuum perturbations
  of kerr black holes},\ }\href@noop {} {\bibfield  {journal} {\bibinfo
  {journal} {Physical Review D}\ }\textbf {\bibinfo {volume} {103}},\ \bibinfo
  {pages} {104018} (\bibinfo {year} {2021})}\BibitemShut {NoStop}%
\bibitem [{\citenamefont {Ansorg}\ and\ \citenamefont
  {Panosso~Macedo}(2016)}]{Ansorg:2016ztf}%
  \BibitemOpen
  \bibfield  {author} {\bibinfo {author} {\bibfnamefont {M.}~\bibnamefont
  {Ansorg}}\ and\ \bibinfo {author} {\bibfnamefont {R.}~\bibnamefont
  {Panosso~Macedo}},\ }\bibfield  {title} {\bibinfo {title} {{Spectral
  decomposition of black-hole perturbations on hyperboloidal slices}},\ }\href
  {https://doi.org/10.1103/PhysRevD.93.124016} {\bibfield  {journal} {\bibinfo
  {journal} {Phys. Rev. D}\ }\textbf {\bibinfo {volume} {93}},\ \bibinfo
  {pages} {124016} (\bibinfo {year} {2016})},\ \Eprint
  {https://arxiv.org/abs/1604.02261} {arXiv:1604.02261 [gr-qc]} \BibitemShut
  {NoStop}%
\bibitem [{\citenamefont {Panosso~Macedo}\ \emph {et~al.}(2018)\citenamefont
  {Panosso~Macedo}, \citenamefont {Jaramillo},\ and\ \citenamefont
  {Ansorg}}]{PanossoMacedo:2018hab}%
  \BibitemOpen
  \bibfield  {author} {\bibinfo {author} {\bibfnamefont {R.}~\bibnamefont
  {Panosso~Macedo}}, \bibinfo {author} {\bibfnamefont {J.~L.}\ \bibnamefont
  {Jaramillo}},\ and\ \bibinfo {author} {\bibfnamefont {M.}~\bibnamefont
  {Ansorg}},\ }\bibfield  {title} {\bibinfo {title} {{Hyperboloidal slicing
  approach to quasi-normal mode expansions: the Reissner-Nordstr\"om case}},\
  }\href {https://doi.org/10.1103/PhysRevD.98.124005} {\bibfield  {journal}
  {\bibinfo  {journal} {Phys. Rev. D}\ }\textbf {\bibinfo {volume} {98}},\
  \bibinfo {pages} {124005} (\bibinfo {year} {2018})},\ \Eprint
  {https://arxiv.org/abs/1809.02837} {arXiv:1809.02837 [gr-qc]} \BibitemShut
  {NoStop}%
\bibitem [{\citenamefont {Jaramillo}\ \emph
  {et~al.}(2021{\natexlab{a}})\citenamefont {Jaramillo}, \citenamefont
  {Macedo},\ and\ \citenamefont {Sheikh}}]{jaramillo2021gravitational}%
  \BibitemOpen
  \bibfield  {author} {\bibinfo {author} {\bibfnamefont {J.~L.}\ \bibnamefont
  {Jaramillo}}, \bibinfo {author} {\bibfnamefont {R.~P.}\ \bibnamefont
  {Macedo}},\ and\ \bibinfo {author} {\bibfnamefont {L.~A.}\ \bibnamefont
  {Sheikh}},\ }\bibfield  {title} {\bibinfo {title} {Gravitational wave
  signatures of black hole quasi-normal mode instability},\ }\href@noop {}
  {\bibfield  {journal} {\bibinfo  {journal} {arXiv preprint arXiv:2105.03451}\
  } (\bibinfo {year} {2021}{\natexlab{a}})}\BibitemShut {NoStop}%
\bibitem [{\citenamefont {Destounis}\ \emph {et~al.}(2021)\citenamefont
  {Destounis}, \citenamefont {Macedo}, \citenamefont {Berti}, \citenamefont
  {Cardoso},\ and\ \citenamefont {Jaramillo}}]{destounis2021pseudospectrum}%
  \BibitemOpen
  \bibfield  {author} {\bibinfo {author} {\bibfnamefont {K.}~\bibnamefont
  {Destounis}}, \bibinfo {author} {\bibfnamefont {R.~P.}\ \bibnamefont
  {Macedo}}, \bibinfo {author} {\bibfnamefont {E.}~\bibnamefont {Berti}},
  \bibinfo {author} {\bibfnamefont {V.}~\bibnamefont {Cardoso}},\ and\ \bibinfo
  {author} {\bibfnamefont {J.~L.}\ \bibnamefont {Jaramillo}},\ }\bibfield
  {title} {\bibinfo {title} {Pseudospectrum of reissner-nordstr$\backslash$" om
  black holes: quasinormal mode instability and universality},\ }\href@noop {}
  {\bibfield  {journal} {\bibinfo  {journal} {arXiv preprint arXiv:2107.09673}\
  } (\bibinfo {year} {2021})}\BibitemShut {NoStop}%
\bibitem [{\citenamefont {Jaramillo}\ \emph
  {et~al.}(2021{\natexlab{b}})\citenamefont {Jaramillo}, \citenamefont
  {Macedo},\ and\ \citenamefont {Al~Sheikh}}]{jaramillo2021pseudospectrum}%
  \BibitemOpen
  \bibfield  {author} {\bibinfo {author} {\bibfnamefont {J.~L.}\ \bibnamefont
  {Jaramillo}}, \bibinfo {author} {\bibfnamefont {R.~P.}\ \bibnamefont
  {Macedo}},\ and\ \bibinfo {author} {\bibfnamefont {L.}~\bibnamefont
  {Al~Sheikh}},\ }\bibfield  {title} {\bibinfo {title} {Pseudospectrum and
  black hole quasinormal mode instability},\ }\href@noop {} {\bibfield
  {journal} {\bibinfo  {journal} {Physical Review X}\ }\textbf {\bibinfo
  {volume} {11}},\ \bibinfo {pages} {031003} (\bibinfo {year}
  {2021}{\natexlab{b}})}\BibitemShut {NoStop}%
\bibitem [{\citenamefont {Gasperin}\ and\ \citenamefont
  {Jaramillo}(2021)}]{gasperin2021physical}%
  \BibitemOpen
  \bibfield  {author} {\bibinfo {author} {\bibfnamefont {E.}~\bibnamefont
  {Gasperin}}\ and\ \bibinfo {author} {\bibfnamefont {J.~L.}\ \bibnamefont
  {Jaramillo}},\ }\bibfield  {title} {\bibinfo {title} {Physical scales in
  black hole scattering pseudospectra: the role of the scalar product},\
  }\href@noop {} {\bibfield  {journal} {\bibinfo  {journal} {arXiv preprint
  arXiv:2107.12865}\ } (\bibinfo {year} {2021})}\BibitemShut {NoStop}%
\bibitem [{\citenamefont {Ripley}(2022)}]{Ripley:2022ypi}%
  \BibitemOpen
  \bibfield  {author} {\bibinfo {author} {\bibfnamefont {J.~L.}\ \bibnamefont
  {Ripley}},\ }\bibfield  {title} {\bibinfo {title} {{Computing the quasinormal
  modes and eigenfunctions for the Teukolsky equation using horizon
  penetrating, hyperboloidally compactified coordinates}},\ }\href@noop {} {\
  (\bibinfo {year} {2022})},\ \Eprint {https://arxiv.org/abs/2202.03837}
  {arXiv:2202.03837 [gr-qc]} \BibitemShut {NoStop}%
\bibitem [{\citenamefont {Teukolsky}(1973)}]{Teukolsky:1973ha}%
  \BibitemOpen
  \bibfield  {author} {\bibinfo {author} {\bibfnamefont {S.~A.}\ \bibnamefont
  {Teukolsky}},\ }\bibfield  {title} {\bibinfo {title} {{Perturbations of a
  rotating black hole. 1. Fundamental equations for gravitational
  electromagnetic and neutrino field perturbations}},\ }\href
  {https://doi.org/10.1086/152444} {\bibfield  {journal} {\bibinfo  {journal}
  {Astrophys. J.}\ }\textbf {\bibinfo {volume} {185}},\ \bibinfo {pages} {635}
  (\bibinfo {year} {1973})}\BibitemShut {NoStop}%
\bibitem [{\citenamefont {Quinn}(2000)}]{Quinn:2000wa}%
  \BibitemOpen
  \bibfield  {author} {\bibinfo {author} {\bibfnamefont {T.~C.}\ \bibnamefont
  {Quinn}},\ }\bibfield  {title} {\bibinfo {title} {{Axiomatic approach to
  radiation reaction of scalar point particles in curved space-time}},\ }\href
  {https://doi.org/10.1103/PhysRevD.62.064029} {\bibfield  {journal} {\bibinfo
  {journal} {Phys. Rev. D}\ }\textbf {\bibinfo {volume} {62}},\ \bibinfo
  {pages} {064029} (\bibinfo {year} {2000})},\ \Eprint
  {https://arxiv.org/abs/gr-qc/0005030} {arXiv:gr-qc/0005030} \BibitemShut
  {NoStop}%
\bibitem [{\citenamefont {Detweiler}\ and\ \citenamefont
  {Whiting}(2003)}]{Detweiler:2002mi}%
  \BibitemOpen
  \bibfield  {author} {\bibinfo {author} {\bibfnamefont {S.~L.}\ \bibnamefont
  {Detweiler}}\ and\ \bibinfo {author} {\bibfnamefont {B.~F.}\ \bibnamefont
  {Whiting}},\ }\bibfield  {title} {\bibinfo {title} {{Selfforce via a Green's
  function decomposition}},\ }\href
  {https://doi.org/10.1103/PhysRevD.67.024025} {\bibfield  {journal} {\bibinfo
  {journal} {Phys. Rev. D}\ }\textbf {\bibinfo {volume} {67}},\ \bibinfo
  {pages} {024025} (\bibinfo {year} {2003})},\ \Eprint
  {https://arxiv.org/abs/gr-qc/0202086} {arXiv:gr-qc/0202086} \BibitemShut
  {NoStop}%
\bibitem [{\citenamefont {Barack}\ and\ \citenamefont
  {Ori}(2000)}]{Barack:1999wf}%
  \BibitemOpen
  \bibfield  {author} {\bibinfo {author} {\bibfnamefont {L.}~\bibnamefont
  {Barack}}\ and\ \bibinfo {author} {\bibfnamefont {A.}~\bibnamefont {Ori}},\
  }\bibfield  {title} {\bibinfo {title} {{Mode sum regularization approach for
  the selfforce in black hole space-time}},\ }\href
  {https://doi.org/10.1103/PhysRevD.61.061502} {\bibfield  {journal} {\bibinfo
  {journal} {Phys. Rev. D}\ }\textbf {\bibinfo {volume} {61}},\ \bibinfo
  {pages} {061502} (\bibinfo {year} {2000})},\ \Eprint
  {https://arxiv.org/abs/gr-qc/9912010} {arXiv:gr-qc/9912010} \BibitemShut
  {NoStop}%
\bibitem [{\citenamefont {Vega}\ and\ \citenamefont
  {Detweiler}(2008)}]{Vega:2007mc}%
  \BibitemOpen
  \bibfield  {author} {\bibinfo {author} {\bibfnamefont {I.}~\bibnamefont
  {Vega}}\ and\ \bibinfo {author} {\bibfnamefont {S.~L.}\ \bibnamefont
  {Detweiler}},\ }\bibfield  {title} {\bibinfo {title} {{Regularization of
  fields for self-force problems in curved spacetime: Foundations and a
  time-domain application}},\ }\href
  {https://doi.org/10.1103/PhysRevD.77.084008} {\bibfield  {journal} {\bibinfo
  {journal} {Phys. Rev. D}\ }\textbf {\bibinfo {volume} {77}},\ \bibinfo
  {pages} {084008} (\bibinfo {year} {2008})},\ \Eprint
  {https://arxiv.org/abs/0712.4405} {arXiv:0712.4405 [gr-qc]} \BibitemShut
  {NoStop}%
\bibitem [{\citenamefont {Barack}\ and\ \citenamefont
  {Golbourn}(2007)}]{Barack:2007jh}%
  \BibitemOpen
  \bibfield  {author} {\bibinfo {author} {\bibfnamefont {L.}~\bibnamefont
  {Barack}}\ and\ \bibinfo {author} {\bibfnamefont {D.~A.}\ \bibnamefont
  {Golbourn}},\ }\bibfield  {title} {\bibinfo {title} {{Scalar-field
  perturbations from a particle orbiting a black hole using numerical evolution
  in 2+1 dimensions}},\ }\href {https://doi.org/10.1103/PhysRevD.76.044020}
  {\bibfield  {journal} {\bibinfo  {journal} {Phys. Rev. D}\ }\textbf {\bibinfo
  {volume} {76}},\ \bibinfo {pages} {044020} (\bibinfo {year} {2007})},\
  \Eprint {https://arxiv.org/abs/0705.3620} {arXiv:0705.3620 [gr-qc]}
  \BibitemShut {NoStop}%
\bibitem [{\citenamefont {Bardeen}\ and\ \citenamefont
  {Press}(1973)}]{Bardeen:1973xb}%
  \BibitemOpen
  \bibfield  {author} {\bibinfo {author} {\bibfnamefont {J.~M.}\ \bibnamefont
  {Bardeen}}\ and\ \bibinfo {author} {\bibfnamefont {W.~H.}\ \bibnamefont
  {Press}},\ }\bibfield  {title} {\bibinfo {title} {{Radiation fields in the
  schwarzschild background}},\ }\href {https://doi.org/10.1063/1.1666175}
  {\bibfield  {journal} {\bibinfo  {journal} {J. Math. Phys.}\ }\textbf
  {\bibinfo {volume} {14}},\ \bibinfo {pages} {7} (\bibinfo {year}
  {1973})}\BibitemShut {NoStop}%
\bibitem [{\citenamefont {Martel}\ and\ \citenamefont
  {Poisson}(2005)}]{Martel:2005ir}%
  \BibitemOpen
  \bibfield  {author} {\bibinfo {author} {\bibfnamefont {K.}~\bibnamefont
  {Martel}}\ and\ \bibinfo {author} {\bibfnamefont {E.}~\bibnamefont
  {Poisson}},\ }\bibfield  {title} {\bibinfo {title} {{Gravitational
  perturbations of the Schwarzschild spacetime: A Practical covariant and
  gauge-invariant formalism}},\ }\href
  {https://doi.org/10.1103/PhysRevD.71.104003} {\bibfield  {journal} {\bibinfo
  {journal} {Phys. Rev. D}\ }\textbf {\bibinfo {volume} {71}},\ \bibinfo
  {pages} {104003} (\bibinfo {year} {2005})},\ \Eprint
  {https://arxiv.org/abs/gr-qc/0502028} {arXiv:gr-qc/0502028} \BibitemShut
  {NoStop}%
\bibitem [{\citenamefont {Barack}\ \emph {et~al.}(2008)\citenamefont {Barack},
  \citenamefont {Ori},\ and\ \citenamefont {Sago}}]{Barack:2008ms}%
  \BibitemOpen
  \bibfield  {author} {\bibinfo {author} {\bibfnamefont {L.}~\bibnamefont
  {Barack}}, \bibinfo {author} {\bibfnamefont {A.}~\bibnamefont {Ori}},\ and\
  \bibinfo {author} {\bibfnamefont {N.}~\bibnamefont {Sago}},\ }\bibfield
  {title} {\bibinfo {title} {{Frequency-domain calculation of the self force:
  The High-frequency problem and its resolution}},\ }\href
  {https://doi.org/10.1103/PhysRevD.78.084021} {\bibfield  {journal} {\bibinfo
  {journal} {Phys. Rev. D}\ }\textbf {\bibinfo {volume} {78}},\ \bibinfo
  {pages} {084021} (\bibinfo {year} {2008})},\ \Eprint
  {https://arxiv.org/abs/0808.2315} {arXiv:0808.2315 [gr-qc]} \BibitemShut
  {NoStop}%
\bibitem [{\citenamefont {Hopper}(2018)}]{Hopper:2017iyq}%
  \BibitemOpen
  \bibfield  {author} {\bibinfo {author} {\bibfnamefont {S.}~\bibnamefont
  {Hopper}},\ }\bibfield  {title} {\bibinfo {title} {{Unbound motion on a
  Schwarzschild background: Practical approaches to frequency domain
  computations}},\ }\href {https://doi.org/10.1103/PhysRevD.97.064007}
  {\bibfield  {journal} {\bibinfo  {journal} {Phys. Rev. D}\ }\textbf {\bibinfo
  {volume} {97}},\ \bibinfo {pages} {064007} (\bibinfo {year} {2018})},\
  \Eprint {https://arxiv.org/abs/1706.05455} {arXiv:1706.05455 [gr-qc]}
  \BibitemShut {NoStop}%
\bibitem [{\citenamefont {Grosch}\ and\ \citenamefont
  {Orszag}(1977)}]{GroschOrszag77}%
  \BibitemOpen
  \bibfield  {author} {\bibinfo {author} {\bibfnamefont {C.~E.}\ \bibnamefont
  {Grosch}}\ and\ \bibinfo {author} {\bibfnamefont {S.~A.}\ \bibnamefont
  {Orszag}},\ }\bibfield  {title} {\bibinfo {title} {Numerical solution of
  problems in unbounded regions: coordinate transforms},\ }\href@noop {}
  {\bibfield  {journal} {\bibinfo  {journal} {Journal of Computational
  Physics}\ }\textbf {\bibinfo {volume} {25}},\ \bibinfo {pages} {273}
  (\bibinfo {year} {1977})}\BibitemShut {NoStop}%
\bibitem [{\citenamefont
  {Zengino{\u{g}}lu}(2011)}]{zenginouglu2011hyperboloidal}%
  \BibitemOpen
  \bibfield  {author} {\bibinfo {author} {\bibfnamefont {A.}~\bibnamefont
  {Zengino{\u{g}}lu}},\ }\bibfield  {title} {\bibinfo {title} {Hyperboloidal
  layers for hyperbolic equations on unbounded domains},\ }\href@noop {}
  {\bibfield  {journal} {\bibinfo  {journal} {Journal of Computational
  Physics}\ }\textbf {\bibinfo {volume} {230}},\ \bibinfo {pages} {2286}
  (\bibinfo {year} {2011})}\BibitemShut {NoStop}%
\bibitem [{\citenamefont {Zengino{\u{g}}lu}(2021)}]{zenginouglu2021null}%
  \BibitemOpen
  \bibfield  {author} {\bibinfo {author} {\bibfnamefont {A.}~\bibnamefont
  {Zengino{\u{g}}lu}},\ }\bibfield  {title} {\bibinfo {title} {A null infinity
  layer for wave scattering},\ }\href@noop {} {\bibfield  {journal} {\bibinfo
  {journal} {arXiv preprint arXiv:2111.14217}\ } (\bibinfo {year}
  {2021})}\BibitemShut {NoStop}%
\bibitem [{\citenamefont {Hopper}\ and\ \citenamefont
  {Evans}(2013)}]{Hopper:2012ty}%
  \BibitemOpen
  \bibfield  {author} {\bibinfo {author} {\bibfnamefont {S.}~\bibnamefont
  {Hopper}}\ and\ \bibinfo {author} {\bibfnamefont {C.~R.}\ \bibnamefont
  {Evans}},\ }\bibfield  {title} {\bibinfo {title} {{Metric perturbations from
  eccentric orbits on a Schwarzschild black hole: I. Odd-parity Regge-Wheeler
  to Lorenz gauge transformation and two new methods to circumvent the Gibbs
  phenomenon}},\ }\href {https://doi.org/10.1103/PhysRevD.87.064008} {\bibfield
   {journal} {\bibinfo  {journal} {Phys. Rev. D}\ }\textbf {\bibinfo {volume}
  {87}},\ \bibinfo {pages} {064008} (\bibinfo {year} {2013})},\ \Eprint
  {https://arxiv.org/abs/1210.7969} {arXiv:1210.7969 [gr-qc]} \BibitemShut
  {NoStop}%
\bibitem [{\citenamefont {Friedrich}(1983)}]{friedrich1983cauchy}%
  \BibitemOpen
  \bibfield  {author} {\bibinfo {author} {\bibfnamefont {H.}~\bibnamefont
  {Friedrich}},\ }\bibfield  {title} {\bibinfo {title} {Cauchy problems for the
  conformal vacuum field equations in general relativity},\ }\href@noop {}
  {\bibfield  {journal} {\bibinfo  {journal} {Communications in Mathematical
  Physics}\ }\textbf {\bibinfo {volume} {91}},\ \bibinfo {pages} {445}
  (\bibinfo {year} {1983})}\BibitemShut {NoStop}%
\bibitem [{\citenamefont {Frauendiener}(2004)}]{frauendiener2004conformal}%
  \BibitemOpen
  \bibfield  {author} {\bibinfo {author} {\bibfnamefont {J.}~\bibnamefont
  {Frauendiener}},\ }\bibfield  {title} {\bibinfo {title} {Conformal
  infinity},\ }\href@noop {} {\bibfield  {journal} {\bibinfo  {journal} {Living
  Reviews in Relativity}\ }\textbf {\bibinfo {volume} {7}},\ \bibinfo {pages}
  {1} (\bibinfo {year} {2004})}\BibitemShut {NoStop}%
\bibitem [{\citenamefont {Cruz-Osorio}\ \emph {et~al.}(2010)\citenamefont
  {Cruz-Osorio}, \citenamefont {Gonz{\'a}lez-Ju{\'a}rez}, \citenamefont
  {Guzm{\'a}n},\ and\ \citenamefont {Lora-Clavijo}}]{cruz2010numerical}%
  \BibitemOpen
  \bibfield  {author} {\bibinfo {author} {\bibfnamefont {A.}~\bibnamefont
  {Cruz-Osorio}}, \bibinfo {author} {\bibfnamefont {A.}~\bibnamefont
  {Gonz{\'a}lez-Ju{\'a}rez}}, \bibinfo {author} {\bibfnamefont
  {F.}~\bibnamefont {Guzm{\'a}n}},\ and\ \bibinfo {author} {\bibfnamefont
  {F.}~\bibnamefont {Lora-Clavijo}},\ }\bibfield  {title} {\bibinfo {title}
  {Numerical solution of the wave equation on particular space-times using cmc
  slices and scri-fixing conformal compactification},\ }\href@noop {}
  {\bibfield  {journal} {\bibinfo  {journal} {Revista mexicana de f{\'\i}sica}\
  }\textbf {\bibinfo {volume} {56}},\ \bibinfo {pages} {456} (\bibinfo {year}
  {2010})}\BibitemShut {NoStop}%
\bibitem [{\citenamefont {Boyd}(2001)}]{Boyd}%
  \BibitemOpen
  \bibfield  {author} {\bibinfo {author} {\bibfnamefont {J.~P.}\ \bibnamefont
  {Boyd}},\ }\href {https://www.springer.com/gp/book/9783540514879} {\emph
  {\bibinfo {title} {Chebyshev \& Fourier Spectral Methods}}},\ Lecture Notes
  in Engineering\ (\bibinfo  {publisher} {Dover Publications; Second Edition,
  Revised},\ \bibinfo {year} {2001})\BibitemShut {NoStop}%
\bibitem [{\citenamefont {Penrose}(1963)}]{Penrose63}%
  \BibitemOpen
  \bibfield  {author} {\bibinfo {author} {\bibfnamefont {R.}~\bibnamefont
  {Penrose}},\ }\bibfield  {title} {\bibinfo {title} {Asymptotic properties of
  fields and space-times},\ }\href {https://doi.org/10.1103/PhysRevLett.10.66}
  {\bibfield  {journal} {\bibinfo  {journal} {Physical Review Letters}\
  }\textbf {\bibinfo {volume} {10}},\ \bibinfo {pages} {66} (\bibinfo {year}
  {1963})}\BibitemShut {NoStop}%
\bibitem [{\citenamefont {Poisson}(2004)}]{Poisson:2004}%
  \BibitemOpen
  \bibfield  {author} {\bibinfo {author} {\bibfnamefont {E.}~\bibnamefont
  {Poisson}},\ }\href {https://doi.org/10.1017/CBO9780511606601} {\emph
  {\bibinfo {title} {A Relativist's Toolkit: The Mathematics of Black-Hole
  Mechanics}}}\ (\bibinfo  {publisher} {Cambridge University Press},\ \bibinfo
  {year} {2004})\BibitemShut {NoStop}%
\bibitem [{\citenamefont {Heffernan}\ \emph {et~al.}(2012)\citenamefont
  {Heffernan}, \citenamefont {Ottewill},\ and\ \citenamefont
  {Wardell}}]{Heffernan:2012su}%
  \BibitemOpen
  \bibfield  {author} {\bibinfo {author} {\bibfnamefont {A.}~\bibnamefont
  {Heffernan}}, \bibinfo {author} {\bibfnamefont {A.}~\bibnamefont
  {Ottewill}},\ and\ \bibinfo {author} {\bibfnamefont {B.}~\bibnamefont
  {Wardell}},\ }\bibfield  {title} {\bibinfo {title} {{High-order expansions of
  the Detweiler-Whiting singular field in Schwarzschild spacetime}},\ }\href
  {https://doi.org/10.1103/PhysRevD.86.104023} {\bibfield  {journal} {\bibinfo
  {journal} {Phys. Rev. D}\ }\textbf {\bibinfo {volume} {86}},\ \bibinfo
  {pages} {104023} (\bibinfo {year} {2012})},\ \Eprint
  {https://arxiv.org/abs/1204.0794} {arXiv:1204.0794 [gr-qc]} \BibitemShut
  {NoStop}%
\bibitem [{\citenamefont {Ansorg}\ \emph {et~al.}(2003)\citenamefont {Ansorg},
  \citenamefont {Kleinwachter},\ and\ \citenamefont {Meinel}}]{Ansorg:2003br}%
  \BibitemOpen
  \bibfield  {author} {\bibinfo {author} {\bibfnamefont {M.}~\bibnamefont
  {Ansorg}}, \bibinfo {author} {\bibfnamefont {A.}~\bibnamefont
  {Kleinwachter}},\ and\ \bibinfo {author} {\bibfnamefont {R.}~\bibnamefont
  {Meinel}},\ }\bibfield  {title} {\bibinfo {title} {{Highly accurate
  calculation of rotating neutron stars: detailed description of the numerical
  methods}},\ }\href {https://doi.org/10.1051/0004-6361:20030618} {\bibfield
  {journal} {\bibinfo  {journal} {Astron. Astrophys.}\ }\textbf {\bibinfo
  {volume} {405}},\ \bibinfo {pages} {711} (\bibinfo {year} {2003})},\ \Eprint
  {https://arxiv.org/abs/astro-ph/0301173} {arXiv:astro-ph/0301173}
  \BibitemShut {NoStop}%
\bibitem [{\citenamefont {Ansorg}(2007)}]{Ansorg:2006gd}%
  \BibitemOpen
  \bibfield  {author} {\bibinfo {author} {\bibfnamefont {M.}~\bibnamefont
  {Ansorg}},\ }\bibfield  {title} {\bibinfo {title} {{Multi-Domain Spectral
  Method for Initial Data of Arbitrary Binaries in General Relativity}},\
  }\href {https://doi.org/10.1088/0264-9381/24/12/S01} {\bibfield  {journal}
  {\bibinfo  {journal} {Class. Quant. Grav.}\ }\textbf {\bibinfo {volume}
  {24}},\ \bibinfo {pages} {S1} (\bibinfo {year} {2007})},\ \Eprint
  {https://arxiv.org/abs/gr-qc/0612081} {arXiv:gr-qc/0612081} \BibitemShut
  {NoStop}%
\bibitem [{\citenamefont {Meinel}\ \emph {et~al.}(2008)\citenamefont {Meinel},
  \citenamefont {Ansorg}, \citenamefont {Kleinwachter}, \citenamefont
  {Neugebauer},\ and\ \citenamefont {Petroff}}]{Meinel:2008kpy}%
  \BibitemOpen
  \bibfield  {author} {\bibinfo {author} {\bibfnamefont {R.}~\bibnamefont
  {Meinel}}, \bibinfo {author} {\bibfnamefont {M.}~\bibnamefont {Ansorg}},
  \bibinfo {author} {\bibfnamefont {A.}~\bibnamefont {Kleinwachter}}, \bibinfo
  {author} {\bibfnamefont {G.}~\bibnamefont {Neugebauer}},\ and\ \bibinfo
  {author} {\bibfnamefont {D.}~\bibnamefont {Petroff}},\ }\href
  {https://doi.org/10.1017/CBO9780511535154} {\emph {\bibinfo {title}
  {{Relativistic Figures of Equilibrium}}}}\ (\bibinfo  {publisher} {Cambridge
  University Press},\ \bibinfo {address} {Cambridge, UK},\ \bibinfo {year}
  {2008})\BibitemShut {NoStop}%
\bibitem [{\citenamefont {Ansorg}(2013)}]{Ansorg13}%
  \BibitemOpen
  \bibfield  {author} {\bibinfo {author} {\bibfnamefont {M.}~\bibnamefont
  {Ansorg}},\ }\href@noop {} {\bibinfo {title} {Lecture notes in spectral
  methods for theoretical physics (in german)}} (\bibinfo {year}
  {2013})\BibitemShut {NoStop}%
\bibitem [{\citenamefont {Canuto}\ \emph {et~al.}(2007)\citenamefont {Canuto},
  \citenamefont {Hussaini}, \citenamefont {Quarteroni},\ and\ \citenamefont
  {Zang}}]{canuto2007spectral}%
  \BibitemOpen
  \bibfield  {author} {\bibinfo {author} {\bibfnamefont {C.}~\bibnamefont
  {Canuto}}, \bibinfo {author} {\bibfnamefont {M.}~\bibnamefont {Hussaini}},
  \bibinfo {author} {\bibfnamefont {A.}~\bibnamefont {Quarteroni}},\ and\
  \bibinfo {author} {\bibfnamefont {T.}~\bibnamefont {Zang}},\ }\href
  {https://books.google.es/books?id=DFJB0kiq0CQC} {\emph {\bibinfo {title}
  {Spectral Methods: Fundamentals in Single Domains}}},\ Scientific
  Computation\ (\bibinfo  {publisher} {Springer Berlin Heidelberg},\ \bibinfo
  {year} {2007})\BibitemShut {NoStop}%
\bibitem [{\citenamefont {Grandcl\'ement}\ and\ \citenamefont
  {Novak}(2007)}]{GraNov07}%
  \BibitemOpen
  \bibfield  {author} {\bibinfo {author} {\bibfnamefont {P.}~\bibnamefont
  {Grandcl\'ement}}\ and\ \bibinfo {author} {\bibfnamefont {J.}~\bibnamefont
  {Novak}},\ }\bibfield  {title} {\bibinfo {title} {Spectral methods for
  numerical relativity},\ }\href@noop {} {\  (\bibinfo {year}
  {2007})}\BibitemShut {NoStop}%
\bibitem [{\citenamefont {Trefethen}(2000)}]{trefethen2000spectral}%
  \BibitemOpen
  \bibfield  {author} {\bibinfo {author} {\bibfnamefont {L.}~\bibnamefont
  {Trefethen}},\ }\href {https://books.google.es/books?id=9Zu4YqPQKocC} {\emph
  {\bibinfo {title} {Spectral Methods in MATLAB}}},\ Software, Environments,
  and Tools\ (\bibinfo  {publisher} {Society for Industrial and Applied
  Mathematics (SIAM, 3600 Market Street, Floor 6, Philadelphia, PA 19104)},\
  \bibinfo {year} {2000})\BibitemShut {NoStop}%
\bibitem [{BHP()}]{BHPToolkit}%
  \BibitemOpen
  \href@noop {} {\bibinfo {title} {{Black Hole Perturbation Toolkit}}},\
  \bibinfo {howpublished}
  {(\href{http://bhptoolkit.org/}{bhptoolkit.org})}\BibitemShut {NoStop}%
\bibitem [{\citenamefont {Hikida}\ \emph {et~al.}(2005)\citenamefont {Hikida},
  \citenamefont {Nakano},\ and\ \citenamefont {Sasaki}}]{Hikida:2004hs}%
  \BibitemOpen
  \bibfield  {author} {\bibinfo {author} {\bibfnamefont {W.}~\bibnamefont
  {Hikida}}, \bibinfo {author} {\bibfnamefont {H.}~\bibnamefont {Nakano}},\
  and\ \bibinfo {author} {\bibfnamefont {M.}~\bibnamefont {Sasaki}},\
  }\bibfield  {title} {\bibinfo {title} {{Self-force regularization in the
  Schwarzschild spacetime}},\ }\href
  {https://doi.org/10.1088/0264-9381/22/15/009} {\bibfield  {journal} {\bibinfo
   {journal} {Class. Quant. Grav.}\ }\textbf {\bibinfo {volume} {22}},\
  \bibinfo {pages} {S753} (\bibinfo {year} {2005})},\ \Eprint
  {https://arxiv.org/abs/gr-qc/0411150} {arXiv:gr-qc/0411150} \BibitemShut
  {NoStop}%
\bibitem [{\citenamefont {Pound}\ \emph {et~al.}(2020)\citenamefont {Pound},
  \citenamefont {Wardell}, \citenamefont {Warburton},\ and\ \citenamefont
  {Miller}}]{Pound:2019lzj}%
  \BibitemOpen
  \bibfield  {author} {\bibinfo {author} {\bibfnamefont {A.}~\bibnamefont
  {Pound}}, \bibinfo {author} {\bibfnamefont {B.}~\bibnamefont {Wardell}},
  \bibinfo {author} {\bibfnamefont {N.}~\bibnamefont {Warburton}},\ and\
  \bibinfo {author} {\bibfnamefont {J.}~\bibnamefont {Miller}},\ }\bibfield
  {title} {\bibinfo {title} {{Second-Order Self-Force Calculation of
  Gravitational Binding Energy in Compact Binaries}},\ }\href
  {https://doi.org/10.1103/PhysRevLett.124.021101} {\bibfield  {journal}
  {\bibinfo  {journal} {Phys. Rev. Lett.}\ }\textbf {\bibinfo {volume} {124}},\
  \bibinfo {pages} {021101} (\bibinfo {year} {2020})},\ \Eprint
  {https://arxiv.org/abs/1908.07419} {arXiv:1908.07419 [gr-qc]} \BibitemShut
  {NoStop}%
\bibitem [{\citenamefont {Warburton}\ \emph {et~al.}(2021)\citenamefont
  {Warburton}, \citenamefont {Pound}, \citenamefont {Wardell}, \citenamefont
  {Miller},\ and\ \citenamefont {Durkan}}]{Warburton:2021kwk}%
  \BibitemOpen
  \bibfield  {author} {\bibinfo {author} {\bibfnamefont {N.}~\bibnamefont
  {Warburton}}, \bibinfo {author} {\bibfnamefont {A.}~\bibnamefont {Pound}},
  \bibinfo {author} {\bibfnamefont {B.}~\bibnamefont {Wardell}}, \bibinfo
  {author} {\bibfnamefont {J.}~\bibnamefont {Miller}},\ and\ \bibinfo {author}
  {\bibfnamefont {L.}~\bibnamefont {Durkan}},\ }\bibfield  {title} {\bibinfo
  {title} {{Gravitational-wave energy flux for compact binaries through second
  order in the mass ratio}},\ }\href@noop {} {\  (\bibinfo {year} {2021})},\
  \Eprint {https://arxiv.org/abs/2107.01298} {arXiv:2107.01298 [gr-qc]}
  \BibitemShut {NoStop}%
\bibitem [{\citenamefont {Schinkel}\ \emph
  {et~al.}(2014{\natexlab{a}})\citenamefont {Schinkel}, \citenamefont
  {Ansorg},\ and\ \citenamefont {Panosso~Macedo}}]{Schinkel:2013zm}%
  \BibitemOpen
  \bibfield  {author} {\bibinfo {author} {\bibfnamefont {D.}~\bibnamefont
  {Schinkel}}, \bibinfo {author} {\bibfnamefont {M.}~\bibnamefont {Ansorg}},\
  and\ \bibinfo {author} {\bibfnamefont {R.}~\bibnamefont {Panosso~Macedo}},\
  }\bibfield  {title} {\bibinfo {title} {{Initial data for perturbed Kerr black
  holes on hyperboloidal slices}},\ }\href
  {https://doi.org/10.1088/0264-9381/31/16/165001} {\bibfield  {journal}
  {\bibinfo  {journal} {Class. Quant. Grav.}\ }\textbf {\bibinfo {volume}
  {31}},\ \bibinfo {pages} {165001} (\bibinfo {year} {2014}{\natexlab{a}})},\
  \Eprint {https://arxiv.org/abs/1301.6984} {arXiv:1301.6984 [gr-qc]}
  \BibitemShut {NoStop}%
\bibitem [{\citenamefont {Schinkel}\ \emph
  {et~al.}(2014{\natexlab{b}})\citenamefont {Schinkel}, \citenamefont
  {Panosso~Macedo},\ and\ \citenamefont {Ansorg}}]{Schinkel:2013tka}%
  \BibitemOpen
  \bibfield  {author} {\bibinfo {author} {\bibfnamefont {D.}~\bibnamefont
  {Schinkel}}, \bibinfo {author} {\bibfnamefont {R.}~\bibnamefont
  {Panosso~Macedo}},\ and\ \bibinfo {author} {\bibfnamefont {M.}~\bibnamefont
  {Ansorg}},\ }\bibfield  {title} {\bibinfo {title} {{Axisymmetric constant
  mean curvature slices in the Kerr space-time}},\ }\href
  {https://doi.org/10.1088/0264-9381/31/7/075017} {\bibfield  {journal}
  {\bibinfo  {journal} {Class. Quant. Grav.}\ }\textbf {\bibinfo {volume}
  {31}},\ \bibinfo {pages} {075017} (\bibinfo {year} {2014}{\natexlab{b}})},\
  \Eprint {https://arxiv.org/abs/1310.4699} {arXiv:1310.4699 [gr-qc]}
  \BibitemShut {NoStop}%
\bibitem [{\citenamefont {Panosso~Macedo}\ and\ \citenamefont
  {Ansorg}(2014)}]{PanossoMacedo:2014dnr}%
  \BibitemOpen
  \bibfield  {author} {\bibinfo {author} {\bibfnamefont {R.}~\bibnamefont
  {Panosso~Macedo}}\ and\ \bibinfo {author} {\bibfnamefont {M.}~\bibnamefont
  {Ansorg}},\ }\bibfield  {title} {\bibinfo {title} {{Axisymmetric fully
  spectral code for hyperbolic equations}},\ }\href
  {https://doi.org/10.1016/j.jcp.2014.07.040} {\bibfield  {journal} {\bibinfo
  {journal} {J. Comput. Phys.}\ }\textbf {\bibinfo {volume} {276}},\ \bibinfo
  {pages} {357} (\bibinfo {year} {2014})},\ \Eprint
  {https://arxiv.org/abs/1402.7343} {arXiv:1402.7343 [physics.comp-ph]}
  \BibitemShut {NoStop}%
\bibitem [{\citenamefont {Canizares}\ \emph {et~al.}(2010)\citenamefont
  {Canizares}, \citenamefont {Sopuerta},\ and\ \citenamefont
  {Jaramillo}}]{Canizares:2010yx}%
  \BibitemOpen
  \bibfield  {author} {\bibinfo {author} {\bibfnamefont {P.}~\bibnamefont
  {Canizares}}, \bibinfo {author} {\bibfnamefont {C.~F.}\ \bibnamefont
  {Sopuerta}},\ and\ \bibinfo {author} {\bibfnamefont {J.~L.}\ \bibnamefont
  {Jaramillo}},\ }\bibfield  {title} {\bibinfo {title} {{Pseudospectral
  Collocation Methods for the Computation of the Self-Force on a Charged
  Particle: Generic Orbits around a Schwarzschild Black Hole}},\ }\href
  {https://doi.org/10.1103/PhysRevD.82.044023} {\bibfield  {journal} {\bibinfo
  {journal} {Phys. Rev. D}\ }\textbf {\bibinfo {volume} {82}},\ \bibinfo
  {pages} {044023} (\bibinfo {year} {2010})},\ \Eprint
  {https://arxiv.org/abs/1006.3201} {arXiv:1006.3201 [gr-qc]} \BibitemShut
  {NoStop}%
\bibitem [{\citenamefont {Pound}(2015)}]{Pound:2015wva}%
  \BibitemOpen
  \bibfield  {author} {\bibinfo {author} {\bibfnamefont {A.}~\bibnamefont
  {Pound}},\ }\bibfield  {title} {\bibinfo {title} {{Second-order perturbation
  theory: problems on large scales}},\ }\href
  {https://doi.org/10.1103/PhysRevD.92.104047} {\bibfield  {journal} {\bibinfo
  {journal} {Phys. Rev. D}\ }\textbf {\bibinfo {volume} {92}},\ \bibinfo
  {pages} {104047} (\bibinfo {year} {2015})},\ \Eprint
  {https://arxiv.org/abs/1510.05172} {arXiv:1510.05172 [gr-qc]} \BibitemShut
  {NoStop}%
\bibitem [{\citenamefont {Blanchet}\ \emph
  {et~al.}(2010{\natexlab{a}})\citenamefont {Blanchet}, \citenamefont
  {Detweiler}, \citenamefont {Le~Tiec},\ and\ \citenamefont
  {Whiting}}]{Blanchet:2009sd}%
  \BibitemOpen
  \bibfield  {author} {\bibinfo {author} {\bibfnamefont {L.}~\bibnamefont
  {Blanchet}}, \bibinfo {author} {\bibfnamefont {S.~L.}\ \bibnamefont
  {Detweiler}}, \bibinfo {author} {\bibfnamefont {A.}~\bibnamefont {Le~Tiec}},\
  and\ \bibinfo {author} {\bibfnamefont {B.~F.}\ \bibnamefont {Whiting}},\
  }\bibfield  {title} {\bibinfo {title} {{Post-Newtonian and Numerical
  Calculations of the Gravitational Self-Force for Circular Orbits in the
  Schwarzschild Geometry}},\ }\href
  {https://doi.org/10.1103/PhysRevD.81.064004} {\bibfield  {journal} {\bibinfo
  {journal} {Phys. Rev. D}\ }\textbf {\bibinfo {volume} {81}},\ \bibinfo
  {pages} {064004} (\bibinfo {year} {2010}{\natexlab{a}})},\ \Eprint
  {https://arxiv.org/abs/0910.0207} {arXiv:0910.0207 [gr-qc]} \BibitemShut
  {NoStop}%
\bibitem [{\citenamefont {Blanchet}\ \emph
  {et~al.}(2010{\natexlab{b}})\citenamefont {Blanchet}, \citenamefont
  {Detweiler}, \citenamefont {Le~Tiec},\ and\ \citenamefont
  {Whiting}}]{Blanchet:2010zd}%
  \BibitemOpen
  \bibfield  {author} {\bibinfo {author} {\bibfnamefont {L.}~\bibnamefont
  {Blanchet}}, \bibinfo {author} {\bibfnamefont {S.~L.}\ \bibnamefont
  {Detweiler}}, \bibinfo {author} {\bibfnamefont {A.}~\bibnamefont {Le~Tiec}},\
  and\ \bibinfo {author} {\bibfnamefont {B.~F.}\ \bibnamefont {Whiting}},\
  }\bibfield  {title} {\bibinfo {title} {{High-Order Post-Newtonian Fit of the
  Gravitational Self-Force for Circular Orbits in the Schwarzschild
  Geometry}},\ }\href {https://doi.org/10.1103/PhysRevD.81.084033} {\bibfield
  {journal} {\bibinfo  {journal} {Phys. Rev. D}\ }\textbf {\bibinfo {volume}
  {81}},\ \bibinfo {pages} {084033} (\bibinfo {year} {2010}{\natexlab{b}})},\
  \Eprint {https://arxiv.org/abs/1002.0726} {arXiv:1002.0726 [gr-qc]}
  \BibitemShut {NoStop}%
\bibitem [{\citenamefont {Dolan}\ \emph {et~al.}(2014)\citenamefont {Dolan},
  \citenamefont {Warburton}, \citenamefont {Harte}, \citenamefont {Le~Tiec},
  \citenamefont {Wardell},\ and\ \citenamefont {Barack}}]{Dolan:2013roa}%
  \BibitemOpen
  \bibfield  {author} {\bibinfo {author} {\bibfnamefont {S.~R.}\ \bibnamefont
  {Dolan}}, \bibinfo {author} {\bibfnamefont {N.}~\bibnamefont {Warburton}},
  \bibinfo {author} {\bibfnamefont {A.~I.}\ \bibnamefont {Harte}}, \bibinfo
  {author} {\bibfnamefont {A.}~\bibnamefont {Le~Tiec}}, \bibinfo {author}
  {\bibfnamefont {B.}~\bibnamefont {Wardell}},\ and\ \bibinfo {author}
  {\bibfnamefont {L.}~\bibnamefont {Barack}},\ }\bibfield  {title} {\bibinfo
  {title} {{Gravitational self-torque and spin precession in compact
  binaries}},\ }\href {https://doi.org/10.1103/PhysRevD.89.064011} {\bibfield
  {journal} {\bibinfo  {journal} {Phys. Rev. D}\ }\textbf {\bibinfo {volume}
  {89}},\ \bibinfo {pages} {064011} (\bibinfo {year} {2014})},\ \Eprint
  {https://arxiv.org/abs/1312.0775} {arXiv:1312.0775 [gr-qc]} \BibitemShut
  {NoStop}%
\bibitem [{\citenamefont {Dolan}\ \emph {et~al.}(2015)\citenamefont {Dolan},
  \citenamefont {Nolan}, \citenamefont {Ottewill}, \citenamefont {Warburton},\
  and\ \citenamefont {Wardell}}]{Dolan:2014pja}%
  \BibitemOpen
  \bibfield  {author} {\bibinfo {author} {\bibfnamefont {S.~R.}\ \bibnamefont
  {Dolan}}, \bibinfo {author} {\bibfnamefont {P.}~\bibnamefont {Nolan}},
  \bibinfo {author} {\bibfnamefont {A.~C.}\ \bibnamefont {Ottewill}}, \bibinfo
  {author} {\bibfnamefont {N.}~\bibnamefont {Warburton}},\ and\ \bibinfo
  {author} {\bibfnamefont {B.}~\bibnamefont {Wardell}},\ }\bibfield  {title}
  {\bibinfo {title} {{Tidal invariants for compact binaries on quasicircular
  orbits}},\ }\href {https://doi.org/10.1103/PhysRevD.91.023009} {\bibfield
  {journal} {\bibinfo  {journal} {Phys. Rev. D}\ }\textbf {\bibinfo {volume}
  {91}},\ \bibinfo {pages} {023009} (\bibinfo {year} {2015})},\ \Eprint
  {https://arxiv.org/abs/1406.4890} {arXiv:1406.4890 [gr-qc]} \BibitemShut
  {NoStop}%
\bibitem [{\citenamefont {Mino}\ \emph {et~al.}(1997)\citenamefont {Mino},
  \citenamefont {Sasaki},\ and\ \citenamefont {Tanaka}}]{Mino:1996nk}%
  \BibitemOpen
  \bibfield  {author} {\bibinfo {author} {\bibfnamefont {Y.}~\bibnamefont
  {Mino}}, \bibinfo {author} {\bibfnamefont {M.}~\bibnamefont {Sasaki}},\ and\
  \bibinfo {author} {\bibfnamefont {T.}~\bibnamefont {Tanaka}},\ }\bibfield
  {title} {\bibinfo {title} {{Gravitational radiation reaction to a particle
  motion}},\ }\href {https://doi.org/10.1103/PhysRevD.55.3457} {\bibfield
  {journal} {\bibinfo  {journal} {Phys. Rev. D}\ }\textbf {\bibinfo {volume}
  {55}},\ \bibinfo {pages} {3457} (\bibinfo {year} {1997})},\ \Eprint
  {https://arxiv.org/abs/gr-qc/9606018} {arXiv:gr-qc/9606018} \BibitemShut
  {NoStop}%
\bibitem [{\citenamefont {Ammon}\ \emph {et~al.}(2016)\citenamefont {Ammon},
  \citenamefont {Leiber},\ and\ \citenamefont {Macedo}}]{Ammon:2016szz}%
  \BibitemOpen
  \bibfield  {author} {\bibinfo {author} {\bibfnamefont {M.}~\bibnamefont
  {Ammon}}, \bibinfo {author} {\bibfnamefont {J.}~\bibnamefont {Leiber}},\ and\
  \bibinfo {author} {\bibfnamefont {R.~P.}\ \bibnamefont {Macedo}},\ }\bibfield
   {title} {\bibinfo {title} {{Phase diagram of 4D field theories with chiral
  anomaly from holography}},\ }\href {https://doi.org/10.1007/JHEP03(2016)164}
  {\bibfield  {journal} {\bibinfo  {journal} {JHEP}\ }\textbf {\bibinfo
  {volume} {03}},\ \bibinfo {pages} {164}},\ \Eprint
  {https://arxiv.org/abs/1601.02125} {arXiv:1601.02125 [hep-th]} \BibitemShut
  {NoStop}%
\bibitem [{\citenamefont {Pynn}\ \emph {et~al.}(2016)\citenamefont {Pynn},
  \citenamefont {Panosso~Macedo}, \citenamefont {Breithaupt}, \citenamefont
  {Palenta},\ and\ \citenamefont {Meinel}}]{Pynn:2016mtw}%
  \BibitemOpen
  \bibfield  {author} {\bibinfo {author} {\bibfnamefont {Y.-C.}\ \bibnamefont
  {Pynn}}, \bibinfo {author} {\bibfnamefont {R.}~\bibnamefont
  {Panosso~Macedo}}, \bibinfo {author} {\bibfnamefont {M.}~\bibnamefont
  {Breithaupt}}, \bibinfo {author} {\bibfnamefont {S.}~\bibnamefont
  {Palenta}},\ and\ \bibinfo {author} {\bibfnamefont {R.}~\bibnamefont
  {Meinel}},\ }\bibfield  {title} {\bibinfo {title} {{Gyromagnetic factor of
  rotating disks of electrically charged dust in general relativity}},\ }\href
  {https://doi.org/10.1103/PhysRevD.94.104035} {\bibfield  {journal} {\bibinfo
  {journal} {Phys. Rev. D}\ }\textbf {\bibinfo {volume} {94}},\ \bibinfo
  {pages} {104035} (\bibinfo {year} {2016})},\ \Eprint
  {https://arxiv.org/abs/1609.08604} {arXiv:1609.08604 [gr-qc]} \BibitemShut
  {NoStop}%
\bibitem [{\citenamefont {{Kalisch}}\ and\ \citenamefont
  {{Ansorg}}(2016)}]{Kalisch2016}%
  \BibitemOpen
  \bibfield  {author} {\bibinfo {author} {\bibfnamefont {M.}~\bibnamefont
  {{Kalisch}}}\ and\ \bibinfo {author} {\bibfnamefont {M.}~\bibnamefont
  {{Ansorg}}},\ }\bibfield  {title} {\bibinfo {title} {{Pseudo-spectral
  construction of non-uniform black string solutions in five and six spacetime
  dimensions}},\ }\href {https://doi.org/10.1088/0264-9381/33/21/215005}
  {\bibfield  {journal} {\bibinfo  {journal} {Classical and Quantum Gravity}\
  }\textbf {\bibinfo {volume} {33}},\ \bibinfo {eid} {215005} (\bibinfo {year}
  {2016})},\ \Eprint {https://arxiv.org/abs/1607.03099} {arXiv:1607.03099
  [gr-qc]} \BibitemShut {NoStop}%
\bibitem [{\citenamefont {Akcay}\ \emph {et~al.}(2012)\citenamefont {Akcay},
  \citenamefont {Barack}, \citenamefont {Damour},\ and\ \citenamefont
  {Sago}}]{Akcay:2012ea}%
  \BibitemOpen
  \bibfield  {author} {\bibinfo {author} {\bibfnamefont {S.}~\bibnamefont
  {Akcay}}, \bibinfo {author} {\bibfnamefont {L.}~\bibnamefont {Barack}},
  \bibinfo {author} {\bibfnamefont {T.}~\bibnamefont {Damour}},\ and\ \bibinfo
  {author} {\bibfnamefont {N.}~\bibnamefont {Sago}},\ }\bibfield  {title}
  {\bibinfo {title} {{Gravitational self-force and the effective-one-body
  formalism between the innermost stable circular orbit and the light ring}},\
  }\href {https://doi.org/10.1103/PhysRevD.86.104041} {\bibfield  {journal}
  {\bibinfo  {journal} {Phys. Rev. D}\ }\textbf {\bibinfo {volume} {86}},\
  \bibinfo {pages} {104041} (\bibinfo {year} {2012})},\ \Eprint
  {https://arxiv.org/abs/1209.0964} {arXiv:1209.0964 [gr-qc]} \BibitemShut
  {NoStop}%
\bibitem [{\citenamefont
  {Zengino{\u{g}}lu}(2010)}]{zenginouglu2010asymptotics}%
  \BibitemOpen
  \bibfield  {author} {\bibinfo {author} {\bibfnamefont {A.}~\bibnamefont
  {Zengino{\u{g}}lu}},\ }\bibfield  {title} {\bibinfo {title} {Asymptotics of
  schwarzschild black hole perturbations},\ }\href@noop {} {\bibfield
  {journal} {\bibinfo  {journal} {Classical and Quantum Gravity}\ }\textbf
  {\bibinfo {volume} {27}},\ \bibinfo {pages} {045015} (\bibinfo {year}
  {2010})}\BibitemShut {NoStop}%
\end{thebibliography}%
\end{document}